\documentclass[english,prd,nofootinbib,preprintnumbers,superscriptaddress]{revtex4-2}
\usepackage[T1]{fontenc}
\usepackage[utf8]{inputenc}
\usepackage{verbatim}
\usepackage{mathrsfs}
\usepackage{amsmath}
\usepackage{ulem}
\usepackage{amssymb}
\usepackage{dsfont}
\usepackage{mathtools}
\usepackage{ifthen}
\usepackage{graphicx}
\usepackage{xfrac}
\usepackage[colorlinks=true,linktocpage=true,linkcolor=blue,citecolor=blue]{hyperref}
\usepackage{babel,oldgerm,combelow,slashed}
\usepackage{xcolor}

\renewcommand{\d}{\mathrm{d}}

\renewcommand{\k}{\mathbf{k}}

\newcommand{\s}{\mathfrak{s}}

\newcommand{\n}{\mathfrak{n}}
\renewcommand{\t}{\mathfrak{t}}
\newcommand{\T}{\mathfrak{T}}

\newcommand{\x}{\mathfrak{x}}

\newcommand{\B}{\mathcal{B}}

\newcommand{\D}{\mathcal{D}}

\newcommand{\F}{\mathcal{F}}
\renewcommand{\H}{\mathcal{H}}

\newcommand{\J}{\mathcal{J}}

\newcommand{\R}{\mathcal{R}}

\renewcommand{\S}{\mathcal{S}}

\newcommand{\W}{\mathcal{W}}

\newcommand{\avg}[1]{\left\langle#1\right\rangle}
\newcommand{\ft}{\widetilde{f}}
\newcommand{\e}{\varepsilon}

\begin{document}

\title{Resummed spin hydrodynamics from quantum kinetic theory}

\author{David Wagner}

\affiliation{Università degli studi di Firenze,\\
Via G. Sansone 1, I-50019 Sesto Fiorentino (Florence), Italy}

\begin{abstract}
In this work, the equations of dissipative relativistic spin hydrodynamics based on quantum kinetic theory are derived. Employing the inverse-Reynolds dominance (IReD) approach, a resummation scheme based on a power counting in Knudsen and inverse Reynolds numbers is constructed, leading to hydrodynamic equations that are accurate to second order.
It is found that the spin dynamics can be characterized by eleven equations: six of them describe the evolution of the components of the spin potential, while the remaining five provide the equation of motion of a dissipative irreducible rank-two tensor. 
For a simple truncation, the first- and second-order transport coefficients are computed explicitly.
\end{abstract}

\maketitle

\section{Introduction and main statement}
It is well known that, if (paramagnetic) matter is subject to a magnetic field, the spins of its constituents will align predominantly with the field, leading to magnetization. Similarly, if matter is under rotation, the spins will align along the rotation axis, which is known as the Barnett effect \cite{Barnett:1935}.

Especially this second effect has attracted much attention in the context of the nonzero global (i.e., momentum-integrated) polarization of $\Lambda$-baryons in heavy-ion collisions \cite{STAR:2017ckg,Adam:2018ivw,ALICE:2019aid}. The idea is that, in noncentral collisions, a large orbital angular momentum is present, that subsequently leads to rotation of the medium and thus to polarization of the emitted particles \cite{Liang:2004ph,Voloshin:2004ha,Betz:2007kg,Becattini:2007sr}.
This simple picture however cannot explain more differential observables, like the so-called local (i.e., angle-dependent) polarization of $\Lambda$ particles, for which additional contributions from fluid-dynamical gradients need to be considered \cite{Becattini:2021iol,Fu:2021pok}.
For this reason, there has been a lot of theoretical efforts in recent years to obtain a more thorough description of spin dynamics \cite{Florkowski:2017ruc,Florkowski:2017dyn,Hidaka:2017auj,Florkowski:2018myy,Weickgenannt:2019dks,Bhadury:2020puc,Weickgenannt:2020aaf,Shi:2020htn,Speranza:2020ilk,Bhadury:2020cop,Singh:2020rht,Bhadury:2021oat,Peng:2021ago,Sheng:2022ssd,Hu:2021pwh,Hu:2022lpi,Fang:2022ttm,Wang:2022yli,Montenegro:2018bcf,Montenegro:2020paq,Gallegos:2021bzp,Hattori:2019lfp,Fukushima:2020ucl,Li:2020eon,She:2021lhe,Wang:2021ngp,Wang:2021wqq,Hongo:2021ona,Singh:2022ltu,Daher:2022xon,Weickgenannt:2022zxs,Weickgenannt:2022jes,Bhadury:2022ulr,Torrieri:2022ogj,Wagner:2022gza,Weickgenannt:2022qvh,Ambrus:2022yzz,Daher:2022wzf, Xie:2023gbo, Shi:2023sxh, Biswas:2023qsw,Wagner:2023cct,Becattini:2023ouz,Weickgenannt:2023btk,Weickgenannt:2023bss,Weickgenannt:2023nge,Bhadury:2023vjx,Daher:2024bah,Florkowski:2024bfw,Banerjee:2024xnd}.

The failure of the argumentation built on the Barnett effect is not too surprising, since it assumes the spins of the emitted particles to be in equilibrium, because only then a direct relation between the polarization of the particles and the rotation of the medium can be inferred. Furthermore, any information on dissipation is discarded. 
Both of these assumptions need a careful reconsideration, since (i) the spin degrees of freedom need a finite time to reach the global-equilibrium state, and (ii) in general dissipative effects may play a role. Indeed, both of these mechanisms are well-known and researched in the nonrelativistic case, in the context of spin relaxation in magnetic fields. In particular, magnetic resonance imaging (MRI) relies fundamentally on nonzero spin-relaxation times, i.e., on point (i). In this context, the spin dynamics are described by the Bloch equations \cite{PhysRev.70.460}. Furthermore, also point (ii) has been considered in the Bloch-Torrey equations \cite{PhysRev.104.563}, which incorporate the effects of diffusion.
In essence, the present efforts are concerned with a viable (i.e., causal and stable) relativistic generalization of these equations.

In the context of point (i), an analysis of ideal-spin hydrodynamics based on quantum kinetic theory has been performed in  Ref. \cite{Wagner:2024fhf}. In such a theory, the spin dynamics are encoded in the so-called spin potential $\Omega_0^{\mu\nu}$, which is the Lagrange multiplier associated to the total angular momentum, and whose evolution equations follow from the conservation of the total angular momentum. In Ref. \cite{Wagner:2024fhf}, it has been found that the components of the spin potential follow damped wave equations, and the damping is determined by nonlocal collisions, which exchange orbital and spin angular momenta and thus can lead to spin equilibration.
Furthermore, the associated relaxation time was found to be potentially large, suggesting that the spin dynamics might play a role in the context of heavy-ion collisions. This result can be compared with, e.g., Refs. \cite{Kapusta:2019sad,Ayala:2019iin,Ayala:2020ndx,Kapusta:2020npk,Hongo:2022izs,Hu:2022xjn,Ayala:2023vgv,Hidaka:2023oze}, which utilized different methods but arrived at similar conclusions.

Regarding the discussion of point (ii), there has been an analysis of the dissipative components of the spin tensor based on quantum kinetic theory in Refs. \cite{Weickgenannt:2022qvh, Weickgenannt:2022zxs}. In these references, it was found that the relaxation times pertaining to dissipative spin degrees of freedom, which are determined by local collisions, are of the same order of magnitude as the timescales associated to standard dissipative quantities. Furthermore, it was found that some of the ``dissipative'' terms do not depend on the absolute strength of the interaction. Lastly, one of these dissipative terms is reminiscent of the so-called ``thermal-shear'' contribution derived in Ref. \cite{Becattini:2021suc} that is crucial to explain the data on local polarization \cite{Becattini:2021iol}. However, further numerical studies are needed to assess whether the quantitative impact is the same.

The main objective of the present paper is to combine the knowledge gained in the aforementioned references and construct a theory of \textit{resummed dissipative spin hydrodynamics}, employing the ``Inverse-Reynolds Dominance'' (IReD) method of Ref. \cite{Wagner:2022ayd}.
The starting point consists in a quantum-kinetic framework that is accurate up to first order in an $\hbar$-gradient expansion and thus captures leading-order spin effects \cite{Weickgenannt:2020aaf,Weickgenannt:2021cuo,Wagner:2022amr,Wagner:2023cct}. From there, an infinite number of coupled partial differential equations for the irreducible moments of the distribution function is derived, containing the information on dissipation in the system.
In order to reduce this infinite-dimensional system of equations to a finite number of fluid-dynamical equations in a controlled manner, a power-counting scheme in a number of dimensionless quantities is developed.
In particular, the Knudsen number (quantifying the separation of microscopic and macroscopic scales) and the inverse Reynolds numbers (encoding the relative magnitude of dissipative effects) are introduced, along with their quantum analogues. 
Assuming that these quantities are small and of the same order of magnitude then allows to derive leading-order relations for the dissipative quantities that can subsequently be used to close the infinite hierarchy of moment equations, obtaining results that are accurate up to second order in Knudsen and inverse Reynolds numbers.
While this procedure may seem rather formal and complicated at first, it results in a clear account of which variables are ``important'' in a hydrodynamic context, and thus leads to valuable simplifications. Most importantly, in comparison to Refs. \cite{Weickgenannt:2022qvh, Weickgenannt:2022zxs}, the number of quantities whose evolution has to be tracked reduces dramatically, from 30 to 11. The equations of motion for these quantities constitute the main result of this work and will be summarized in the following, with the notation being relegated to the end of the introduction section.

In standard fluid dynamics based on kinetic theory, the dissipative currents, i.e., the bulk viscous pressure $\Pi$, the particle diffusion current $n^\mu$, and the shear-stress tensor $\pi^{\mu\nu}$, fulfill coupled relaxation-type equations of the form \cite{Denicol:2012cn, Denicol:2018rbw,Denicol:2019iyh, denicolMicroscopicFoundationsRelativistic2021}
\begin{subequations}
\begin{align}
\tau_\Pi \dot{\Pi}+ \Pi &= -\zeta \theta -\ell_{\Pi n} \nabla_\mu n^\mu - 
 \tau_{\Pi n} n_\mu \dot{u}^\mu - \delta_{\Pi\Pi} \Pi \theta 
 - \lambda_{\Pi n} n_\mu I^\mu + 
 \lambda_{\Pi \pi} \pi^{\mu\nu} \sigma_{\mu\nu}\;,\\
\tau_n \dot{n}^{\langle\mu\rangle} + n^\mu &= \varkappa I^\mu  -\tau_n n_\nu \omega^{\nu\mu} - \delta_{nn} n^\mu \theta 
 - \ell_{n\Pi} \nabla^\mu \Pi + \ell_{n\pi} \Delta^{\mu\nu} \nabla_\lambda \pi^\lambda{}_\nu + \tau_{n\Pi} \Pi \dot{u}^\mu \nonumber\\
 &\quad - 
 \tau_{n\pi} \pi^{\mu\nu} F_\nu -\lambda_{nn} n_\nu \sigma^{\mu\nu} + 
 \lambda_{n \Pi} \Pi I^\mu - \lambda_{n\pi} \pi^{\mu\nu} I_\nu \;,\\
\tau_\pi \dot{\pi}^{\langle\mu\nu\rangle} + \pi^{\mu\nu} &= 2 \eta \sigma^{\mu\nu} + 2\tau_\pi\pi^{\langle\mu}_\lambda \omega^{\nu\rangle \lambda} - 
 \delta_{\pi\pi} \pi^{\mu\nu} \theta - 
 \tau_{\pi\pi} \pi^{\lambda\langle \mu}\sigma^{\nu\rangle}_\lambda + 
 \lambda_{\pi \Pi} \Pi \sigma^{\mu\nu} \nonumber\\
 &\quad - 
 \tau_{\pi n} n^{\langle \mu} \dot{u}^{\nu \rangle} + \ell_{\pi n} \nabla^{\langle \mu} n^{\nu \rangle} 
 + \lambda_{\pi n} n^{\langle \mu} I^{\nu \rangle}\;.
\end{align}
\end{subequations}
Here, the first-order transport coefficients, i.e., the bulk viscosity $\zeta$, the thermal conductivity $\varkappa$, and the shear viscosity $\eta$, determine the proportionality between dissipative quantities  and fluid-dynamical gradients at asymptotically long times. On the other hand, the relaxation times $\tau_\Pi$, $\tau_n$, and $\tau_\pi$, which are typically of the same order of magnitude in terms of the mean free path $\lambda_{\mathrm{mfp}}$, quantify the timescales on which these asymptotic values are approached, while the remaining second-order transport coefficients give rise to various couplings between dissipative quantities and fluid-dynamical gradients.
In the ultrarelativistic limit for a system of particles interacting with a constant cross-section, the explicit values of these coefficients can be found in Ref. \cite{Denicol:2012cn}, as well as in Refs. \cite{Wagner:2022ayd, Wagner:2023joq}. For massless scalar particles subject to a quartic interaction, they are given in Ref. \cite{deBrito:2023vzv}.

The main result of this work is the following additional set of hydrodynamic equations for the components of the spin potential $\omega_0^\mu\coloneqq \sfrac{1}{2}\epsilon^{\mu\nu\alpha\beta}u_\nu \Omega_{0,\alpha\beta}$ and $\kappa_0^\mu\coloneqq u_\nu \Omega_0^{\nu\mu}$, as well as for a traceless symmetric rank-two tensor $\t^{\mu\nu}$,
\begin{subequations}
\label{eqs:eom_omegakappat}
\begin{align}
\tau_\omega \dot{\omega}_0^{\langle\mu\rangle}+\omega_0^\mu &= - \beta_0 \omega^\mu + \epsilon^{\mu\nu\alpha\beta}u_\nu \left(\ell_{\omega\kappa}\nabla_\alpha \kappa_{0,\beta}
-\tau_{\omega}\dot{u}_\alpha \kappa_{0,\beta}
+\lambda_{\omega\kappa}I_\alpha \kappa_{0,\beta}
+\ell_{\omega n}\nabla_\alpha n_{\beta}
+\tau_{\omega n}\dot{u}_\alpha n_{\beta}
+\lambda_{\omega n}I_\alpha n_{\beta}\right)\nonumber\\
&\quad \qquad \quad \; +\delta_{\omega\omega}\omega_0^\mu \theta 
+ \lambda_{\omega\omega}\sigma^{\mu\nu}\omega_{0,\nu}
+\lambda_{\omega \t} \t^{\mu\nu}\omega_\nu \;,\label{eq:eom_omega_hydro}\\
\tau_\kappa \dot{\kappa}_0^{\langle\mu\rangle}+\kappa_0^\mu &= -\beta_0 \dot{u}^\mu +\mathfrak{b}I^\mu + \epsilon^{\mu\nu\alpha\beta}u_\nu \left(\frac{\tau_\kappa}{2}\nabla_\alpha \omega_{0,\beta}+\tau_{\kappa}\dot{u}_\alpha \omega_{0,\beta}+\lambda_{\kappa \omega}I_\alpha \omega_{0,\beta}\right)+\delta_{\kappa\kappa}\kappa_0^\mu \theta 
+\left(\lambda_{\kappa\kappa} \sigma^{\mu\nu}
+\frac{\tau_\kappa}{2} \omega^{\mu\nu}\right)\kappa_{0,\nu}
\nonumber\\
&\quad \qquad \qquad \qquad
+\delta_{\kappa n}n^\mu \theta
+\left(\lambda_{\kappa n} \sigma^{\mu\nu}
+\tau_{\kappa n} \omega^{\mu\nu}\right)n_{\nu}
+\ell_{\kappa n} \dot{n}^{\langle\mu\rangle}
+\t^{\mu\nu}\left(\tau_{\kappa \t}\dot{u}_\nu 
+\lambda_{\kappa\t}I_\nu\right)
+\ell_{\kappa \t}\Delta^\mu_\lambda \nabla_\nu \t^{\nu\lambda}  \;,\label{eq:eom_kappa_hydro}\\
\tau_\t \dot{\t}^{\langle\mu\nu\rangle}+\t^{\mu\nu} &= \mathfrak{d}\beta_0\sigma^{\mu\nu} + \delta_{\t\t} \t^{\mu\nu} \theta 
+\lambda_{\t\t}\t_\lambda{}^{\langle\mu}\sigma^{\nu\rangle\lambda}
+\frac53 \tau_{\t}\t_\lambda{}^{\langle\mu}\omega^{\nu\rangle\lambda}
+\ell_{\t \kappa} \nabla^{\langle\mu} \kappa_0^{\nu\rangle}
+\lambda_{\t\kappa} I^{\langle\mu}\kappa_0^{\nu\rangle}
+\ell_{\t n} \nabla^{\langle\mu} n^{\nu\rangle}
\nonumber\\
&\quad \qquad \quad \; \,
+\tau_{\t n} \dot{u}^{\langle\mu}n^{\nu\rangle}
+\lambda_{\t n} I^{\langle\mu}n^{\nu\rangle} +\tau_{\t\omega} \omega^{\langle\mu}\omega_{0}^{\nu\rangle}
+\lambda_{\t\omega} \sigma_\lambda{}^{\langle\mu}\epsilon^{\nu\rangle\lambda\alpha\beta} u_\alpha \omega_{0,\beta} \;.\label{eq:eom_t_hydro}
\end{align} 
\end{subequations}
These equations completely characterize the evolution of the hydrodynamic degrees of freedom that are relevant for the polarization of particles. 
Note that $\omega_0^\mu$ and $\kappa_0^\mu$ can be considered ``ideal'', in the sense that they constitute the components of the spin potential, which is a local-equilibrium quantity. On the other hand, the tensor $\t^{\mu\nu}$ is a genuinely dissipative quantity that must vanish in local equilibrium.
For a specific interaction in a simple truncation discussed in Sec. \ref{sec:applications}, the values of the transport coefficients appearing in Eqs. \eqref{eq:eom_omega_hydro}, \eqref{eq:eom_kappa_hydro}, and \eqref{eq:eom_t_hydro} are listed in the ultra- and nonrelativistic limits in Tables \ref{tab:coeff_omega}, \ref{tab:coeff_kappa}, and \ref{tab:coeff_t}, respectively, while their behavior for any regime is plotted in Figs. \ref{fig:rel_times}--\ref{fig:coeffs_t}.
In particular, it is interesting to note that the relaxation time $\tau_\omega$ (in units of $\lambda_{\mathrm{mfp}}$) can become much larger than $\tau_\kappa$ and $\tau_\t$, which has been noticed also in Ref. \cite{Wagner:2024fhf}. Furthermore, in the ultrarelativistic limit all couplings to the tensor $\t^{\mu\nu}$ vanish, such that the spin dynamics can be described through $\omega_0^\mu$ and $\kappa_0^\mu$ only.

This paper is structured as follows: Sec. \ref{sec:spinhydro} introduces the conservation equations underlying spin hydrodynamics, while Sec. \ref{sec:kin_description} discusses the kinetic description of the macroscopic currents. Section \ref{sec:moments} describes the method of moments for a system of particles with spin, and Sec. \ref{sec:cons_eq_power_counting} presents the reformulated conservation equations for the variables of ideal-spin hydrodynamics. Section \ref{sec:powercounting} introduces the formal power counting that will be used for carrying out the resummation procedure. In Sec. \ref{sec:mom_eq}, the equations of motion of the irreducible moments are provided, and Sec. \ref{sec:coll} deals with the (linearized) collision terms appearing therein. Section \ref{sec:resum} explains the resummation procedure, finally leading to Eqs. \eqref{eqs:eom_omegakappat}. Lastly, in Sec. \ref{sec:applications} polarization observables are discussed and the values for all transport coefficients in a simple truncation are listed.

Throughout this work, I set $c=k_B=1$, but explicitly keep the reduced Planck constant $\hbar$ as a power-counting parameter around the classical limit. The metric tensor is given by $g^{\mu\nu}=\mathrm{diag}(1,-1,-1,-1)^{\mu\nu}$, and scalar products are denoted by a dot, $A^\mu B_\mu\eqqcolon A\cdot B$. Symmetrization (Antisymmetrization) is indicated by round (square) brackets, i.e., $A^{(\mu}B^{\nu)}\coloneqq A^\mu B^\nu+A^\nu B^\mu$ ($A^{[\mu}B^{\nu]}\coloneqq A^\mu B^\nu-A^\nu B^\mu$).
The fluid four-velocity $u^\mu$ is normalized, $u^2\coloneqq u\cdot u =1$, the projector onto the three-space orthogonal to it is denoted by $\Delta^{\mu\nu}\coloneqq g^{\mu\nu}-u^\mu u^\nu$, and the fourth-rank symmetric traceless projector reads $\Delta^{\mu\nu}_{\alpha\beta}\coloneqq \sfrac{1}{2} \Delta^{(\mu}_\alpha \Delta^{\nu)}_\beta -\sfrac{1}{3}\Delta^{\mu\nu}\Delta_{\alpha\beta}$. A tensor projected onto the traceless subspace orthogonal to $u^\mu$ is denoted with angular brackets, e.g., $A^{\langle\mu\rangle}\coloneqq \Delta^{\mu\nu}A_\nu$ and $A^{\langle\mu\nu\rangle}\coloneqq \Delta^{\mu\nu}_{\alpha\beta}A^{\alpha\beta}$. Similarly, the projector onto the three-space orthogonal to the (on-shell) four-momentum $k^\mu$ is defined as $K^{\mu\nu}\coloneqq g^{\mu\nu}-k^\mu k^\nu/m^2$, and $K^{\mu\nu}_{\alpha\beta}\coloneqq \sfrac{1}{2} K^{(\mu}_\alpha K^{\nu)}_\beta- \sfrac{1}{3} K^{\mu\nu}K_{\alpha\beta}$.
The comoving derivative of a quantity is denoted by a dot, $\dot{A}\coloneqq u\cdot \partial A$. Conversely, the spacelike gradient is defined as $\nabla^\mu \coloneqq \Delta^{\mu\nu}\partial_\nu$. The derivative of the fluid four-velocity can be decomposed as $\partial^\mu u^\nu = \dot{u}^\mu u^\nu + \omega^{\mu\nu}+ \sigma^{\mu\nu}+ \sfrac{1}{3} \theta \Delta^{\mu\nu}$, with the vorticity tensor $\omega^{\mu\nu}\coloneqq \sfrac{1}{2} \nabla^{[\mu} u^{\nu]}$, the shear tensor $\sigma^{\mu\nu}\coloneqq \nabla^{\langle\mu} u^{\nu\rangle}$, and the expansion scalar $\theta\coloneqq \nabla\cdot u$. The vorticity vector of the fluid is defined as $\omega^\mu \coloneqq \sfrac{1}{2}\epsilon^{\mu\nu\alpha\beta} u_\nu \omega_{\alpha\beta}$. The thermal vorticity of the fluid is defined as $\varpi^{\mu\nu}\coloneqq \sfrac{1}{2} \partial^{[\nu} \beta_0 u^{\mu]}$, with the inverse temperature $\beta_0\equiv 1/T$.
Finally, the ratio of chemical potential and temperature reads $\alpha_0\equiv \mu/T$, and its gradient is denoted as $I^\mu \coloneqq \nabla^\mu \alpha_0$.

\section{Hydrodynamics with spin}
\label{sec:spinhydro}
The subject of this work will be a fluid with a single species of particle whose constituents have spin $\sigma\in \{0,\sfrac{1}{2},1\}$.  
The aim of spin hydrodynamics is then to determine the evolution of the conserved currents, i.e., the particle four-current $N^\mu$, the energy-momentum tensor $T^{\mu\nu}$, and the total angular-momentum tensor $J^{\lambda\mu\nu}=\hbar S^{\lambda\mu\nu}+L^{\lambda\mu\nu}$, with $S^{\lambda\mu\nu}$ and $L^{\lambda\mu\nu}\coloneqq T^{\lambda[\nu}x^{\mu]}$ being the spin and the orbital angular momentum tensor, respectively.
These currents fulfill the following conservation equations,
\begin{subequations}
\begin{align}
\partial_\lambda N^\lambda &=0\;,\label{eq:dN}\\
\partial_\lambda T^{\lambda\mu}&=0\;,\label{eq:dT1}\\
\hbar \partial_\lambda S^{\lambda\mu\nu}&= T^{[\nu\mu]}\;.\label{eq:dS}
\end{align}
\end{subequations}
If the fluid is not strongly polarized, one may assume the spin tensor to be small, $S^{\lambda\mu\nu}\sim \mathcal{O}(\hbar)$. Then, it follows from Eq. \eqref{eq:dS} that $T^{[\mu\nu]}\sim\mathcal{O}(\hbar^2)$, which will in the following section also be confirmed explicitly.
As I will work up to first order in quantum corrections, the antisymmetric part of the energy-momentum tensor can be omitted from Eq. \eqref{eq:dT1}, which takes the form
\begin{equation}
\frac12 \partial_\lambda T^{(\lambda\mu)}=0\;.\label{eq:dT2}
\end{equation}
This approximation ignores the feedback of the spin on the evolution of the fluid flow, and as such neglects phenomena due to strong polarization, such as the Einstein-de Haas effect \cite{dehaas:1915}. On the other hand, the effects of the fluid flow on the spin, such as the Barnett effect \cite{Barnett:1935}, are captured to leading order.
It also renders the equations easier to implement, as the spin dynamics are determined on top of a given fluid flow.

In general, the $1+4+6=11$ conservation equations are not enough to uniquely determine the evolution of the conserved currents, which have $4+16+24=44$ components.
This already hints at the necessity to employ a more detailed microscopic description of the fluid, to obtain both the equation of state as well as the evolution of the viscous degrees of freedom.

\section{Kinetic description of conserved currents}
\label{sec:kin_description}
The kinetic description of a system of massive particles with spin $\sigma$ and mass $m>0$ can be derived from quantum field theory via the Wigner-function formalism \cite{Weickgenannt:2019dks,Weickgenannt:2021cuo,Sheng:2021kfc,Wagner:2022amr,Wagner:2022eec,Wagner:2023cct}. In such a formulation, the main dynamical quantity is the single-particle distribution function $f(x,k,\s)$. In contrast to classical kinetic theory, it depends on a ``spin'' variable $\s^\mu$ that encodes the various components of the underlying (matrix-valued) Green function.
To first order in an $\hbar$-expansion around the classical limit, the currents of interest can then be written as
\begin{subequations}
\label{eqs:def_kin_NTS}
\begin{align}
 N^\lambda &=\int \d \Gamma \, k^\lambda f(x,k,\s)\;,\label{eq:def_N}\\
\frac12 T^{(\lambda\mu)}&=\int \d \Gamma \, k^\lambda k^\mu f(x,k,\s) \;,\label{eq:def_T}\\
S^{\lambda\mu\nu}&= \sigma \int \d \Gamma \, k^\lambda \Sigma_\s^{\mu\nu} f(x,k,\s)\;,\label{eq:def_S}
\end{align}
\end{subequations}
where $\Sigma_\s^{\mu\nu}\coloneqq -(1/m) \epsilon^{\mu\nu\alpha\beta}k_\alpha \s_\beta$, and $\d \Gamma\coloneqq \d K\,\d S(k)$ constitutes the combined measure in extended phase space, with $\d K\coloneqq \d^3 k/[(2\pi\hbar)^3k^0]$ and $\d S(k)\coloneqq S_0[m/(\varsigma \pi)] \d^4 \s \, \delta(\s^2 +\varsigma^2)\delta (k\cdot \s)$ being the measures in momentum and spin space, respectively. Depending on the spin of the particles, the constants $S_0$ and $\varsigma$ are chosen differently (see Appendix \ref{app:spin}), such that the first nonvanishing integrals over spin space read
\begin{equation}
\int \d S\, = g\;,\qquad \int \d S \, \s^\mu \s^\nu = -2 K^{\mu\nu}\;,
\label{eq:spin_ints}
\end{equation}
where $g\coloneqq 2\sigma +1$ is the spin degeneracy factor. The integral over an odd number of spin vectors vanishes by symmetry.
Note that the specific form of the spin tensor \eqref{eq:def_S} is related, but not identical, to the so-called Hilgevoord-Wouthuysen (HW) pseudogauge, cf., e.g., Ref. \cite{Weickgenannt:2022jes}.\footnote{In particular, compared to the HW pseudogauge, terms of the form $C\int \d \Gamma k^\lambda k^{[\mu}\partial^{\nu]} f$, with some constant $C$, are omitted. These terms do not contribute to the equations of motion for the spin tensor because of momentum conservation, and they can be removed by a redefinition $S^{\lambda\mu\nu}\to S^{\lambda\mu\nu}+\partial_\rho Z^{\mu\nu\lambda\rho}$, with $Z^{\mu\nu\lambda\rho}=C\int \d \Gamma k^{[\lambda}g^{\rho][\mu}k^{\nu]}f$.}

The time evolution of the single-particle distribution function is determined by the Boltzmann equation,\footnote{Note that here a possible Vlasov-type term emerging from a self-consistent mean field has been neglected \cite{Mrowczynski:1992hq}.}
\begin{equation}
k\cdot \partial f(x,k,\s)=C[f]\;,\label{eq:Boltzmann}
\end{equation}
where
\begin{align}
C[f]&\coloneqq \frac12 \int \d \Gamma_1\, \d \Gamma_2\, \d \Gamma' \, \d \bar{S}(k) (2\pi\hbar)^4 \delta^{(4)}(k+k'-k_1-k_2) \mathcal{W} \nonumber\\
&\quad\times \Big[ f(x+\Delta_1-\Delta,k_1,\s_1)f(x+\Delta_2-\Delta,k_2,\s_2)\ft(x+\Delta'-\Delta,k',\s')\ft(x,k,\bar{\s})\nonumber\\
&\qquad- \ft(x+\Delta_1-\Delta,k_1,\s_1)\ft(x+\Delta_2-\Delta,k_2,\s_2)f(x+\Delta'-\Delta,k',\s')f(x,k,\bar{\s})  \Big]\;.\label{eq:C}
\end{align}
Here, $\mathcal{W}$ is the transition rate, whereas $\Delta^\mu, \Delta'^\mu, \Delta_1^\mu,\Delta_2^\mu$ constitute spacetime shifts that provide the mechanism for the exchange of spin and orbital angular momentum \cite{Weickgenannt:2020aaf,Weickgenannt:2021cuo,Wagner:2022amr,Wagner:2023cct}. These quantities depend on the matrix elements of the underlying quantum field theory, and are listed explicitly in Appendix \ref{app:defs}. Furthermore, $\ft\coloneqq 1-af$ encodes (for $a=1$) the Pauli-blocking and (for $a=-1$) the Bose-enhancement factors, while $a=0$ represents classical statistics.

By using the Boltzmann equation in the equation of motion for the spin tensor, the antisymmetric part of the energy-momentum tensor can be expressed as
\begin{equation}
T^{[\mu\nu]} = \frac{\hbar \sigma}{2} \int \d K \Sigma_\s^{\mu\nu} C[f]\;.\label{eq:T_antisymm_explicit}
\end{equation}

It is crucial to keep in mind that the collision term \eqref{eq:C} is to be understood in a perturbative way, i.e., it can be split into ``local'' and ``nonlocal'' parts, $C[f]=C_{\ell}[f]+C_{n\ell}[f]$, where
\begin{subequations}
\begin{align}
C_{\ell}[f] &\coloneqq  \frac12 \int \d \Gamma_1\, \d \Gamma_2\, \d \Gamma' \, \d \bar{S}(k) (2\pi\hbar)^4 \delta^{(4)}(k+k'-k_1-k_2) \mathcal{W} \left( f_1 f_2 \ft' \ft- \ft_1 \ft_2 f' f\right)\;,\label{eq:C_l}\\
C_{n\ell}[f] &\coloneqq \frac12 \int \d \Gamma_1\, \d \Gamma_2\, \d \Gamma' \, \d \bar{S}(k) (2\pi\hbar)^4 \delta^{(4)}(k+k'-k_1-k_2) \mathcal{W} \Big\{ \left(\Delta_1^\mu -\Delta^\mu\right) \left[(\partial_\mu f_1)f_2 \ft' \ft-(\partial_\mu \ft_1)\ft_2 f' f\right]\nonumber\\
&\quad+\left(\Delta_2^\mu -\Delta^\mu\right) \left[f_1(\partial_\mu f_2) \ft' \ft-\ft_1(\partial_\mu \ft_2) f' f\right]
+\left(\Delta'^\mu -\Delta^\mu\right) \left[f_1 f_2 (\partial_\mu \ft') \ft-\ft_1\ft_2 (\partial_\mu f') f\right]
\Big\}\;,\label{eq:C_nl}
\end{align}
\end{subequations}
with $f_1\equiv f(x,k_1,\s_1)$ and similarly for $f_2$, $f'$, and $f$.
The single-particle distribution function $f=f_\mathrm{eq}+\delta f$ can then be split into a local-equilibrium part $f_\mathrm{eq}$, and a non-equilibrium contribution $\delta f$, where the former is defined by demanding that $C_{\ell}[f_\mathrm{eq}]=0$. Note that, in contrast to the spinless case, $C[f_\mathrm{eq}]\neq 0$ due to the existence of nonlocal collisions \cite{Weickgenannt:2021cuo,Wagner:2024fhf}.
The local-equilibrium distribution function reads explicitly
\begin{equation}
f_\mathrm{eq}(x,k,\s)\coloneqq \left[\exp\left(-\alpha_0+\beta_0 E_\k - \frac{\sigma \hbar}{2} \Omega_{0,\mu\nu}\Sigma_\s^{\mu\nu}\right)+a\right]^{-1}\approx f_{0\k}+f_{0\k}\ft_{0\k}  \frac{\sigma \hbar}{2} \Omega_{0,\mu\nu}\Sigma_\s^{\mu\nu}\;,
\label{eq:f0}
\end{equation}
where the local-equilibrium distribution function without spin was introduced, $f_{0\k}\coloneqq [\exp(-\alpha_0+\beta_0 E_\k)+a]^{-1}$, and $E_\k\coloneqq k\cdot  u$ is the energy in the fluid-rest frame. The quantity $\beta_0\equiv 1/T$ is the inverse temperature, while $\alpha_0 \equiv \mu/T$ can be interpreted as the ratio of chemical potential and temperature. The second-rank tensor $\Omega_0^{\mu\nu}=-\Omega_0^{\nu\mu}$ finally denotes the spin potential, which, in a thermodynamic context, is the intensive quantity conjugate to the total angular momentum of the system, and can be decomposed as
\begin{equation}
\Omega_0^{\mu\nu}=u^{[\mu}\kappa_0^{\nu]}+\epsilon^{\mu\nu\alpha\beta}u_\alpha \omega_{0,\beta}\;,
\end{equation}
with the inverse relations
\begin{equation}
\kappa_0^\mu = u_\nu \Omega_0^{\nu\mu}\;, \qquad \omega_0^\mu = \frac12 \epsilon^{\mu\nu\alpha\beta}u_\nu \Omega_{0,\alpha\beta}\;.
\end{equation}

Then, the currents \eqref{eqs:def_kin_NTS} can be decomposed as
\begin{equation}
N^\lambda = N_0^\lambda + \delta N^\lambda\;,\qquad T^{\lambda\mu}= T_0^{\lambda\mu}+ \delta T^{\lambda\mu}\;,\qquad S^{\lambda\mu\nu}=S_0^{\lambda\mu\nu}+\delta S^{\lambda\mu\nu}\;,
\end{equation}
where
\begin{subequations}
\label{eqs:kin_desc_NTS2}
\begin{alignat}{3}
N^\lambda_0 &\coloneqq\int \d \Gamma \, k^\lambda f_{\mathrm{eq}}(x,k,\s)\;, \qquad \quad &\delta N^\lambda &\coloneqq\int \d \Gamma \, k^\lambda \delta f(x,k,\s)\;,\\
\frac12 T^{(\lambda\mu)}_0&\coloneqq\int \d \Gamma \, k^\lambda k^\mu f_{\mathrm{eq}}(x,k,\s)\;, \qquad \quad &\frac12\delta T^{(\lambda\mu)}&\coloneqq\int \d \Gamma \, k^\lambda k^\mu \delta f(x,k,\s)\;,\\
S_0^{\lambda\mu\nu}&\coloneqq \sigma \int \d \Gamma \, k^\lambda \Sigma_\s^{\mu\nu} f_\mathrm{eq}(x,k,\s)\;, \qquad \quad  &\delta S^{\lambda\mu\nu}&\coloneqq \sigma \int \d \Gamma \, k^\lambda \Sigma_\s^{\mu\nu} \delta f(x,k,\s)\;.
\end{alignat}
\end{subequations}
Note that, to render the system of equations well-posed, the local-equilibrium quantities $\alpha_0, \beta_0, u^\mu$, and $\Omega_0^{\mu\nu}$ need to be defined through the choice of a hydrodynamic frame. In this work, I choose one natural extension of the Landau frame to the case of a fluid with spin,
\begin{equation}
u_\lambda \delta N^\lambda =0\;,\qquad u_\lambda \delta T^{(\lambda\mu)} =0\;,\qquad u_\lambda \delta S^{\lambda\mu\nu}=0\;,
\label{eq:matching_conditions}
\end{equation}
which is similar, but not identical, to the frame used in Ref. \cite{Weickgenannt:2022zxs}.

Inserting the local-equilibrium distribution function \eqref{eq:f0} and defining the thermodynamic integrals
\begin{equation}
I_{nq}(\alpha_0,\beta_0)\coloneqq \frac{1}{(2q+1)!!} \int \d \Gamma \,E_\k^{n-2q}(E_\k^2-m^2)^q f_{0\k}\;,
\label{eq:def_Inq}
\end{equation}
which are computed in Appendix \ref{app:therm_int}, as well as their derivatives $J_{nq}\coloneqq \partial I_{nq}/\partial \alpha_0$, the respective currents read
\begin{subequations}
\label{eq:NTS_0}
\begin{align}
N_0^\lambda &= I_{10} u^\lambda \;,\label{eq:N0}\\
\frac12T_0^{(\lambda\mu)}&= I_{20} u^\lambda u^\mu -I_{21} \Delta^{\lambda\mu}\;,\label{eq:T0}\\
S_0^{\lambda\mu\nu} &=  \frac{2\sigma^2\hbar}{g m^2} \left[-2 u^{\lambda} u^{[\mu} \kappa_0^{\nu]}J_{31} +u^\lambda \epsilon^{\mu\nu\alpha\beta}u_\alpha \omega_{0,\beta}(J_{30}-J_{31}) -u^{[\mu}\epsilon^{\nu]\lambda\alpha\beta}u_\alpha \omega_{0,\beta} J_{31}+\Delta^{\lambda[\mu} \kappa_0^{\nu]}J_{31} \right] \;.\label{eq:S0}
\end{align}
From Eq. \eqref{eq:N0}, the particle-number density $n_0\coloneqq u\cdot N_0=I_{10}$ can be identified, while Eq. \eqref{eq:T0} yields the energy density and isotropic pressure as $\e_0\coloneqq u_\mu u_\nu T_0^{\mu\nu}=I_{20}$ and $P_0\coloneqq -\sfrac{1}{3} \Delta_{\mu\nu}T_0^{\mu\nu}=I_{21}$, respectively.

Similarly, the local-equilibrium contribution of the antisymmetric part of the energy-momentum tensor is found to be
\begin{align}
T_0^{[\mu\nu]}&=\hbar^2\bigg\{\Gamma^{(\omega)} \epsilon^{\mu\nu\alpha\beta}u_\alpha\left(\omega_{0,\sigma}+\beta_0\omega_{\sigma}\right) - \Gamma^{(\kappa)}u^{[\mu}\left[\kappa_0^{\nu]}+\frac{1}{2}\left(\beta_0\dot{u}^{\nu]}-\nabla^{\nu]}\beta_0\right)\right]\nonumber\\
&\qquad+\Gamma^{(I)}u^{[\mu}I^{\nu]} + \frac{\Gamma^{(a)}}{2}u^{[\mu}\left(\beta_0\dot{u}^{\nu]}+\nabla^{\nu]}\beta_0\right)  \bigg\}\;,\label{eq:TA0}
\end{align}
\end{subequations}
with the gradient of chemical potential over temperature being denoted by $I^\mu\coloneqq \nabla^\mu \alpha_0$, respectively. For future use, note the covariant Gibbs-Duhem relation
\begin{equation}
\nabla^\mu \beta_0 = \frac{I^\mu}{h_0} - \beta_0\frac{F^\mu}{\e_0+P_0}\;,
\label{eq:relation_nabla_beta}
\end{equation}
with the enthalpy per particle $h_0\coloneqq (\e_0+P_0)/n_0$ and the pressure gradient $F^\mu\coloneqq \nabla^\mu P_0$. This relation will be continuously used throughout the remainder of this work to replace gradients of the inverse temperature.
The coefficients appearing in Eq. \eqref{eq:TA0} read
\begin{subequations}
\label{eqs:Gammas}
\begin{align}
    \Gamma^{(\omega)}&\coloneqq  \frac{1}{12\hbar} \Delta_{\nu\beta} \Delta_{\mu\alpha} \int_f \Delta^{[\mu}k^{\nu]}\left(\Sigma_{\s_1}^{\alpha\beta}+\Sigma_{\s_2}^{\alpha\beta}-\Sigma_{\s'}^{\alpha\beta}-\Sigma_{\s}^{\alpha\beta}\right)\;,\\
    \Gamma^{(\kappa)}&\coloneqq \frac{1}{6\hbar} u_\mu u_\beta \Delta_{\nu\alpha} \int_f \Delta^{[\mu}k^{\nu]}\left(\Sigma_{\s_1}^{\alpha\beta}+\Sigma_{\s_2}^{\alpha\beta}-\Sigma_{\s'}^{\alpha\beta}-\Sigma_{\s}^{\alpha\beta}\right)\;,\\
    \Gamma^{(I)}&\coloneqq \frac{1}{6\hbar} \Delta_{\alpha\mu}u_\nu \int_f \Delta^\alpha\left(\Sigma_{\s_1}^{\mu\nu}+\Sigma_{\s_2}^{\mu\nu}-\Sigma_{\s'}^{\mu\nu}-\Sigma_{\s}^{\mu\nu}\right)\;,\\
   \Gamma^{(a)}&\coloneqq  \frac{1}{6\hbar}u_\mu u_\beta \Delta_{\nu\alpha} \int_f \Delta^{(\alpha}k^{\beta)}\left(\Sigma_{\s_1}^{\mu\nu}+\Sigma_{\s_2}^{\mu\nu}-\Sigma_{\s'}^{\mu\nu}-\Sigma_{\s}^{\mu\nu}\right)\;,
\end{align}
\end{subequations}
and the coefficients $\Gamma^{(\omega)}$, $\Gamma^{(\kappa)}$, and $\Gamma^{(a)}$ have already been obtained in Ref. \cite{Wagner:2024fhf}, while the coefficient $\Gamma^{(I)}$ appears additionally due to considering a chemical potential. Further calculations are provided in Appendix \ref{app:Tmunu}.
Furthermore, in the expressions above, the abbreviation \cite{Molnar:2013lta}
\begin{equation}
\int_f \coloneqq \frac12 \int \d \Gamma_1\, \d \Gamma_2 \, \d \Gamma' \, \d \Gamma \,\d \bar{S}(k) (2\pi\hbar)^4 \delta^{(4)}(k+k'-k_1-k_2) \mathcal{W} f_{0\k} f_{0\k'} \ft_{0\k_1} \ft_{0\k_2}\;.
\end{equation}
has been introduced. Note that, in contrast to $N_0^\lambda$, $T_0^{(\lambda\mu)}$, and $S_0^{\lambda\mu\nu}$, the antisymmetric part of the energy-momentum tensor in local equilibrium is essentially determined by the nonlocal part of the collision term \cite{Wagner:2024fhf}.

\section{Moment expansion}
\label{sec:moments}
For a system composed of particles of spin $\sigma\leq 1$, the distribution function is at most bilinear in the spin vector $\s^\mu$ \cite{Wagner:2023cct}. Thus, the deviation from local equilibrium can be written as
\begin{equation}
\delta f_{\k\s}= f_{0\k}\widetilde{f}_{0\k} \left(\phi_\k -\s_\mu \zeta^\mu_\k +\s_\alpha \s_\beta K_{\mu\nu}^{\alpha\beta} \xi_\k^{\mu\nu}\right)\;,\label{eq:delta_f_exp_1}
\end{equation}  
where $\phi_\k$, $\zeta_\k$, and $\xi_\k$ are functions of the momentum only.
Upon considering Eqs. \eqref{eqs:kin_desc_NTS2}, it becomes clear that the parts of the distribution function which are bilinear in the spin vector do not contribute directly to any macroscopic current of interest. These parts are related to the tensor polarization of spin-1 particles \cite{Wagner:2022gza}, and they will induce corrections to the transport coefficients of standard dissipative fluid dynamics \cite{hessKineticTheoryDilute1968,hessKineticTheoryDilute1971}
. However, these issues will be treated in future works, and this paper is dedicated to describing the dynamics of the spin tensor. For this reason, I will from now on set $\xi_\k=0$, which is trivially true for spin-0 or spin-$\sfrac{1}{2}$ particles, but an assumption for particles of higher spin.

Orienting on Refs. \cite{Denicol:2012cn,Weickgenannt:2022zxs}, one may further expand\footnote{Since the expansion in irreducible tensors in momentum space is essentially the same as an expansion in spherical harmonics \cite{andersonpolynomials}, the expanded functions are assumed to be square integrable. Recently, it has been shown in the case of Bjorken flow \cite{deBrito:2024qow} that this assumption may not always be satisfied.}
\begin{subequations}
\label{eqs:expansion_k_irred}
\begin{align}
\phi_\k &= \sum_{\ell=0}^\infty \lambda^{\mu_1\cdots \mu_\ell} k_{\langle\mu_1}\cdots k_{\mu_\ell\rangle}\;,\\
\zeta^\mu_\k &= \sum_{\ell=0}^\infty \eta^{\mu,\mu_1\cdots \mu_\ell} k_{\langle\mu_1}\cdots k_{\mu_\ell\rangle}\;,
\end{align}
\end{subequations}
where the tensors $\lambda$ and $\eta$ are functions of the energy in the fluid-rest frame, $E_\k$, only. Equations \eqref{eqs:expansion_k_irred} constitute expansions in terms of the set of irreducible tensors
\begin{equation}
1,\, k^{\langle\mu\rangle},\,k^{\langle\mu}k^{\nu\rangle},\, k^{\langle\mu}k^\nu k^{\lambda\rangle},\,\cdots\;,
\end{equation}
which form a complete and orthogonal basis. Specifically, for any function $F$ depending on $E_\k$ only, there is the orthogonality relation
\begin{equation}
\int \d K k^{\langle\mu_1}\cdots k^{\mu_n \rangle} k_{\langle\nu_1} \cdots k_{\nu_m\rangle} F(E_\k)= \frac{m!\delta_{mn}}{(2m+1)!!} \Delta^{\mu_1\cdots \mu_m}_{\nu_1\cdots\nu_m} \int \d K \left(m^2-E_\k^2\right)^mF(E_\k)\;.
\label{eq:orthogonality}
\end{equation}
The tensors introduced in Eqs. \eqref{eqs:expansion_k_irred} can be further expressed as 
\begin{subequations}
\label{eqs:expansion_coeffs}
\begin{align}
\lambda^{\mu_1\cdots \mu_\ell} &= \sum_{n=0}^\infty c_n^{\mu_1\cdots \mu_\ell}P_{\k n}^{(\ell)}\;,\\
\eta^{\mu,\mu_1\cdots \mu_\ell} &= \sum_{n=0}^\infty d_n^{\mu,\mu_1\cdots \mu_\ell}P_{\k n}^{(\ell)}\;,
\end{align}
\end{subequations}
where the energy polynomials $P_{\k n}^{(\ell)}=\sum_{r=0}^n a^{(\ell)}_{nr} E_\k^r$ fulfill
\begin{equation}
\int \d K \omega^{(\ell)} P^{(\ell)}_{\k m} P^{(\ell)}_{\k n} =\delta_{mn} \;, \qquad \omega^{(\ell)}\coloneqq g \frac{W^{(\ell)}}{(2\ell+1)!!} (m^2-E_\k^2)^\ell f_{0\k}\widetilde{f}_{0\k}\;,
\label{eq:orthogonality_P}
\end{equation}
and the coefficients $a^{(\ell)}_{nr}$ are constructed via Gram-Schmidt orthonormalization, see e.g. Ref. \cite{Denicol:2012cn}.
Setting $P^{(\ell)}_{\k 0}=1$, the normalization is found to be $W^{(\ell)}=(-1)^\ell/J_{2\ell,\ell}$.

Employing the orthogonality relation \eqref{eq:orthogonality} as well as the spin-space integrals \eqref{eq:spin_ints}, the tensors introduced in Eqs. \eqref{eqs:expansion_coeffs} can be expressed as integrals over $\delta f_{\k\s}$,
\begin{subequations}
\label{eqs:cde_integral}
\begin{align}
c_n^{\mu_1\cdots \mu_\ell}&= \frac{W^{(\ell)}}{\ell!} \int \d \Gamma P_{\k n}^{(\ell)} k^{\langle\mu_1}\cdots k^{\mu_\ell\rangle} \delta f_{\k \s} \;,\\
d_n^{\mu,\mu_1\cdots \mu_\ell}&=\frac{g}{2} \frac{W^{(\ell)}}{\ell!} \int \d \Gamma P_{\k n}^{(\ell)} \s^\mu k^{\langle\mu_1}\cdots k^{\mu_\ell\rangle} \delta f_{\k \s} \;.
\end{align}
\end{subequations}
Defining the set of irreducible moments of $\delta f_{\k\s}$,
\begin{subequations}
\label{eqs:def_moments}
\begin{align}
\rho_r^{\mu_1\cdots \mu_\ell}&\coloneqq \int \d \Gamma E_\k^r k^{\langle\mu_1}\cdots k^{\mu_\ell\rangle} \delta f_{\k\s}\;,\label{eq:def_rho}\\
\tau_r^{\mu,\mu_1\cdots \mu_\ell}&\coloneqq \int \d \Gamma E_\k^r \s^\mu  k^{\langle\mu_1}\cdots k^{\mu_\ell\rangle} \delta f_{\k\s}\;,\label{eq:def_tau}
\end{align}
\end{subequations}
the functions \eqref{eqs:expansion_k_irred} can be expressed as
\begin{subequations}
\label{eqs:expr_phizetaxi_moments}
\begin{align}
\phi_\k &= \sum_{\ell=0}^\infty \sum_{n=0}^{N_\ell^{(0)}} \mathcal{H}_{\k n}^{(0,\ell)} \rho_n^{\mu_1\cdots \mu_\ell} k_{\langle\mu_1} \cdots k_{\mu_\ell\rangle}\;,\\
\zeta^\mu_\k &= \sum_{\ell=0}^\infty \sum_{n=0}^{N_\ell^{(1)}} \mathcal{H}_{\k n}^{(1,\ell)} \tau_n^{\mu,\mu_1\cdots \mu_\ell} k_{\langle\mu_1} \cdots k_{\mu_\ell\rangle}\;,
\end{align}
\end{subequations}
with
\begin{equation}
\mathcal{H}_{\k n}^{(j,\ell)} \coloneqq \widetilde{g}^{(j)} \frac{W^{(\ell)}}{\ell!} \sum_{m=0}^{N_\ell^{(j)}} P^{(\ell)}_{\k m} a_{mn}^{(\ell)} \;,\qquad \widetilde{g}^{(j)}\coloneqq \begin{cases} 
1\;&,\; j=0\;,\\
g/2\; &, \; j=1 \;.
\end{cases}
\end{equation}
The quantity $N_\ell^{(j)}$ denotes the truncation of the employed basis.\footnote{As was done in Ref. \cite{Weickgenannt:2022zxs}, the employed basis can in principle by indexed by an arbitrary subset of the natural numbers $\mathbb{S}^{(j)}_\ell$ (and not only of the first $N_\ell^{(j)}$ numbers). For the sake of simplicity, here I consider the latter choice.}

Note that, when combining Eqs. \eqref{eq:delta_f_exp_1}, \eqref{eqs:def_moments}, and  \eqref{eqs:expr_phizetaxi_moments}, any moment can be expressed in terms of all others with the same tensor-rank in momentum and spin,
\begin{subequations}
\label{eqs:completeness_moments}
\begin{align}
\rho_r^{\mu_1\cdots\mu_\ell} &= \sum_{n=0}^{N_\ell^{(0)}} \F_{-r,n}^{(0,\ell)} \rho_n^{\mu_1\cdots \mu_\ell}\;,\\
\tau_r^{\mu,\mu_1\cdots\mu_\ell} &= \sum_{n=0}^{N_\ell^{(1)}} \F_{-r,n}^{(1,\ell)} \tau_n^{\mu,\mu_1\cdots \mu_\ell}\;,
\end{align}
\end{subequations}
with the thermodynamic integrals
\begin{equation}
\F_{rn}^{(j,\ell)}\coloneqq \frac{1}{\widetilde{g}^{(j)}} \frac{\ell!}{(2\ell+1)!!} \int \d \Gamma f_{0\k} \widetilde{f}_{0\k}  E_\k^{-r} \mathcal{H}_{\k n}^{(j,\ell)} \left(m^2-E_\k^2\right)^\ell \;.
\end{equation}
The expressions \eqref{eqs:completeness_moments} are exact if the irreducible moment on the left-hand side is included in the basis, and an approximation in the case that it is not. In particular, Eqs. \eqref{eqs:completeness_moments} can be used to approximate moments with $r<0$.

Not all components of the irreducible moments of nonzero spin-rank are independent due to the spin vector fulfilling $k\cdot \s=0$. To see this, consider the projection of the moments of spin-rank one onto the four-velocity,
\begin{equation}
u_\mu \tau^{\mu,\mu_1\cdots \mu_\ell}_r = - \int \d \Gamma E_\k^{r-1} \s_\mu k^{\langle\mu\rangle} k^{\langle\mu_1}\cdots k^{\mu_\ell\rangle} \delta f_{\k\s} \;,
\label{eq:contract_tau_u}
\end{equation}
and in particular for the first few tensor ranks
\begin{subequations}
\label{eqs:contract_tau_u}
\begin{align}
u_\mu \tau^{\mu}_r&=-\tau_{r-1}^{\mu,}{}_\mu \;,\\
u_\mu \tau^{\mu,\nu}_r&=-\tau_{r-1}^{\mu,\nu}{}_\mu -\frac13 \left(m^2\tau_{r-1}^{\langle\nu\rangle}-\tau_{r+1}^{\langle\nu\rangle}\right)\;,\\
u_\mu \tau_r^{\mu,\nu\lambda}&= -\tau_{r-1}^{\mu,\nu\lambda}{}_\mu -\frac25 \left(m^2 \tau_{r-1}^{\langle\nu,\lambda\rangle}-\tau_{r+1}^{\langle\nu,\lambda\rangle}\right)\;.
\end{align}
\end{subequations}
The expressions above show that the component of $\tau_r^{\mu,\mu_1\cdots\mu_\ell}$ that is parallel to the fluid four-velocity in its first index is not an independent quantity \cite{Weickgenannt:2022zxs}.
Since $\zeta_\k^\mu$ can be taken to be orthogonal to the four-momentum, it holds that $
\s_\mu \zeta_\k^\mu = \s_\mu \Xi^\mu{}_\nu \zeta_\k^{\langle\nu\rangle}$, with the tensor
\begin{equation}
\Xi^{\mu\nu} \coloneqq \Delta^{\mu\nu} +\frac{k^{\langle\mu\rangle} k^{\langle\nu\rangle}}{E_\k^2}\;.
\label{eq:def_Xi}
\end{equation}
In this way, only the independent components of the irreducible moments are retained in the expansion of the single-particle distribution function,
\begin{align}
\delta f_{\k \s} &= f_{0\k} \widetilde{f}_{0\k}\sum_{\ell=0}^\infty k_{\langle\mu_1}\cdots k_{\mu_\ell\rangle} \left(\sum_{n=0}^{N_\ell^{(0)}} \mathcal{H}_{\k n}^{(0,\ell)} \rho_n^{\mu_1\cdots \mu_\ell}-\s^\mu \Xi_{\mu\nu} \sum_{n=0}^{N_\ell^{(1)}}  \mathcal{H}_{\k n}^{(1,\ell)} \tau_n^{\langle\nu\rangle,\mu_1\cdots \mu_\ell}\right)\;.\label{eq:delta_f_expl_2}
\end{align}
Note that, for $\ell>0$, the moments $\tau_r^{\langle\mu\rangle,\mu_1\cdots \mu_\ell}$ are not irreducible, due to the additional Lorentz index coming from the spin vector. It will turn out to be useful to rewrite the moments of tensor-ranks one and two in momentum in terms of irreducible quantities. Specifically, one can decompose
\begin{subequations}
\label{eqs:decomp_tau_23}
\begin{align}
\tau_r^{\langle\mu\rangle,\nu}&= \frac13 \Delta^{\mu\nu}\tau_r^\alpha{}_\alpha + \tau_r^{\langle\mu,\nu\rangle}+\frac12 \epsilon^{\mu\nu\alpha\beta}u_\alpha w_{r,\beta}\;,\label{eq:decomp_tau2}\\
\tau_r^{\langle\mu\rangle,\nu\lambda} &= \frac35 \Delta^{\mu\langle\nu}t_r^{\lambda\rangle}-\frac23 t_{r,\rho}{}^{\langle\lambda}\epsilon^{\nu\rangle\mu\alpha\rho}u_\alpha+\tau_r^{\langle\mu,\nu\lambda\rangle}\;,\label{eq:decomp_tau3}
\end{align}
\end{subequations}
with the irreducible vectors, pseudovectors, and tensors
\begin{equation}
w_r^\mu\coloneqq \epsilon^{\mu\nu\alpha\beta}u_\nu \tau_{r,\alpha,\beta} \;, \qquad t_r^\mu \coloneqq \tau_r^{\alpha,\mu}{}_\alpha \;,\qquad t_r^{\mu\nu}\coloneqq \tau_{r,\alpha,\beta}{}^{\langle\mu}\epsilon^{\nu\rangle\alpha\beta\rho}u_\rho\;.\label{eq:def_w_t_t}
\end{equation}
Finally, the matching conditions \eqref{eq:matching_conditions} read in terms of irreducible moments
\begin{subequations}
\label{eq:matching_moments}
\begin{equation}
\rho_1=\rho_2=0\;,\qquad \rho_1^\mu =0\;,\qquad \tau_1^{[\mu,\nu]}=u^{[\mu}\tau_2^{\nu]} \;.
\end{equation}
Contracting the last matching condition with $u_\mu$ and $\sfrac{1}{2}\epsilon_{\alpha\beta\mu\nu}u^\beta$, respectively, one obtains
\begin{equation}
\frac23 \tau_2^{\langle\mu\rangle}+\frac{m^2}{3}\tau_0^{\langle\mu\rangle}+t_0^\mu=0\;,\qquad w_1^\mu=0\;.
\end{equation}
\end{subequations}

Employing the expansion \eqref{eq:delta_f_expl_2} and the matching conditions \eqref{eq:matching_moments}, the dissipative contributions to the particle four-current and the symmetric part of the energy-momentum tensor become
\begin{subequations}
\label{eq:deltaNTS}
\begin{align}
\delta N^\lambda &= \rho_0^\lambda\;,\label{eq:delta_N}\\
\frac12\delta T^{(\lambda\mu)} &= -\frac{m^2}{3} \Delta^{\lambda\mu} \rho_0+ \rho_0^{\lambda\mu}\;.\label{eq:delta_T}
\end{align}
At this point, one can identify the familiar forms of the particle diffusion $n^\lambda \equiv \rho_0^\lambda$, the bulk viscous pressure $\Pi\equiv -(m^2/3)\rho_0$, and the shear-stress tensor $\pi^{\lambda\mu}\equiv \rho_0^{\lambda\mu}$. 
Additionally invoking the decompositions \eqref{eqs:decomp_tau_23} and the relations \eqref{eqs:contract_tau_u} allows to express the dissipative contributions to the spin tensor and the antisymmetric part of the energy-momentum tensor as
\begin{align}
\delta S^{\lambda\mu\nu} &= 
\frac{\sigma}{m}\left(\epsilon^{\lambda\mu\nu\alpha}u_\alpha \S
-\epsilon^{\langle\lambda\rangle\mu\nu\alpha} \S_\alpha
-\epsilon^{\mu\nu\alpha\beta}u_\alpha \S_\beta{}^{\lambda}
+\t^{\lambda[\mu}u^{\nu]}\right)\;,\label{eq:delta_S}\\
\delta T^{[\mu\nu]}&=\hbar \left[\epsilon^{\mu\nu\alpha\beta}u_\alpha\left(\sum_{n=0}^{N_0^{(1)}}\gamma_n^{(0)}\tau_{n,\beta}+\sum_{n=0}^{N_2^{(1)}}\gamma_n^{(2)}t_{n,\beta}\right)-u^{[\mu}\sum_{n=0}^{N_1^{(1)}}\gamma_n^{(1)}w^{\nu]}_n\right]\;,\label{eq:delta_TA}
\end{align}
\end{subequations}
with the details of the calculations regarding Eq. \eqref{eq:delta_TA} relegated to Appendix \ref{app:Tmunu} and the definitions
\begin{equation}
\S\coloneqq \frac{m^2}{3} \tau_{-1}^\alpha{}_\alpha \;,\quad 
\S^\mu\coloneqq \frac{m^2}{2}\tau_0^{\langle\mu\rangle} \;,\quad 
\S^{\mu\nu} \coloneqq \tau_{-1}^{\alpha,\mu\nu}{}_\alpha +\frac{2m^2}{5} \tau_{-1}^{\langle\mu,\nu\rangle}+\frac{3}{5}\tau_{1}^{\langle\mu,\nu\rangle}\;,\quad
\t^{\mu\nu}\coloneqq t_0^{\mu\nu}\;.
\end{equation}
Note that, while $\S$, $\S^\mu$, and $\S^{\mu\nu}$ are a pseudoscalar, pseudovector, and (traceless symmetric) pseudotensor, respectively, $\t^{\mu\nu}$ is a tensor. 
Furthermore, in the expression for $\delta T^{[\mu\nu]}$, terms originating from the nonlocal part of the collision kernel have been neglected, which will be discussed further in Sec. \ref{sec:coll}.

\section{Conservation equations}
\label{sec:cons_eq_power_counting}
From the conservation laws \eqref{eq:dN} and \eqref{eq:dT2}, evolution equations for $\alpha_0$, $\beta_0$, and $u^\mu$ can be obtained,
\begin{subequations}
\label{eqs:eom_alphabetau}
\begin{align}
\dot{\alpha}_0&=\mathcal{H}\theta +\frac{J_{20}\Pi \theta - J_{30} \partial_\mu n^\mu}{D_{20}}- \frac{J_{20}}{D_{20}}\pi^{\mu\nu}\sigma_{\mu\nu}   \;,\label{eq:eom_alpha}\\
\dot{\beta}_0&= \overline{\mathcal{H}}\theta +\frac{J_{10}\Pi \theta - J_{20} \partial_\mu n^\mu}{D_{20}}- \frac{J_{10}}{D_{20}}\pi^{\mu\nu}\sigma_{\mu\nu}\;,\label{eq:eom_beta}\\
\dot{u}^\mu  &= \frac{1}{\e_0+P_0}\left(F^\mu+\nabla^\mu \Pi -\Pi \dot{u}^\mu+\pi^{\mu\nu}\dot{u}_\nu-\Delta^{\mu\nu}\nabla^\alpha \pi_{\nu\alpha}\right)\;,\label{eq:eom_u}
\end{align}
\end{subequations}
where the explicit kinetic descriptions \eqref{eq:N0}, \eqref{eq:T0}, \eqref{eq:delta_N}, and \eqref{eq:delta_T} have been used. Here, $D_{nq}$ is defined as
\begin{equation}
D_{nq}\coloneqq J_{n+1,q}J_{n-1,q}-J_{nq}^2\;, \label{eq:def_D}
\end{equation}
and the quantities
\begin{equation}
\mathcal{H}\coloneqq \frac{J_{20}(\e_0+P_0)-J_{30}n_0}{D_{20}}\;,\qquad \overline{\mathcal{H}}\coloneqq \frac{J_{10}(\e_0+P_0)-J_{20}n_0}{D_{20}}\label{eq:def_HHbar}
\end{equation}
have been introduced.

Similarly, by inserting Eqs. \eqref{eq:S0}, \eqref{eq:TA0}, \eqref{eq:delta_S}, and \eqref{eq:delta_TA}, the conservation law \eqref{eq:dS} can be mapped into equations of motion for $\kappa_0^\mu$ and $\omega_0^\mu$. The equation for $\kappa_0^\mu$ reads
\begin{subequations}
\label{eqs:eom_Omega_expl_final}
\begin{align}
&\quad\frac{4\sigma^2}{gm^2}J_{31}\dot{\kappa}_0^{\langle\mu\rangle} +\Gamma^{(\kappa)} \kappa_0^\mu+ \sum_{n=0}^{N_1^{(1)}}\gamma^{(1)}_n w_n^\mu \nonumber\\
&= - \Gamma^{(\kappa)}\beta_0 \dot{u}^\mu + \widetilde{\Gamma}^{(I)}I^\mu   +\frac{4\sigma^2}{gm^2}\bigg\{-\kappa_0^\mu\left(K_{31}\dot{\alpha}_0-K_{41}\dot{\beta}_0+\frac43 J_{31} \theta\right)\nonumber\\
&\quad -\frac12 \epsilon^{\mu\nu\alpha\beta}u_\nu\left[J_{30}\dot{u}_\alpha \omega_{0,\beta}-J_{31} \nabla_\alpha \omega_{0,\beta}- \omega_{0,\beta} \left(K_{31}I_\alpha-K_{41}\nabla_\alpha \beta_0 \right)  \right]+\frac12 J_{31}\left(\sigma^{\mu\nu}+\omega^{\mu\nu}\right)\kappa_{0,\nu}\bigg\}\nonumber\\
&\quad- \frac{\sigma}{\hbar m}\left[2 \omega^\mu \S -\epsilon^{\mu\nu\alpha\beta}u_\nu \left(\nabla_\alpha -\dot{u}_\alpha\right)\S_\beta-\epsilon^{\mu\nu\alpha\beta}u_\nu\left(\sigma_{\lambda\alpha}+\omega_{\lambda\alpha}\right)\S_\beta{}^\lambda +\Delta^\mu_\lambda\nabla_\nu \t^{\nu\lambda} -\dot{u}_\nu \t^{\nu\mu}  \right]  \;,\label{eq:eom_kappa_expl_final}
\end{align}
with $K_{nq}\coloneqq \partial J_{nq}/\partial \alpha_0$ and $\widetilde{\Gamma}^{(I)}\coloneqq \Gamma^{(I)}+(\Gamma^{(a)}+\Gamma^{(\kappa)})/(2h_0)$.
Importantly, in Eq. \eqref{eq:eom_kappa_expl_final}, the equation of motion for the four-velocity  \eqref{eq:eom_u} has been used to replace the pressure gradient appearing in Eq. \eqref{eq:TA0}, with the dissipative terms being neglected. The reason for this approximation is that $T_0^{[\mu\nu]}$ is essentially an integral over the nonlocal collision term, and I will not consider dissipative contributions originating from these type of terms, cf. also Eq. \eqref{eq:delta_TA}. Under the same approximation, the equation of motion for $\omega_0^\mu$ is obtained as
\begin{align}
&\quad \frac{2\sigma^2}{gm^2}(J_{30}-J_{31})\dot{\omega}_0^{\langle\mu\rangle}+\Gamma^{(\omega)} \omega_0^\mu +\sum_{n=0}^{N_0^{(1)}}\gamma^{(0)}_n \tau_{n}^\mu +\sum_{n=0}^{N_2^{(1)}}\gamma^{(2)}_n t_{n}^\mu\nonumber\\
&=-\beta_0 \Gamma^{(\omega)} \omega^\mu   -\frac{2\sigma^2}{gm^2}\bigg\{\left[(K_{30}-K_{31})\dot{\alpha}_0-(K_{40}-K_{41})\dot{\beta}_0+\left(J_{30}-\frac13 J_{31}\right)\theta \right]\omega_0^\mu\nonumber\\
&\quad + \epsilon^{\mu\nu\alpha\beta} u_\nu \left[J_{31}\nabla_\alpha \kappa_{0,\beta} + \kappa_{0,\beta} (K_{31}I_\alpha-K_{41}\nabla_\alpha \beta_0)-3J_{31}\dot{u}_\alpha \kappa_{0,\beta}\right]-J_{31} (\sigma^{\mu\nu}+\omega^{\mu\nu})\omega_{0,\nu}\bigg\}\nonumber\\
&\quad -\frac{\sigma}{\hbar m} \left[\dot{u}^\mu \S- \nabla^\mu \S -\left(\sigma^{\mu\nu}+\omega^{\mu\nu}\right)\S_\nu+\frac23 \theta \S^\mu - \Delta^\mu_\lambda\nabla_\nu \S^{\nu\lambda}+\S^{\mu\nu}\dot{u}_\nu - \t^{\mu\nu}\omega_\nu  +\epsilon^{\mu\nu\alpha\beta}u_\nu \t^\lambda{}_\alpha \sigma_{\beta\lambda}\right]  \;.\label{eq:eom_omega_expl_final}
\end{align}
\end{subequations}
In contrast to Eqs. \eqref{eqs:eom_alphabetau}, the equations of motion \eqref{eq:eom_kappa_expl_final} and \eqref{eq:eom_omega_expl_final} constitute coupled relaxation equations. This is the case because the local-equilibrium state \eqref{eq:f0} does not make the nonlocal collision term vanish, and this type of collisions will continue to drive the spin potential $\Omega_0^{\mu\nu}$ towards the thermal vorticity $\varpi^{\mu\nu}\coloneqq \sfrac{1}{2} \partial^{[\nu} \beta_0 u^{\mu]}$.
This behavior has already been observed in Ref. \cite{Wagner:2024fhf} in the case of ideal-spin hydrodynamics, i.e., when $\tau_n^\mu=t_n^\mu=w_n^\mu=0$. The main change when considering dissipative corrections lies in the additional terms linear in the moments of spin-rank one, which fulfill their own equations of motion.

\section{Power-counting scheme}
\label{sec:powercounting}
As is known from the spinless case already \cite{Denicol:2012cn}, the moment equations do not form a closed system, due to the coupling of different irreducible moments to each other. In order to obtain a hydrodynamic theory with a finite number of equations, some sort of resummation has to be performed, which has to be based on a specific type of power-counting, which I will now discuss.
There are four length scales in the problem: the interaction length scale $\ell_{\mathrm{int}}$, the (reduced) Compton wavelength $\lambda_\mathrm{C}\coloneqq \hbar/m$, the mean free path $\lambda_{\mathrm{mfp}}$, and the hydrodynamic length scale $L_\mathrm{hydro}$. For a fluid that can be described via kinetic theory, it must hold that both $\ell_{\mathrm{int}}$ and $\lambda_\mathrm{C}$ are much smaller than the mean free path, such that the quasiparticles may be considered free between collisions. In the following, I will take $\lambda_\mathrm{C}\approx \ell_{\mathrm{int}}$ for simplicity. The ordering of scales is then
\begin{equation}
\lambda_\mathrm{C}\ll \lambda_{\mathrm{mfp}}\ll L_{\mathrm{hydro}}\;,
\end{equation}
where the second inequality describes the fact that collisions inside a fluid cell have to occur frequently enough that it can be considered close to local equilibrium. From these length scales, two independent ratios may be constructed: the standard Knudsen number $\mathrm{Kn}\coloneqq \lambda_\mathrm{mfp}/L_{\mathrm{hydro}}$, and the ``quantum Knudsen number'' $\mathrm{Kn}_Q\coloneqq \lambda_\mathrm{C}/L_{\mathrm{hydro}}\ll \mathrm{Kn}$ 
\cite{Weickgenannt:2022zxs,Weickgenannt:2022qvh,Wagner:2024fhf}. Note that in principle, one can consider an additional length scale $\ell_{\mathrm{vort}}$ related to the vorticity, cf. Refs. \cite{Weickgenannt:2022zxs,Weickgenannt:2022qvh}. In this work, however, I will not consider such a scale; in other words, $\ell_{\mathrm{vort}} \equiv L_\mathrm{hydro}$, such that all fluid-dynamical gradients are taken to be of the same order.

Furthermore, a set of dimensionless quantities which, in a hydrodynamic regime, can be taken to be small, are the inverse Reynolds numbers $\mathrm{Re}^{-1}_\Pi\coloneqq |\Pi|/P_0$, $\mathrm{Re}^{-1}_n\coloneqq |n^\mu|/(\beta_0 P_0)$, and $\mathrm{Re}^{-1}_\pi \coloneqq |\pi^{\mu\nu}|/P_0$. While these numbers need not be the same, in the following I will take $\mathrm{Re}^{-1}_\Pi\approx \mathrm{Re}^{-1}_n \approx \mathrm{Re}^{-1}_\pi \equiv \mathrm{Re}^{-1}$. 
Similar to the quantum Knudsen number, it is sensible to also introduce a set of ``quantum inverse Reynolds numbers'' that describe how large the quantum corrections to the dissipative terms are. Again, I will take $|\tau_r^\mu|/(\beta_0^{2-r}P_0)\approx |\tau_r^{\mu,\nu}|/(\beta_0^{1-r}P_0)\approx |\tau_r^{\mu,\nu\lambda}|/(\beta_0^{-r}P_0)\equiv \mathrm{Re}_Q^{-1}$. 

Lastly, considering that also the equilibrium distribution function \eqref{eq:f0} is expanded perturbatively around the classical limit, and the components of the spin potential follow the relaxation-type equations \eqref{eq:eom_kappa_expl_final} and \eqref{eq:eom_omega_expl_final}, it makes sense to also assign a dimensionless quantity to the magnitude of the spin potential. Alluding to nonrelativistic fluid dynamics, one may refer to this quantity as the ``quantum inverse Rossby number'',\footnote{The analogy is only approximate, as the Rossby number traditionally refers to the Coriolis forces from planetary rotation, and the spin potential is only rigidly tied to the fluid's rotation in global equilibrium.} i.e., set $\mathrm{Ro}_Q^{-1}\coloneqq \beta_0^{-1}\lambda_\mathrm{C}|\Omega_0^{\mu\nu}|$.

In the following, I will place several assumptions that relate the dimensionless quantities introduced above. First, the standard Knudsen and inverse Reynolds number should be of the same magnitude, $\mathrm{Kn}\approx \mathrm{Re}^{-1}$. Second,  a similar relation should hold between the quantum Knudsen and inverse Reynolds number, $\mathrm{Kn}_Q\approx \mathrm{Re}_Q^{-1}$. Third, also the quantum inverse Rossby number should be of the same order of magnitude as the quantum inverse Reynolds number (and thus as the quantum Knudsen number), $\mathrm{Ro}_Q^{-1}\approx \mathrm{Re}_Q^{-1}\approx \mathrm{Kn}_Q$. Considering the definition of $\mathrm{Ro}_Q^{-1}$, this amounts to assigning $|\Omega_0^{\mu\nu}|\sim \mathcal{O}(\beta_0 L^{-1}_{\mathrm{hydro}})\sim |\varpi^{\mu\nu}|$, relating the orders of magnitude of the spin potential and the thermal vorticity. 

With the power counting clarified, the resummation can be performed according to the Inverse-Reynolds Dominance (IReD) method \cite{Wagner:2022ayd}, which is based on the nonrelativistic work by Struchtrup \cite{Struchtrup.2004} and also known as the ``order-of-magnitude approximation'' \cite{Fotakis:2022usk,Rocha:2023hts}. The aim is to derive hydrodynamic equations of motion that are accurate to order $\mathcal{O}(\mathrm{Kn}_Q \mathrm{Re}^{-1})\sim \mathcal{O}(\mathrm{Kn} \mathrm{Re}^{-1}_Q)\sim \mathcal{O}(\mathrm{Re}^{-1}_Q \mathrm{Re}^{-1})$. To achieve this, one first derives asymptotic relations allowing to approximately connect different quantities that are of first order in $\mathrm{Re}_Q^{-1}$. Then, neglecting terms of third and higher orders, these relations are used inside the second-order terms, closing the system.

While the actual resummation procedure will be shown in Sec. \ref{sec:resum}, one can already limit the set of tensors that contribute to the hydrodynamic equations by symmetry considerations. In order for a certain quantity to be important for second-order hydrodynamics, it needs to have a Navier-Stokes value that is of first order in fluid-dynamical gradients, i.e., of order $\mathcal{O}(\mathrm{Kn})$ or $\mathcal{O}(\mathrm{Kn}_Q)$, depending on whether it is a classical or quantum term. The set of these gradients on the other hand is rather limited, consisting of scalar-valued ($\theta$), vector-valued ($\dot{u}^\mu,I^\mu$), axial vector-valued ($\omega^\mu$), and tensor-valued ($\sigma^{\mu\nu}$) terms. 
Given that there is no way to construct a scalar from the irreducible moments $\tau_r^{\mu,\mu_1\cdots\mu_\ell}$ (note that $\tau_r^\alpha{}_\alpha$ is a pseudoscalar, and the interactions are assumed to conserve parity), only vectors, axial vectors, and traceless symmetric tensors can have first-order Navier-Stokes values.
In particular, one can anticipate that, up to first order, one will be able to relate
\begin{equation}
\omega_0^\mu \sim \omega^\mu \;,\quad \tau_r^{\langle\mu\rangle}\sim \omega^\mu \;,\quad t_r^\mu \sim \omega^\mu\;,\qquad \kappa_0^\mu \sim \dot{u}^\mu ,\, I^\mu \;,\quad w_r^\mu \sim \dot{u}^\mu,\,I^\mu \;, \qquad t_r^{\mu\nu}\sim \sigma^{\mu\nu}\;,
\label{eq:rels_axialvectors_tensors}
\end{equation}
with the precise relations being derived in Sec. \ref{sec:resum}. Here and in the following, the symbol ``$\sim$'' shall imply proportionality up to first order.
In contrast, all other quantities do not have Navier-Stokes values of first order, and thus approximately vanish,
\begin{equation}
\tau_r^\alpha{}_\alpha\simeq 0\;,\quad  \tau_r^{\langle\mu,\nu\lambda\rangle}\simeq 0\;,\quad \tau_r^{\mu,\mu_1\cdots\mu_\ell}\simeq 0 \;\forall \; \ell >2\;,\label{eq:asymp_rest}
\end{equation}
with the symbol ``$\simeq$'' denoting equality up to first order.
In particular, this means that the terms $\S$ and $\S^{\mu\nu}$ in Eqs. \eqref{eqs:eom_Omega_expl_final} can be neglected, as they constitute third-order contributions. 
Furthermore, two more contributions in the last line of Eq. \eqref{eq:eom_omega_expl_final} can be neglected to this order, namely the terms $\omega^{\mu\nu}\S_\nu \sim \omega^{\mu\nu}\omega_\nu=0$ as well as $\epsilon^{\mu\nu\alpha\beta}u_\nu \t^\lambda{}_{\alpha} \sigma_{\beta\lambda} \sim \epsilon^{\mu\nu\alpha\beta}u_\nu \sigma^\lambda{}_{\alpha} \sigma_{\beta\lambda}=0$.
In the following, the relations \eqref{eq:rels_axialvectors_tensors} and \eqref{eq:asymp_rest} will be of use to simplify several equations.

\section{Moment equations}
\label{sec:mom_eq}
The evolution equations for the irreducible moments can be obtained from the Boltzmann equation \cite{Denicol:2012cn,Weickgenannt:2022zxs}, which can be rewritten as
\begin{equation}
\delta\dot{f}_{\k\s}= E_\k^{-1} C[f]- \dot{f}_{\mathrm{eq}}- E_\k^{-1} k^{\langle\mu\rangle}\nabla_\mu \left(f_\mathrm{eq}+\delta f_{\k \s}\right)\;.
\label{eq:Boltzmann_decomp}
\end{equation}
Since this work is concerned with the hydrodynamic description of the spin tensor, only the evolution equations for the moments $\tau_r^{\mu,\mu_1\cdots\mu_\ell}$ up to $\ell=2$ are needed. Furthermore, according to the symmetry discussion in the last section, not all components of each irreducible moment are needed. Instead, it is only necessary to obtain the equations of motion for $\tau_r^{\langle\mu\rangle}$, $t_r^\mu$, $w_r^\mu$, and $t_r^{\mu\nu}$. These can be obtained from applying the appropriate projections listed in Eq. \eqref{eq:def_w_t_t}. Explicitly, one has to compute
\begin{subequations}
\label{eqs:contractions_twtdot}
\begin{align}
\dot{t}_r^{\langle\mu\rangle}&\simeq \Delta_{\lambda\nu}\dot{\tau}_r^{\langle\lambda\rangle,\langle\mu\nu\rangle}\;,\\
\dot{w}_r^{\langle\mu\rangle}&=\epsilon^{\mu\nu\alpha\beta}u_\nu \dot{\tau}_{r,\alpha,\beta}-\epsilon^{\mu\nu\alpha\beta}\dot{u}_\nu u_\alpha \left[t_{r-1,\beta}+\frac13 \left(m^2 \tau_{r-1,\beta}-\tau_{r+1,\beta}\right)\right]\;,\\
\dot{t}_r^{\langle\mu\nu\rangle} &\simeq \dot{\tau}_r^{\langle\alpha\rangle,\langle\beta\gamma\rangle} \Delta_\gamma{}^{\delta,\mu\nu}\epsilon_{\delta\alpha\beta\lambda} u^\lambda \;,
\end{align}
\end{subequations}
where the relations \eqref{eq:asymp_rest} have been employed.
With more detailed comments relegated to Appendix \ref{app:mom}, for the moments of rank one in spin and rank zero in momentum, one finds
\begin{align}
\dot{\tau}^{\langle\mu\rangle}_r-C^{\langle\mu\rangle}_{r-1}&\simeq  
\frac{2\sigma\hbar}{gm}\bigg\{\omega_0^\mu  \theta \left[K_{r+1,0}\H-K_{r+2,0}\overline{\H}+J_{r+1,0}+\left(r-\frac23\right)J_{r+1,1}\right]+J_{r+1,1}\sigma^{\mu\nu}\omega_{0,\nu}\nonumber\\
&\quad+J_{r+1,0}\dot{\omega}_0^{\langle\mu\rangle}+J_{r+1,1}\epsilon^{\mu\nu\alpha\beta}u_\nu \nabla_\alpha \kappa_{0,\beta}-\epsilon^{\mu\nu\alpha\beta}u_\alpha \kappa_{0,\beta}\left[J_{r+1,0}\dot{u}_\nu+\left(K_{r+1,1}-\frac{K_{r+2,1}}{h_0}\right)I_\nu\right]\bigg\}\nonumber\\
&\quad+ \frac{1}{2} \epsilon^{\mu\nu\alpha\beta} u_\nu \left(\nabla_\alpha -r\dot{u}_\alpha \right) w_{r-1,\beta} +\frac35(r-1)\sigma^{\mu\nu} t_{r-2,\nu}
-\frac13\left[(r+2)\tau_r^{\langle\mu\rangle}
-(r-1)m^2\tau_{r-2}^{\langle\mu\rangle} \right]\theta \nonumber\\
&+\left(\sigma^{\mu\nu}+\frac{\theta}{3}\Delta^{\mu\nu}\right)\left[t_{r-2,\nu}+\frac13 \left(m^2 \tau_{r-2,\nu}-\tau_{r,\nu}\right)\right]\;.
\label{eq:taudot_0}
\end{align}
Note that the evolution equation for $\alpha_0$, $\beta_0$, and $u^\mu$, i.e., Eqs. \eqref{eq:eom_alpha}, \eqref{eq:eom_beta}, and \eqref{eq:eom_u}, have been used and terms of third or higher order have been neglected. Furthermore, terms of the form $\omega^{\mu\nu}Y_{\nu}$, with $Y^\mu$ being an axial vector (i.e., $\omega_0^\mu$, $\tau_r^{\langle\mu\rangle}$, or $t_r^\mu$) have been omitted, because to second order they are equal to zero, $\omega^{\mu\nu}Y_\nu\sim \omega^{\mu\nu}\omega_\nu=0$, cf. Eq. \eqref{eq:rels_axialvectors_tensors}. Similarly, terms of the form $\epsilon^{\mu\nu\alpha\beta}u_\nu t_{r,\alpha}{}^\rho \sigma_{\rho\beta} \sim \epsilon^{\mu\nu\alpha\beta}u_\nu \sigma_{\alpha}{}^\rho \sigma_{\rho\beta}=0$ have been omitted for symmetry reasons. 
Under the same approximations, the axial vectors $t_r^\mu$ evolve according to
\begin{align}
\dot{t}_r^{\langle\mu\rangle} -C_{t,r-1}^{\mu}
&\simeq \frac{4\sigma \hbar}{gm} \bigg[ \frac56 \left(K_{r+3,2}-\frac{K_{r+4,2}}{h_0}\right) \epsilon^{\mu\alpha\beta\gamma}u_\alpha I_\beta\kappa_{0,\gamma}+\frac56K_{r+3,2}\epsilon^{\mu\alpha\beta\gamma}u_\alpha \nabla_\beta \kappa_{0,\gamma}\nonumber\\
&\quad 
-\frac59 K_{r+3,2}\theta \omega_0^\mu
+\left(\frac56 K_{r+3,2}-\beta_0 K_{r+4,2}\right) \sigma^{\mu\nu}\omega_{0,\nu} \bigg]\nonumber\\
&\quad+\frac{1}{6}\epsilon^{\mu\alpha\beta\gamma}u_\alpha \dot{u}_\beta\left[r m^2w_{r-1,\gamma}
-(r+5) w_{r+1,\gamma}\right]
+\frac{1}{6}\epsilon^{\mu\alpha\beta\gamma}u_\alpha \nabla_\beta\left(w_{r+1,\gamma}
-m^2 w_{r-1,\gamma}\right)\nonumber\\
&\quad+ \frac13\left[(r-1)m^2t_{r-2}^{\mu}-(r+4)t_r^{\mu} \right]\theta
 +\frac{1}{10} \sigma^{\mu\nu}\left[2(r-1)m^2t_{r-2,\nu}-(2r+5)t_{r,\nu} \right] 
 \nonumber\\
&\quad-t_r^{\mu\nu}\omega_\nu+\frac{2}{15} \sigma^{\mu\nu}\left[ (r-1)m^4 \tau_{r-2,\nu}-(2r+3)m^2\tau_{r,\nu}
+(r+4)\tau_{r+2,\nu}\right]\nonumber\\
&\quad - \frac25 \left(\frac16 \sigma^{\mu\nu} +\frac59 \theta \Delta^{\mu\nu}\right)\left[t_{r,\nu}-m^2 t_{r-2,\nu}-\frac13 \left(m^4 \tau_{r-2,\nu}-2m^2 \tau_{r,\nu}+\tau_{r+2,\nu}\right)\right]  \;.
\label{eq:eom_t_axial}
\end{align}
The vectors $w_r^\mu$ fulfill the following evolution equation,
\begin{align}
\dot{w}_r^{\langle\mu\rangle}-C_{w,r-1}^\mu &\simeq-\frac{2\sigma \hbar}{gm} \bigg\{ J_{r+2,1} \left(2\dot{\kappa}_{0}^{\langle\mu\rangle}-2\epsilon^{\mu\nu\alpha\beta}u_\nu\dot{u}_\alpha \omega_{0,\beta}\right)-\left(K_{r+2,1}-\frac{K_{r+3,1}}{h_0}\right)\epsilon^{\mu\nu\alpha\beta}u_\nu I_\alpha \omega_{0,\beta} \nonumber\\
&\quad -J_{r+2,1}\epsilon^{\mu\nu\alpha\beta}u_\nu\nabla_\alpha \omega_{0,\beta}
-\left[\left(2\beta_0 K_{r+3,2}-J_{r+2,1}\right)\sigma^{\mu\nu}+J_{r+2,1}\omega^{\mu\nu}\right]\kappa_{0,\nu}\nonumber\\
&\quad  +2\kappa_0^\mu \theta \left[K_{r+2,1}\H-K_{r+3,1} \overline{\H}+\frac13\left( 5\beta_0 K_{r+3,2}-J_{r+2,1}\right)\right]\bigg\}\nonumber\\
&\quad+\frac12 \omega^{\mu\nu}w_{r,\nu}+\frac13\left[ (r-1)m^2w_{r-2}^{\mu}
-(r+3)w_r^{\mu}\right]\theta +\frac{1}{10}\left[(2r-2)m^2w_{r-2,\nu}
-(2r+3)w_{r,\nu} \right] \sigma^{\mu\nu}\nonumber\\
&\quad +r t_{r-1}^{\mu\nu}\dot{u}_\nu-\nabla_\nu t_{r-1}^{\nu\mu}
- \epsilon^{\mu\nu\alpha\beta}u_\nu \dot{u}_{\alpha} \left\{\frac13\left[ m^2 (r-1) \tau_{r-1,\beta}-(r+2)\tau_{r+1,\beta} \right]-\frac{r+2}{2}t_{r-1,\beta}\right\}\nonumber\\
&\quad+\epsilon^{\mu\nu\alpha\beta}u_\nu \nabla_\alpha \left[\frac13\left( m^2 \tau_{r-1,\beta}-\tau_{r+1,\beta}\right)-\frac12 t_{r-1,\beta}\right]
\;.
\label{eq:eom_w}
\end{align}
Lastly, for the the traceless symmetric tensors $t_r^{\mu\nu}$ it holds that
\begin{align}
\dot{t}_r^{\langle\mu\nu\rangle} -C_{t,r-1}^{\mu\nu}
&\simeq \frac{4\sigma \hbar}{gm} \bigg\{ \frac32\left(K_{r+3,2}-\frac{K_{r+4,2}}{h_0}\right) I^{\langle\mu}\kappa_0^{\nu\rangle}+\frac32 K_{r+3,2}\nabla^{\langle\mu}\kappa_{0}^{\nu\rangle}
+\frac32 K_{r+3,2} \omega_0^{\langle\mu}\omega^{\nu\rangle}
\nonumber\\
&\quad +\left(\frac32 K_{r+3,2}-\beta_0 K_{r+4,2}\right) \sigma_\lambda{}^{\langle\mu}\epsilon^{\nu\rangle\alpha\beta\lambda}u_\alpha \omega_{0,\beta} \bigg\}
+\frac{3}{10}\left[r m^2w_{r-1}^{\langle\mu}
-(r+5) w_{r+1}^{\langle\mu}\right]\dot{u}^{\nu\rangle}
\nonumber\\
&\quad 
+\frac{3}{10}\nabla^{\langle\mu}\left(w_{r+1}^{\nu\rangle}
-m^2 w_{r-1}^{\nu\rangle}\right)
+ \frac13\left[(r-1)m^2t_{r-2}^{\mu\nu}-(r+4)t_r^{\mu\nu} \right]\theta-\frac{9}{10} \omega^{\langle\mu}t_r^{\nu\rangle}+\frac53 t_{r,\lambda}{}^{\langle\mu}\omega^{\nu\rangle\lambda}
 \nonumber\\
&\quad -\frac{1}{10}\sigma_\lambda{}^{\langle\mu}\epsilon^{\nu\rangle\alpha\beta\lambda}u_\alpha \left[2(r-1)m^2t_{r-2,\beta}-(2r+5)t_{r,\beta} \right] 
+\frac{1}{7} \sigma_\lambda{}^{\langle\mu}\left[2(r-1)m^2t_{r-2}^{\nu\rangle\lambda}-(2r+5)t_r^{\nu\rangle\lambda} \right]
\nonumber\\
&\quad
+\frac{2}{15} \sigma_\lambda{}^{\langle\mu}\epsilon^{\nu\rangle\alpha\beta\lambda}u_\alpha\left[ (r-1)m^4 \tau_{r-2,\beta}-(2r+3)m^2\tau_{r,\beta}
+(r+4)\tau_{r+2,\beta}\right]\nonumber\\
&\quad +\frac15 \sigma_\lambda{}^{\langle\mu} \epsilon^{\nu\rangle\alpha\beta\lambda}u_\alpha \left[t_{r,\beta}-m^2 t_{r-2,\beta}-\frac13 \left(m^4 \tau_{r-2,\beta}-2m^2 \tau_{r,\beta}+\tau_{r+2,\beta}\right)\right]\nonumber\\
&\quad+\frac35 \omega^{\langle\mu} \left[t_{r}^{\nu\rangle}-m^2 t_{r-2}^{\nu\rangle}-\frac13 \left(m^4 \tau_{r-2}^{\nu\rangle}-2m^2 \tau_{r}^{\nu\rangle}+\tau_{r+2}^{\nu\rangle}\right)\right]\;.
\label{eq:eom_t_tensor}
\end{align}
The collision terms appearing on the left-hand sides of Eqs. \eqref{eq:eom_t_axial}--\eqref{eq:eom_t_tensor} are defined as
\begin{equation}
C_{t,r-1}^\mu\coloneqq \Delta_{\lambda\nu} C_{r-1}^{\langle\lambda\rangle,\langle\mu\nu\rangle} \;, \qquad
C_{w,r-1}^\mu\coloneqq \epsilon^{\mu\nu\alpha\beta} u_\nu C_{r-1,\alpha,\beta}\;, \qquad 
C_{t,r-1}^{\mu\nu}\coloneqq C_{r-1,\alpha,\beta}{}^{\langle\mu}\epsilon^{\nu\rangle\alpha\beta\rho}u_\rho\;,
\end{equation}
with the irreducible moments of the collision term being
\begin{equation}
C^{\langle\mu\rangle,\langle\mu_1\cdots \mu_\ell\rangle}_{r-1}\coloneqq \int \d \Gamma E_\k^{r-1} \s^{\langle\mu\rangle} k^{\langle\mu_1}\cdots k^{\mu_\ell\rangle}  C[f]\;.\label{eq:def_C_spin1}
\end{equation}
Upon using Eqs. \eqref{eq:C_l} and \eqref{eq:C_nl}, they can be split into local and nonlocal contributions,
\begin{equation}
C^{\langle\mu\rangle,\langle\mu_1\cdots \mu_\ell\rangle}_{r-1} = C^{\langle\mu\rangle,\langle\mu_1\cdots \mu_\ell\rangle}_{\ell,r-1}+ C^{\langle\mu\rangle,\langle\mu_1\cdots \mu_\ell\rangle}_{n\ell,r-1}\;,
\end{equation}
where
\begin{subequations}
\label{eqs:moments_C_l_nl}
\begin{align}
C^{\langle\mu\rangle,\langle\mu_1\cdots \mu_\ell\rangle}_{\ell,r-1}&\coloneqq \int \d \Gamma E_\k^{r-1} \s^{\langle\mu\rangle} k^{\langle\mu_1}\cdots k^{\mu_\ell\rangle}  C_\ell[f] \;,\\
C^{\langle\mu\rangle,\langle\mu_1\cdots \mu_\ell\rangle}_{n\ell,r-1}&\coloneqq \int \d \Gamma E_\k^{r-1} \s^{\langle\mu\rangle} k^{\langle\mu_1}\cdots k^{\mu_\ell\rangle}  C_{n\ell}[f]\;.
\end{align}
\end{subequations}
The moment equations \eqref{eq:taudot_0}--\eqref{eq:eom_t_tensor} are derived analogously to the equations of motion for the moments of spin-rank zero, $\rho_r^{\mu_1\cdots\mu_\ell}$, which can be found, e.g., in Refs. \cite{Denicol:2012cn,deBrito:2024vhm}.

\section{Collision terms}
\label{sec:coll}
In order to be able to carry out the resummation procedure, it is necessary to analyze the structure of the collision terms. In this work, I will consider only those contributions that are at most linear in the deviation from local equilibrium, i.e., linear in $\phi_\k$ and $\zeta^\mu_\k$. Note that, due to the fact that the collision term is accurate only to first order in quantum corrections \cite{Weickgenannt:2021cuo,Wagner:2022amr,Wagner:2023cct}, it is not an additional approximation to linearize in $\zeta_\k^\mu$; however, terms bilinear in $\zeta_\k^\mu$ and $\phi_\k$ as well as of higher order in $\phi_\k$ can (and in general do) appear. I leave these contributions, which can be treated as shown in Ref. \cite{Molnar:2013lta}, to future work.

Upon inserting the local-equilibrium distribution function \eqref{eq:f0} and the deviation \eqref{eq:delta_f_expl_2}, the moments \eqref{eqs:moments_C_l_nl} of the linearized local collision term become
\begin{equation}
C^{\langle\mu\rangle,\langle\mu_1\cdots \mu_\ell\rangle}_{\ell,r-1}=-L^{\langle\mu\rangle,\langle\mu_1\cdots \mu_\ell\rangle}_{\ell,r-1}\coloneqq -\int_f \mathcal{W} E_\k^{r-1} \s^{\langle\mu\rangle} k^{\langle\mu_1}\cdots k^{\mu_\ell\rangle} \left(\s_{1,\alpha} \zeta_{\k_1}^\alpha+\s_{2,\alpha}\zeta_{\k_2}^\alpha-\bar{s}_\alpha\zeta_{\k}^\alpha-\s'_\alpha\zeta_{\k'}^\alpha\right)\;.\label{eq:def_L_l}
\end{equation}
In this expression, only the parts of the distribution function that are linear in the spin vector enter, which is based on the assumption that the spin-space integral of the transition rate weighted with one spin vector vanishes, i.e.,
\begin{equation}
\int [\d S]\,\d  \bar{S}(k)\s_i^\mu \mathcal{W}=0\;,
\end{equation}
where $[\d S]\coloneqq \d S_1(k_1) \, \d S_2(k_2) \, \d S'(k') \, \d S(k)$ and $\s_i\in \{\s_1,\s_2,\s',\s,\bar{\s}\}$.
Upon inserting the moment expansion \eqref{eq:delta_f_expl_2} into Eq. \eqref{eq:def_L_l}, one obtains
\begin{equation}
L^{\langle\mu\rangle,\langle\mu_1\cdots \mu_\ell\rangle}_{\ell,r-1}=\sum_{\ell'=0}^\infty \sum_{n=0}^{N_{\ell'}^{(1)}} \left(\B^{(\ell\ell')}_{rn}\right)^{\mu,\mu_1\cdots\mu_\ell}_{\nu,\nu_1\cdots \nu_{\ell'}} \tau_n^{\langle\nu\rangle,\nu_1\cdots \nu_{\ell'}}\;,
\label{eq:L_moms}
\end{equation}
with the tensors
\begin{align}
\left(\B^{(\ell\ell')}_{rn}\right)^{\mu,\mu_1\cdots\mu_\ell}_{\nu,\nu_1\cdots \nu_{\ell'}}&\coloneqq  \int_f \mathcal{W} E_\k^{r-1} \s^{\langle\mu\rangle} k^{\langle\mu_1}\cdots k^{\mu_\ell\rangle} \left(\mathcal{H}^{(1,\ell')}_{\k_1 n}\s_{1}^{\alpha} \Xi_{1,\alpha\nu}  k_{1,\langle\nu_1}\cdots k_{1,\nu_{\ell'}\rangle}
+\mathcal{H}^{(1,\ell')}_{\k_2 n}\s_{2}^{\alpha} \Xi_{2,\alpha\nu}  k_{2,\langle\nu_1}\cdots k_{2,\nu_{\ell'}\rangle}\right.\nonumber\\
&\left.\hspace{4.1cm}
-\mathcal{H}^{(1,\ell')}_{\k n}\bar{\s}^{\alpha} \Xi_{\alpha\nu}  k_{\langle\nu_1}\cdots k_{\nu_{\ell'}\rangle}
-\mathcal{H}^{(1,\ell')}_{\k n}\s'^{\alpha} \Xi'_{\alpha\nu}  k'_{\langle\nu_1}\cdots k'_{\nu_{\ell'}\rangle}\right)\;,
\end{align}
where $\Xi_i^{\mu\nu}$ denotes the tensor defined in Eq. \eqref{eq:def_Xi}, with the momentum replaced by $k_i\in\{k_1,k_2,k',k\}$.

As stated before, the collision term does not vanish exactly in local equilibrium, which follows from the existence of nonlocal collisions. Indeed, the moments of the nonlocal collision term become
\begin{align}
C^{\langle\mu\rangle,\langle\mu_1\cdots \mu_\ell\rangle}_{n\ell,r-1}&\simeq \int_f \mathcal{W}E_\k^{r-1} \s^{\langle\mu\rangle} k^{\langle\mu_1}\cdots k^{\mu_\ell\rangle}\bigg[-\frac{\sigma\hbar}{2m}\left(\widetilde{\Omega}_{0,\alpha\beta}-\widetilde{\varpi}_{\alpha\beta}\right)\left(k_1^\alpha \s_1^\beta +k_2^\alpha \s_2^\beta -k^\alpha \s^\beta -k'^\alpha \s'^\beta\right) \nonumber\\
&\qquad +(\partial_\alpha \alpha_0)\left(\Delta_1^\alpha +\Delta_2^\alpha -\Delta^\alpha -\Delta'^\alpha\right) -\frac12 \left(\Delta_1^\alpha k_1^\beta+\Delta_2^\alpha k_2^\beta-\Delta^\alpha k^\beta -\Delta'^\alpha k'^\beta\right)\partial_{(\alpha}\beta_{0}u_{\beta)}\bigg]\;,\label{eq:C_0_def}
\end{align}
with 
\begin{equation}
\widetilde{\Omega}_0^{\mu\nu}\coloneqq \epsilon^{\mu\nu\alpha\beta} \Omega_{0,\alpha\beta}=-2u^{[\mu} \omega_0^{\nu]} + 2\epsilon^{\mu\nu\alpha\beta}u_\alpha \kappa_{0,\beta}\;.
\label{eq:def_Omegatilde}
\end{equation}
As has been found in Refs. \cite{Weickgenannt:2022qvh,Weickgenannt:2022zxs}, these collision terms act as Navier--Stokes-type contributions; in other words, they provide the leading-order mechanism that lets the spin-related dissipative quantities respond to fluid-dynamical gradients. Note that the nonlocal contributions inside $T^{[\mu\nu]}_0$ play exactly the same role in Eqs. \eqref{eqs:eom_Omega_expl_final}.

In principle, the nonlocal collision term also contains dissipative contributions. I will neglect these terms in the following, in accordance with the treatment of $T^{[\mu\nu]}$ [cf. Eqs. \eqref{eq:delta_TA} and \eqref{eqs:eom_Omega_expl_final}], but still discuss their structure shortly. Since the spacetime shifts are already a first-order quantum correction, there will not be any contribution from the dissipative terms $\zeta_\k^\mu$, so only $\phi_\k$ will appear. Due to the derivatives appearing in the nonlocal collision term \eqref{eq:C_nl}, the contributions will be twofold: first, there will be terms which are bilinear in irreducible moments $\rho_r^{\mu_1\cdots\mu_\ell}$ and fluid-dynamical gradients, and second, there will be contributions containing the derivative of these irreducible moments. Since the dynamical quantities related to the spin do not appear in these terms, and there is no backreaction of the spin on the standard fluid-dynamical quantities, one can interpret these contributions as higher-order corrections to the Navier--Stokes-type terms \eqref{eq:C_0_def}, and speculate that they will not influence the evolution drastically.

Evaluating the collision terms appearing in Eqs. \eqref{eq:taudot_0}--\eqref{eq:eom_t_tensor}, one finds
\begin{subequations}
\label{eqs:C_1_full}
\begin{align}
C^{\langle\mu\rangle}_{r-1}&=g_r^{(0)}\left(\omega_0^\mu +\beta_0 \omega^\mu\right)-\sum_{n=0}^{N_0^{(1)}}\mathcal{B}^{(0)}_{rn}\tau^{\langle\mu\rangle}_n-\sum_{n=0}^{N_2^{(1)}}\mathcal{B}^{(02)}_{rn}t^\mu_n\;,\label{C_1_0_full}\\
C_{t,r-1}^{\mu}&=g_r^{(2)} \left(\omega_0^\mu +\beta_0 \omega^\mu\right) -\sum_{n=0}^{N_0^{(1)}}\mathcal{B}^{(20)}_{rn}\tau^{\langle\mu\rangle}_n- \sum_{n=0}^{N_2^{(1)}}\mathcal{B}^{(2)}_{rn}t^{\mu}_n\;,\label{C_2_trace}\\
C_{w,r-1}^{\mu}&=g_r^{(\kappa)}\left(\kappa_0^\mu+ \beta_0 \dot{u}^\mu \right) +g_r^{(I)} I^\mu -\sum_{n=0}^{N_1^{(1)}}\mathcal{B}^{(1)}_{rn}w^{\mu}_n\;,\label{C_1_1_dual}\\
C_{t,r-1}^{\mu\nu}&=h_r^{(2)}\beta_0 \sigma^{\mu\nu}-\sum_{n=0}^{\overline{N}_2^{(1)}} \overline{\B}_{rn}^{(2)}t_n^{\mu\nu}\;.\label{C_2_dual}
\end{align}
\end{subequations}
Additional remarks on the derivation of Eqs. \eqref{eqs:C_1_full} as well as the definitions of the coefficients appearing therein are provided in Appendix \ref{app:coll}.
Note that an additional number $\overline{N}_2^{(1)}$ has been introduced, which is sensible to enable different truncation orders in the axial vectors $t_r^\mu$ and the tensors $t_r^{\mu\nu}$.
Interestingly, the collision terms couple the equations for  the axial vectors $\omega_0^\mu$, $\tau_r^{\langle\mu\rangle}$, and $t_r^\mu$ at the linear level. Similarly, they also connect the evolution equations for the vector-valued quantities $\kappa_0^\mu$ and $w_r^\mu$. In the next section, this fact will be of great importance for carrying out the resummation procedure.

\section{Resummation}
\label{sec:resum}
Manifestly, the equations of motion for the axial vector-valued quantities $\omega_0^\mu$, $\tau_r^{\langle\mu\rangle}$, and $t_r^\mu$, i.e., Eqs. \eqref{eq:eom_omega_expl_final}, \eqref{eq:taudot_0}, and \eqref{eq:eom_t_axial}, form a coupled system that has to be treated together. More specifically, neglecting all terms of orders $\mathcal{O}(\mathrm{Kn}_Q \mathrm{Re}^{-1})\sim \mathcal{O}(\mathrm{Kn} \mathrm{Re}^{-1}_Q)\sim \mathcal{O}(\mathrm{Re}^{-1}_Q \mathrm{Re}^{-1})$, one finds
\begin{equation}
\sum_n \begin{pmatrix}
\Gamma^{(\omega)}\delta_{n0}& \gamma_n^{(0)} & \gamma_n^{(2)} \\
-g_r^{(0)}\delta_{n0}&\B^{(0)}_{rn} & \B^{(02)}_{rn}\\
-g_r^{(2)}\delta_{n0}&\B^{(20)}_{rn} & \B^{(2)}_{rn}
\end{pmatrix} \begin{pmatrix}
\omega_0^\mu\\
\tau^{\langle\mu\rangle}_n\\
t^\mu_n
\end{pmatrix} \simeq -\begin{pmatrix}
\Gamma^{(\omega)}\\
-g_r^{(0)}\\
-g_r^{(2)}
\end{pmatrix}\beta_0 \omega^\mu\label{eq:NS_tau_t}\;.
\end{equation}
Inverting the $(3+N^{(1)}_0+N^{(1)}_2)$-dimensional matrix
\begin{equation}
\begin{pmatrix}
\T_A^{(\omega)} & \vec{\T}_{A}^{(\omega \tau)}  &  \vec{\T}_{A}^{(\omega t)}\\
\vec{\T}_{A}^{(\tau\omega)} &\T_{A}^{(\tau)} & \T_{A}^{(\tau t)}\\
\vec{\T}_{A}^{(t\omega)}&\T_{A}^{(t\tau)} & \T_{A}^{(t)}
\end{pmatrix}_{rn} \coloneqq \begin{pmatrix}
\Gamma^{(\omega)}& \vec{\gamma}^{(0)} & \vec{\gamma}^{(2)} \\
-\vec{g}^{\,(0)}&\B^{(0)} & \B^{(02)}\\
-\vec{g}^{\,(2)}&\B^{(20)} & \B^{(2)}
\end{pmatrix}^{-1}_{rn}\;,
\end{equation}
where $\vec{g}^{\,(j)}\coloneqq (g_0^{(j)},\cdots , g_r^{(j)})$, and analogous for $\vec{\gamma}^{(j)}$ and $\vec{\T}_A^{(j)}$, one arrives at the relations
\begin{equation}
\omega_0^\mu \simeq - \beta_0 \omega^\mu \;,\qquad \tau_r^{\langle\mu\rangle} \simeq 0 \;,\qquad 
t_r^{\mu} \simeq 0 \;.\label{eq:asymp_tau_t}
\end{equation}
Interestingly, the dissipative quantities $\tau_r^{\langle\mu\rangle}$ and $t_r^\mu$ are equal to zero to first order. This follows from the fact that the vector on the right-hand side of Eq. \eqref{eq:NS_tau_t} is the same as the first column of the matrix on the left-hand side, or equivalently, from the fact that the magnetic part of the spin potential always appears in the combination $\omega_0^\mu+\beta \omega^\mu$ inside the collision terms.
The asymptotic relations \eqref{eq:asymp_tau_t} can now be used to close the system of equations in terms of $\omega_0^\mu$ alone.

Similarly, the equation of motion for the component $\kappa_0^\mu$ of the spin potential, Eq. \eqref{eq:eom_kappa_expl_final}, has to be combined with the one for the vector-valued moments $w_r^\mu$, Eq. \eqref{eq:eom_w}, yielding
\begin{equation}
\sum_n \begin{pmatrix}
\Gamma^{(\kappa)}\delta_{n0} & \gamma_n^{(1)}\\
-g_r^{(\kappa)}\delta_{n0} & \B^{(1)}_{rn} 
\end{pmatrix}\begin{pmatrix}
\kappa^{\mu}\\
w_n^\mu
\end{pmatrix} \simeq -\begin{pmatrix}
\Gamma^{(\kappa)}\\
-g_r^{(\kappa)}
\end{pmatrix}\beta_0 \dot{u}^\mu + \begin{pmatrix}
\widetilde{\Gamma}^{(I)}\\
g_r^{(I)}
\end{pmatrix}I^\mu \label{eq:NS_eq_w}\;.
\end{equation}
Inverting the $(2+N^{(1)}_1)$-dimensional collision matrix
\begin{equation}
\begin{pmatrix}
\T_V^{(\kappa)} & \vec{\T}_{V}^{(\kappa w)} \\
\vec{\T}_{V}^{(w\kappa)} &\T_{V}^{(w)}
\end{pmatrix}_{rn} \coloneqq \begin{pmatrix}
\Gamma^{(\kappa)}& \vec{\gamma}^{(1)}  \\
-\vec{g}^{\,(\kappa)} & \B^{(1)}
\end{pmatrix}^{-1}_{rn}\;,
\end{equation}
one arrives at
\begin{equation}
\kappa_0^\mu \simeq - \beta_0 \dot{u}^\mu  +\mathfrak{b}I^\mu \;,\qquad
w_r^\mu \simeq  \mathfrak{b}^{(1)}_r I^\mu\;,
\label{eq:NS_w_kappa}
\end{equation}
with
\begin{equation}
\qquad \mathfrak{b} \coloneqq \T_{V}^{(\kappa)}\widetilde{\Gamma}^{(I)}+\sum_{n=0}^{N_1^{(1)}}  \T_{V,n}^{(\kappa w)} g_n^{(I)} \;,\qquad
\mathfrak{b}^{(1)}_r \coloneqq \T_{V,r}^{(w\kappa)}\widetilde{\Gamma}^{(I)}+\sum_{n=0}^{N_1^{(1)}}  \T_{V,rn}^{(w)} g_n^{(I)} \;.
\end{equation}
Note that the coefficients $\mathfrak{b}$ and $\mathfrak{b}^{(1)}_r$ are essentially given by ratios of different collision integrals, and thus the coupling strength of the microscopic interaction will drop out.
To express the moments $w_r^\mu$ in terms of known quantities, one can use the Navier-Stokes relation for the particle diffusion $n^\mu \simeq \varkappa I^\mu$ (with the thermal conductivity $\varkappa$) to obtain
\begin{equation}
w_r^\mu \simeq  B_r n^\mu \;,\label{eq:asymp_w}
\end{equation}
where $B_r\coloneqq \mathfrak{b}_r^{(1)}/\varkappa$ has been defined. As has been discussed in Refs. \cite{Wagner:2022ayd,Wagner:2023jgq}, employing a relation such as Eq. \eqref{eq:asymp_w} that does not contain quantities of first order in the (quantum) Knudsen number on the right-hand side avoids the emergence of potentially problematic terms of second order in the Knudsen number in the hydrodynamic equations, improving the accuracy of the resulting evolution \cite{Wagner:2023jgq, Ambrus:2024qsa}.
Furthermore, in order to implement the matching condition \eqref{eq:matching_moments}, one has to set $B_1\to 0$.

For the traceless symmetric tensors $t_r^{\mu\nu}$, which are not connected with any other quantities, the equations of motion \eqref{eq:eom_t_tensor} truncated at first order are
\begin{equation}
\sum_{n=0}^{\overline{N}_2^{(1)}} \overline{\B}^{(2)}_{rn}t_n^{\mu\nu} \simeq h_r^{(2)} \beta_0 \sigma^{\mu\nu}\;.
\end{equation}
Defining the inverse
\begin{equation}
\T^{(t)}_{T,rn}\coloneqq \left(\overline{\B}^{(2)}\right)^{-1}_{rn}\;,
\end{equation}
one arrives at
\begin{equation}
t_r^{\mu\nu} \simeq \mathfrak{d}_r \beta_0\sigma^{\mu\nu}\;,\label{eq:NS_tmunu}
\end{equation}
with
\begin{equation}
\mathfrak{d}_r\coloneqq \sum_{n=0}^{\overline{N}_2^{(1)}}  \T_{T,rn}^{(t)} h_n^{(2)} \;.
\end{equation}
Again, the coefficients $\mathfrak{d}_r$ are given by ratios of collision integrals, and thus independent of the absolute strength of the microscopic interaction.
In this case, one cannot relate the tensors $t_r^{\mu\nu}$ to any component of the spin potential. Instead, orienting on the spin tensor \eqref{eq:delta_S}, I take $\t^{\mu\nu}$ to be the dynamical quantity, define $D_r\coloneqq \mathfrak{d}_r/\mathfrak{d}$ with $\mathfrak{d}\coloneqq \mathfrak{d}_0$, and relate
\begin{equation}
t_r^{\mu\nu}\simeq  D_r \t^{\mu\nu}\;.\label{eq:asymp_t}
\end{equation}

In addition to Eq. \eqref{eq:asymp_rest}, the asymptotic relations \eqref{eq:asymp_tau_t}, \eqref{eq:asymp_w}, and \eqref{eq:asymp_t} are the key relations that will be used to perform the resummation, as they allow to express all moments appearing in the second-order terms in terms of a very limited set of dynamical quantities, namely (besides the diffusion current $n^\mu$) in terms of $\omega_0^\mu$, $\kappa_0^\mu$, and $\t^{\mu\nu}$. This then reduces the infinite number of moment equations (and the conservation equation of the total angular momentum) to a system of $3+3+5=11$ relaxation-type equations.

After applying the asymptotic relations in the axial-vector valued equations \eqref{eq:eom_omega_expl_final}, \eqref{eq:taudot_0}, and \eqref{eq:eom_t_axial}, one arrives at the hydrodynamic equation for $\omega_0^\mu$. Likewise, from the vector-valued equations \eqref{eq:eom_kappa_expl_final} and \eqref{eq:eom_w} a hydrodynamic evolution equation for $\kappa_0^\mu$ can be  obtained. Lastly, the tensor-valued equations \eqref{eq:eom_t_tensor} lead to an equation of motion for $\t^{\mu\nu}$. 
The resulting set of equations reads
\begin{subequations}
\label{eqs:eom_omegakappat_full}
\begin{align}
\tau_\omega \dot{\omega}_0^{\langle\mu\rangle}+\omega_0^\mu &= -\beta_0\omega^\mu + \J_\omega^\mu +\D_\omega^\mu+ \R_\omega^\mu \;,\\
\tau_\kappa \dot{\kappa}_0^{\langle\mu\rangle} +\kappa_0^\mu &= -\beta_0 \dot{u}^\mu +\mathfrak{b}I^\mu + \J_\kappa^\mu+\D_\kappa^\mu + \R_\kappa^\mu \;,\\
\tau_\t \dot{\t}^{\langle\mu\nu\rangle}+\t^{\mu\nu} &= \mathfrak{d}\beta_0\sigma^{\mu\nu} + \J_\t^{\mu\nu}+\D_\t^{\mu\nu} + \R_\t^{\mu\nu} \;.
\end{align} 
\end{subequations}
In these equations, the terms $\J_\omega^\mu$, $\J_\kappa^\mu$, and $\J^{\mu\nu}_\t$ collect the following contributions,
\begin{subequations}
\label{eqs:def_J}
\begin{align}
\label{eq:def_Jomega}
\J_\omega^\mu &=\epsilon^{\mu\nu\alpha\beta}u_\nu \left(\ell_{\omega\kappa}\nabla_\alpha \kappa_{0,\beta}
+\tau_{\omega\kappa}\dot{u}_\alpha \kappa_{0,\beta}
+\lambda_{\omega\kappa}I_\alpha \kappa_{0,\beta}
+\ell_{\omega n}\nabla_\alpha n_{\beta}
+\tau_{\omega n}\dot{u}_\alpha n_{\beta}
+\lambda_{\omega n}I_\alpha n_{\beta}\right)\nonumber\\
&\quad +\delta_{\omega\omega}\omega_0^\mu \theta 
+ \lambda_{\omega\omega}\sigma^{\mu\nu}\omega_{0,\nu}
+\lambda_{\omega \t} \t^{\mu\nu}\omega_\nu\;,\\
\label{eq:def_Jkappa}
\J_\kappa^\mu &=\epsilon^{\mu\nu\alpha\beta}u_\nu \left(\ell_{\kappa\omega}\nabla_\alpha \omega_{0,\beta}+\tau_{\kappa\omega}\dot{u}_\alpha \omega_{0,\beta}+\lambda_{\kappa \omega}I_\alpha \omega_{0,\beta}\right)+\delta_{\kappa\kappa}\kappa_0^\mu \theta 
+\lambda_{\kappa\kappa} \sigma^{\mu\nu}\kappa_{0,\nu}
+\tau_{\kappa\kappa} \omega^{\mu\nu}\kappa_{0,\nu}\nonumber\\
&\quad 
+\delta_{\kappa n}n^\mu \theta
+\lambda_{\kappa n} \sigma^{\mu\nu}n_{\nu}
+\tau_{\kappa n} \omega^{\mu\nu}n_{\nu}
+\ell_{\kappa n} \dot{n}^{\langle\mu\rangle}
+\tau_{\kappa \t}\t^{\mu\nu}\dot{u}_\nu 
+\lambda_{\kappa\t}\t^{\mu\nu}I_\nu
+\ell_{\kappa \t}\Delta^\mu_\lambda \nabla_\nu \t^{\nu\lambda}\;,\\
\label{eq:def_Jt}
\J_\t^{\mu\nu} &=\delta_{\t\t} \t^{\mu\nu} \theta 
+\lambda_{\t\t}\t_\lambda{}^{\langle\mu}\sigma^{\nu\rangle\lambda}
+\tau_{\t\t}\t_\lambda{}^{\langle\mu}\omega^{\nu\rangle\lambda}
+\ell_{\t \kappa} \nabla^{\langle\mu} \kappa_0^{\nu\rangle}
+\lambda_{\t\kappa} I^{\langle\mu}\kappa_0^{\nu\rangle}
+\ell_{\t n} \nabla^{\langle\mu} n^{\nu\rangle}
+\tau_{\t n} \dot{u}^{\langle\mu}n^{\nu\rangle}\nonumber\\
&\quad
+\lambda_{\t n} I^{\langle\mu}n^{\nu\rangle} +\tau_{\t\omega} \omega^{\langle\mu}\omega_{0}^{\nu\rangle}
+\lambda_{\t\omega} \sigma_\lambda{}^{\langle\mu}\epsilon^{\nu\rangle\lambda\alpha\beta} u_\alpha \omega_{0,\beta}
\;.
\end{align}
\end{subequations}
The transport coefficients appearing here are listed in Appendix \ref{app:coeff}. It turns out that some of them are exactly related to the respective relaxation times. In particular, one has $\tau_{\omega\kappa}=-\tau_\omega$, as well as $\ell_{\kappa\omega}=\tau_{\kappa\omega}/2=\tau_{\kappa\kappa}=\tau_{\kappa}/2$ and $\tau_{\t\t}=5\tau_\t/3$.
Note that the terms which are proportional to the particle diffusion $n^\mu$ are of order $\mathcal{O}(\mathrm{Kn}_Q \mathrm{Re}^{-1})$, whereas all other terms are of order $\mathcal{O}(\mathrm{Kn}\mathrm{Re}^{-1}_Q)$.

The quantities $\D_\omega^\mu$, $\D_\kappa^\mu$, and $\D^{\mu\nu}_\t$ on the other hand contain the dissipative contributions arising from the nonlocal part of the collision term,
\begin{subequations}
\label{eqs:def_D}
\begin{align}
\D_\omega^\mu &=\psi_{\omega \pi}\pi^{\mu\nu}\omega_\nu 
+ \epsilon^{\mu\nu\alpha\beta}u_\nu\left(\psi_{\omega n u} \dot{u}_\alpha n_\beta
+\psi_{\omega n I} I_\alpha n_\beta 
+\psi_{\omega n}\nabla_\alpha n_\beta \right)
\;,\\
\D_\kappa^\mu &= \psi_{\kappa\Pi u} \Pi \dot{u}^\mu+\psi_{\kappa\Pi I} \Pi I^\mu  
+\psi_{\kappa n \theta} \theta n^\mu 
+\psi_{\kappa n \sigma} \sigma^{\mu\nu}n_\nu
+\psi_{\kappa n \omega} \omega^{\mu\nu}n_\nu
+\psi_{\kappa n} \dot{n}^{\langle\mu\rangle} \nonumber\\
&\quad +\psi_{\kappa \pi u} \pi^{\mu\nu}\dot{u}_\nu
+\psi_{\kappa \pi I}\pi^{\mu\nu}I_\nu
+\psi_{\kappa \pi}\Delta^\mu_\lambda \nabla_\nu \pi^{\nu\lambda}
  \;,\\
\D_\t^{\mu\nu} &= \psi_{\t \Pi} \Pi \sigma^{\mu\nu}
+\psi_{\t \pi \theta} \theta \pi^{\mu\nu}
+\psi_{\t \pi \sigma} \pi_\lambda{}^{\langle\mu} \sigma^{\nu\rangle\lambda}
+\psi_{\t \pi \omega} \pi_\lambda{}^{\langle\mu} \omega^{\nu\rangle\lambda}
+\psi_{\t n u} \dot{u}^{\langle\mu}n^{\nu\rangle}
+\psi_{\t n I} I^{\langle\mu}n^{\nu\rangle}
+\psi_{\t n} \nabla^{\langle\mu}n^{\nu\rangle}
\;.
\end{align}
\end{subequations}
Note that all of these quantities are of order $\mathcal{O}(\mathrm{Re}^{-1}\mathrm{Kn}_Q)$. Furthermore, when comparing to Eqs. \eqref{eqs:def_J}, one can see that some tensor structures, namely all those that involve the particle diffusion $n^\mu$, appear both in the $\J$- and the $\D$-terms. However, it is still sensible to separate them, as their origins are quite different, with the former coming from using the asymptotic relation \eqref{eq:asymp_w}, and the latter originating from the dissipative parts of the nonlocal collision term. 

Lastly, the quantities $\R_\omega^\mu$, $\R_\kappa^\mu$, and $\R^{\mu\nu}_\t$ denote the nonlinear contributions from the local collision term,
\begin{subequations}
\label{eqs:def_R}
\begin{align}
\R_\omega^\mu &=\varphi_{\omega \Pi}\Pi \omega_0^\mu 
+ \varphi_{\omega \pi} \pi^{\mu\nu}\omega_{0,\nu}
+ \varphi_{\omega n}\epsilon^{\mu\nu\alpha\beta}u_\alpha  n_\alpha \kappa_{0,\beta} 
\;,\\
\R_\kappa^\mu &= \varphi_{\kappa \Pi} \Pi \kappa_0^\mu 
+\varphi_{\kappa \t n} \t^{\mu\nu} n_\nu 
+\varphi_{\kappa \pi} \pi^{\mu\nu}\kappa_{0,\nu}
+ \varphi_{\kappa \omega n}\epsilon^{\mu\nu\alpha\beta}u_\alpha n_\alpha \omega_{0,\beta}\;,\\
\R_\t^{\mu\nu} &=\varphi_{\t\Pi}\Pi \t^{\mu\nu}
+\varphi_{\t\pi} \t_\lambda{}^{\langle\mu} \pi^{\nu\rangle\lambda} 
+\varphi_{\t n}n^{\langle\mu}\kappa_0^{\nu\rangle}
+\varphi_{\t\omega\pi} \pi_\lambda{}^{\langle\mu}\epsilon^{\nu\rangle\lambda\alpha\beta} u_\alpha \omega_{0,\beta}
\;.
\end{align}
\end{subequations}
These terms are of order $\mathcal{O}(\mathrm{Re}^{-1}\mathrm{Re}_Q^{-1})$. 
Since there is no back-reaction of the spin degrees of freedom onto the fluid evolution, the second-order terms \eqref{eqs:def_D} will merely act as higher-order corrections to the Navier-Stokes values. On the other hand, the terms of second order in inverse Reynolds numbers, Eqs. \eqref{eqs:def_R}, contain the spin degrees of freedom themselves and may thus alter the evolution more significantly. In future works, they can be computed along the lines of Ref. \cite{Molnar:2013lta}. 
According to the approximations introduced in Sec. \ref{sec:coll}, the terms \eqref{eqs:def_D} and \eqref{eqs:def_R} are neglected in this work. Omitting them in Eqs. \eqref{eqs:eom_omegakappat_full} then leads to Eqs. \eqref{eqs:eom_omegakappat} put forward in the introduction.

\section{Applications: A simple truncation}
\label{sec:applications}
The transport coefficients appearing in Eqs
. \eqref{eqs:eom_omegakappat} depend on $4$ numbers that determine the truncation, cf. Appendix \ref{app:coeff}. While $N_0^{(1)}$ and $N_2^{(1)}$ fix how many contributions coming from the axial vectors $\tau_r^{\langle\mu\rangle}$ and $t_r^\mu$ are included that modify the evolution of $\omega_0^\mu$, the number $N_1^{(1)}$ determines how many dissipative corrections coming from the vectors $w_r^\mu$ that influence $\kappa_0^\mu$ are kept. 
Lastly, $\overline{N}_2^{(1)}$ controls the accuracy of the coefficients appearing in the equation of motion for $\t^{\mu\nu}$.
The simplest nontrivial truncation (in the sense that it includes dissipative effects) is obtained by merely neglecting all dissipative corrections to $\omega_0^\mu$ and $\kappa_0^\mu$ and keeping only one tensor-valued moment, i.e., by (symbolically) setting $N_0^{(1)}=N_2^{(1)}=N_1^{(1)}=-1$, and choosing $\overline{N}_2^{(1)}=0$.
Even though this truncation formally implies reducing the transport coefficients related to $\omega_0^\mu$ and $\kappa_0^\mu$ to their ideal-spin counterparts, the system of equations \eqref{eqs:eom_omegakappat} still consists of coupled relaxation-type equations, due to nonlocal collisions driving the relaxation of the spin potential towards the thermal vorticity.

\subsection{Polarization}
The momentum-dependent local polarization is given by the following expression \cite{Weickgenannt:2022zxs},
\begin{equation}
\label{eq:PL_1}
S^\mu(k)=\frac{\sigma}{N(k)} \int \d \Sigma_\lambda k^\lambda \int \d S(k) \s^\mu f(x,k,\s)\;,
\end{equation}
where $\Sigma$ is a spacelike hypersurface and
\begin{equation}
N(k)\coloneqq \int \d \Sigma_\lambda k^\lambda \int \d S(k) f(x,k,\s)\;.
\end{equation}
After inserting the local-equilibrium distribution function \eqref{eq:f0} and performing the spin integration, the local-equilibrium part of the local polarization reads
\begin{subequations}
\begin{align}
S_0^\mu &\coloneqq  \frac{\sigma}{N(k)} \int \d \Sigma_\lambda k^\lambda \int \d S(k) \s^\mu f_\mathrm{eq}(x,k,\s)\nonumber\\
&=\frac{2\sigma^2 \hbar}{N(k) m} \int \d \Sigma_\lambda k^\lambda \left(u^\mu \omega_0^\nu k_\nu - E_\k \omega_0^\mu +\epsilon^{\mu\nu\alpha\beta} u_\nu k_\alpha \kappa_{0,\beta}\right) f_0\ft_0\;.
\end{align} 
On the other hand, by making use of the expansion \eqref{eq:delta_f_expl_2} and inserting the relations 
\eqref{eq:asymp_rest}, \eqref{eq:asymp_tau_t}, \eqref{eq:asymp_w}, and \eqref{eq:asymp_t}, the nonequilibrium contribution to the local polarization is given by
\begin{align}
\delta S^\mu  &\coloneqq  \frac{\sigma}{N(k)} \int \d \Sigma_\lambda k^\lambda \int \d S(k) \s^\mu \delta f_{\k\s}\nonumber\\
&=  -\frac{2\sigma}{N(k)} \int \d \Sigma_\lambda k^\lambda \left( \x_n\epsilon^{\mu\beta\rho\sigma}u_\beta k_{\rho}  n_\sigma+ \x_\t \t_\rho{}^{\beta}\epsilon^{\gamma\mu\sigma\rho}u_\sigma k_{\beta}k_{\gamma} \right) f_0 \ft_0\;,
\end{align}
\end{subequations}
where the coefficients
\begin{equation}
\x_n\coloneqq \frac12 \sum_{n=0}^{N_1^{(1)}} \mathcal{H}_{\k n}^{(1,1)} B_n\;,\qquad
\x_\t\coloneqq \frac23\sum_{n=0}^{\overline{N}_2^{(1)}} \mathcal{H}_{\k n}^{(1,2)} D_n
\end{equation}
have been defined. Note that, due to the nature of the asymptotic relations used, this expression is accurate to first order. In the simple truncation put forward earlier, one has $\x_n=0$, while $\x_\t=(2/3) \mathcal{H}_{\k 0}^{(1,2)}=g/(6J_{42})$. This implies that the contribution due to the tensor $\t^{\mu\nu}$ survives, which is similar to the ``thermal-shear contribution'' found in Refs. \cite{Becattini:2021iol,Fu:2021pok} that can explain the local polarization observed in experiment. However, further (numerical) studies are needed to assess whether the contribution of the tensor $\t^{\mu\nu}$ can have the same effect.

The global polarization on the other hand is given by the integrated version of Eq. \eqref{eq:PL_1},
\begin{equation}
\overline{S}^\mu \coloneqq \frac{\sigma}{\overline{N}} \int \d \Gamma \int \d \Sigma_\lambda k^\lambda \s^\mu f(x,k,\s)\;,
\end{equation}
where $\overline{N}\coloneqq \int \d K \, N(k)$. The local-equilibrium contribution to the global polarization is then obtained as
\begin{subequations}
\begin{equation}
\overline{S}^\mu_0 = -\frac{2\sigma^2 \hbar}{\overline{N}g m} \int \d \Sigma_\lambda \left(J_{21} u^\mu \omega_0^\lambda +J_{20} \omega_0^\mu u^\lambda + J_{21}\epsilon^{\mu\nu\lambda\beta}u_\nu \kappa_{0,\beta} \right)\;,
\end{equation}
while the (first-order accurate) nonequilibrium part reads
\begin{equation}
\delta \overline{S}^\mu = \frac{\sigma}{\overline{N}}  \frac12\int \d \Sigma_\lambda \, B_0 \epsilon^{\mu\lambda\alpha\beta}u_\alpha  n_{\beta}\;.
\end{equation}
\end{subequations}
Note that, after integration, the dependence on the tensor $\t^{\mu\nu}$ vanishes, as has already been noticed in Ref. \cite{Weickgenannt:2022zxs}. In the simple truncation introduced above, the nonequilibrium contributions to the global polarization vanish.

\subsection{Transport coefficients}
\label{subsec:transport_coeffs}
\begin{figure}
\includegraphics[scale=1]{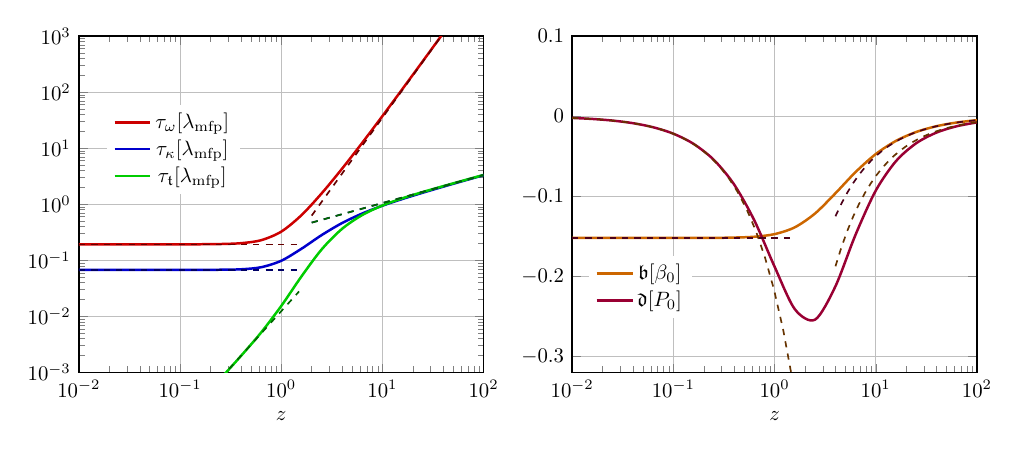}
\caption{The relaxation times for $\omega_0^\mu$, $\kappa_0^\mu$, and $\t^{\mu\nu}$, as well as the first-order coefficients $\mathfrak{b}$ and $\mathfrak{d}$. The ultra- and nonrelativistic limits are indicated by dashed lines. It can be observed that for large $z$ the relaxation time $\tau_\omega$ increases faster than $\tau_\kappa$ and $\tau_\t$ by a factor of $z^2$, in agreement with the results of Ref. \cite{Wagner:2024fhf}.}
\label{fig:rel_times}
\end{figure}

\begin{figure}
\includegraphics[scale=1]{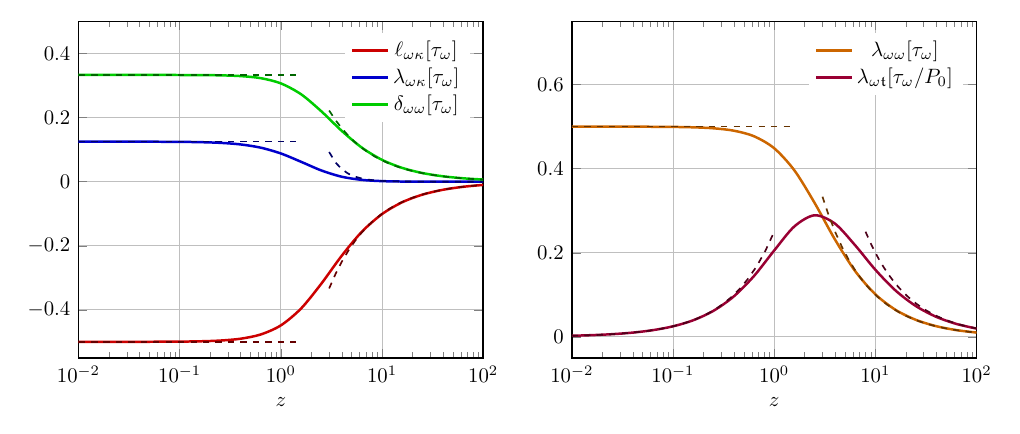}
\caption{The coefficients appearing in the equation for the magnetic part of the spin potential $\omega_0^\mu$ which are not constant in $z$. The ultra- and nonrelativistic limits are indicated by dashed lines.}
\label{fig:coeffs_omega}
\end{figure}

\begin{figure}
\includegraphics[scale=1]{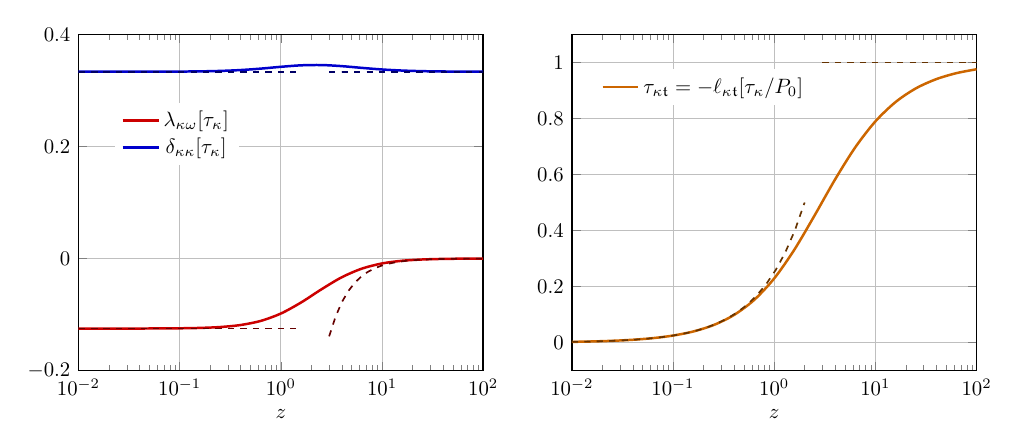}
\caption{The coefficients appearing in the equation for the electric part of the spin potential $\kappa_0^\mu$ which are not constant in $z$. The ultra- and nonrelativistic limits are indicated by dashed lines.}
\label{fig:coeffs_kappa}
\end{figure}

\begin{figure}
\includegraphics[scale=1]{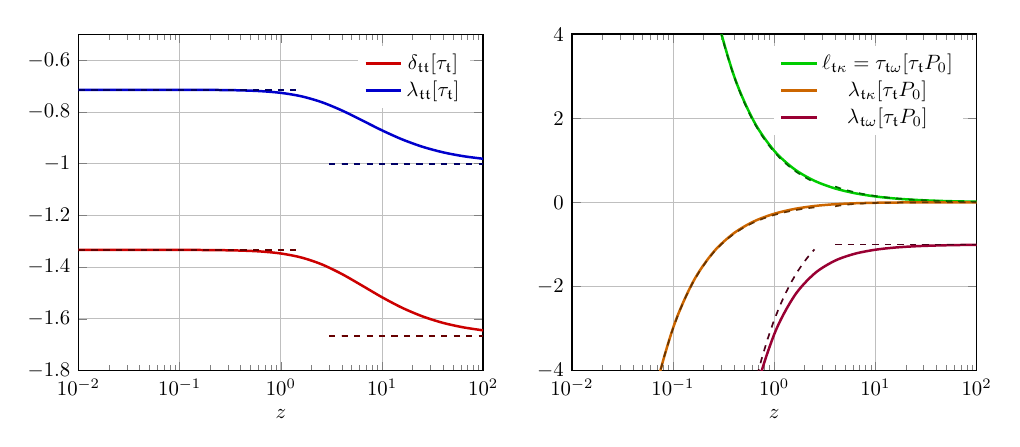}
\caption{The coefficients appearing in the equation for the tensor $\t^{\mu\nu}$ which are not constant in $z$. The ultra- and nonrelativistic limits are indicated by dashed lines.}
\label{fig:coeffs_t}
\end{figure}

In the simple truncation discussed at the beginning of this section, a number of transport coefficients vanish identically. In particular, $\ell_{\omega n}=\tau_{\omega n}=\lambda_{\omega n}=0$ in the equation for $\omega_0^\mu$, while $\delta_{\kappa n}=\lambda_{\kappa n}=\tau_{\kappa n}=\ell_{\kappa n}=\lambda_{\kappa \t}=0$ in the equation for $\kappa_0^\mu$. Finally, in the equation for $\t^{\mu\nu}$ one also has $\ell_{\t n}=\tau_{\t n}=\lambda_{\t n}=0$.
The other (nonvanishing) transport coefficients are computed for a system of $\sigma=\sfrac{1}{2}$-particles subject to a quartic interaction, $\mathcal{L}_\mathrm{int}=G(\overline{\psi}\psi)^2$, as in Ref. \cite{Wagner:2024fhf}.\footnote{For this interaction, the total cross-section is given by $\sigma_T=G^2 (6 m^4 + 6 m^2 P^2 + 5 P^4))/[(12 (m^2 + P^2)\pi \hbar^2]$, which can be found by using, e.g., the real part of Eq. (C2) of Ref. \cite{Wagner:2022amr} and integrating over the solid angle. Using that the typical kinetic momentum is of the order of the temperature \cite{degrootRelativisticKineticTheory1980}, $P\approx T$, one finds $\sigma_T \approx (G^2 T^2/\hbar^2)[6 (z^4+z^2)+5]/[12 \pi (z^2+1)]$. This is then used to compute the mean free path $\lambda_{\text{mfp}}=1/(n_0\sigma_T)$. Note that the factors $(G^2/\hbar^2)$ occur in the relaxation times as well and are thus canceled when expressing the relaxation times in terms of the mean free path.} 
There are two interesting limits that can be considered, namely the ultrarelativistic limit, where $z\coloneqq m\beta_0\to 0$, and the nonrelativistic one, where $z\to \infty$. In Tables \ref{tab:coeff_omega}--\ref{tab:coeff_t}, the transport coefficients are listed in both of these limits. To capture the asymptotic behavior of vanishing or diverging coefficients, I always list the leading contribution in $z$ or $1/z$, respectively.
The relaxation times $\tau_\omega$, $\tau_\kappa$, and $\tau_\t$ as well as the first-order coefficients $\mathfrak{b}$ and $\mathfrak{d}$ are shown in Fig. \ref{fig:rel_times}, and the remaining second-order coefficients which are not constant in $z$ are plotted in Figs. \ref{fig:coeffs_omega}--\ref{fig:coeffs_t}.  For simplicity, all computations are done in the case of Boltzmann statistics. The evaluation of the transport coefficients has been carried out in a  \textsc{Mathematica} notebook \cite{Mathematica,MERTIG1991345,SHTABOVENKO2016432,SHTABOVENKO2020107478}, which can be found in the ancillary files.

A very peculiar behavior can be observed for the ultrarelativistic limit: In the equations for $\kappa_0^\mu$ and $\omega_0^\mu$, the coefficients $\lambda_{\omega\t}$, $\tau_{\kappa \t}$, and $\ell_{\kappa\t}$ vanish, while the other terms stay finite. This implies that the dynamics of the spin potential decouple from the dissipative tensor $\t^{\mu\nu}$ in this limit, and follow the ideal-spin hydrodynamic equations that have been analyzed for a static fluid in Ref. \cite{Wagner:2024fhf}. On the other hand, in the equation of motion for $\t^{\mu\nu}$, even though the coefficients $\lambda_{\t\kappa}$, $\tau_{\t\omega}$, and $\lambda_{\t\omega}$ (in units of $\tau_\t P_0$) diverge as $1/z$, the relaxation time $\tau_\t$ itself vanishes with $z^2$, such that all transport coefficients vanish, and one simply has $\t^{\mu\nu}=0$. These limits suggest that, while the spin relaxation that is responsible for the relaxation of the spin potential towards thermal vorticity (encoded in the relaxation times $\tau_\omega$ and $\tau_\kappa$) still takes place, the dissipative components of the spin tensor are suppressed. In other words, in the terminology employed here and in Ref. \cite{Wagner:2024fhf}, one can say that the ultrarelativistic spin fluid is also an ideal-spin fluid.

In closing, it should be mentioned that the form of the ultrarelativistic limit raises some further questions. Since the quantum kinetic theory used as the microscopic foundation of spin hydrodynamics assumes the particles to be massive, the ultrarelativistic limit cannot easily be taken at any point during the calculation, e.g., due to factors of $1/m$ appearing in the local-equilibrium distribution function \eqref{eq:f0}. However, in the fluid-dynamical theory discussed above, the limit is well-behaved. What remains to be seen is whether the hydrodynamic theory emerging from a microscopic model of massless particles, e.g., chiral kinetic theory, matches the ultrarelativistic limit of the equations derived here. I leave these investigations for future work.

\begin{table}
\begin{tabular}{|c|c|c|c|c|c|c|c|}
\hline
&$\tau_\omega[\lambda_{\text{mfp}}]$ &$\ell_{\omega\kappa}[\tau_\omega]$  &$\lambda_{\omega\kappa}[\tau_\omega]$ &$\delta_{\omega \omega}[\tau_\omega]$ &$\lambda_{\omega \omega}[\tau_\omega]$ &$\lambda_{\omega \t}[\tau_\omega/P_0]$ \\ \hline
$z\to 0$&$5/26$&$-1/2$&$1/8$&$1/3$&$1/2$&$z/4$\\\hline
$z\to \infty$&$\sqrt{\pi} z^{5/2}/16$&$-1/z$&$5/(2z^3)$&$2/(3z)$&$1/z$&$2/z$ \\\hline
\end{tabular}
\caption{The nonzero first-and second-order coefficients appearing in the equation for the magnetic part of the spin potential $\omega_0^\mu$, in the ultrarelativistic ($z\to 0$) and nonrelativistic ($z\to \infty$) limits.}\label{tab:coeff_omega}
\end{table}

\begin{table}
\begin{tabular}{|c|c|c|c|c|c|c|c|c|}
\hline
&$\mathfrak{b}[\beta_0]$&$\tau_\kappa[\lambda_{\text{mfp}}]$ &$\lambda_{\kappa\omega}[\tau_\kappa]$ &$\delta_{\kappa\kappa}[\tau_\kappa]$ &$\lambda_{\kappa\kappa}[\tau_\kappa]$ &$\tau_{\kappa\t}[\tau_\kappa/P_0]$ & $\ell_{\kappa\t}[\tau_\kappa/P_0]$ \\ \hline
$z\to 0$&$-45/296$&$5/74$&$-1/8$&$1/3$&$1/2$&$z/4$&$-z/4$\\\hline
$z\to \infty$&$-1/(2z)$&$3\sqrt{\pi} \sqrt{z}/16$&$-5/(4z^2)$&$1/3$&$1/2$&$1$&$-1$ \\\hline
\end{tabular}
\caption{The nonzero first-and second-order coefficients appearing in the equation for the electric part of the spin potential $\kappa_0^\mu$, in the ultrarelativistic ($z\to 0$) and nonrelativistic ($z\to \infty$) limits. The coefficient $\lambda_{\kappa\kappa}$ (in units of $\tau_\kappa$) is independent of $z$.}
\label{tab:coeff_kappa}
\end{table}

\begin{table}
\begin{tabular}{|c|c|c|c|c|c|c|c|c|c|c|}
\hline
&$\mathfrak{d}[P_0]$&$\tau_\t[\lambda_{\text{mfp}}]$ &$\delta_{\t\t}[\tau_\t]$&$\lambda_{\t\t}[\tau_\t]$&$\ell_{\t\kappa}[\tau_\t P_0]$&$\lambda_{\t\kappa}[\tau_\t P_0]$&$\tau_{\t\omega}[\tau_\t P_0]$&$\lambda_{\t\omega}[\tau_\t P_0]$ \\ \hline
$z\to 0$&$-148z/675$&$z^2/81

$&$-4/3$&$-5/7$&$6/(5z)$&$-3/(10z)$&$6/(5z)$&$-14/(5z)$\\\hline
$z\to \infty$&$-3/(4 z)$&$3\sqrt{\pi} \sqrt{z}/16$&$-5/3$&$-1$&$3/(2z)$&$-3/(2z^2)$&$3/(2z)$&$-1$ \\\hline
\end{tabular}
\caption{The nonzero first-and second-order coefficients appearing in the equation for the tensor $\t^{\mu\nu}$, in the ultrarelativistic ($z\to 0$) and nonrelativistic ($z\to \infty$) limits.}
\label{tab:coeff_t}
\end{table}

\section*{Acknowledgments}
I thank F. Becattini, L. Gavassino, D. H. Rischke, M. Shokri, and E. Speranza for fruitful discussions. 
I acknowledge support by the project PRIN2022 Advanced Probes of the Quark Gluon Plasma funded by ”Ministero dell’Università e della Ricerca”.
This work is supported by the Deutsche Forschungsgemeinschaft (DFG, German Research Foundation) through the CRC-TR 211 ``Strong-interaction matter under extreme conditions'' -- project number 315477589 -- TRR 211, and by the State of Hesse within the Research Cluster ELEMENTS (Project ID 500/10.006).

\appendix

\section{Useful integrals}
The purpose of this appendix is to compute some spin-and momentum-space integrals used in the main text. In particular, the construction of arbitrary spin-space integrals is shown in Subsec. \ref{app:spin}, and the thermodynamic integrals $I_{nq}$ are treated in Subsec. \ref{app:therm_int}.
\subsection{Spin-space integrals}
\label{app:spin}
Consider a general integral over spin space,
\begin{equation}
I^{\mu_1\cdots \mu_\n}\coloneqq \int \d S(k)\, \s^{\mu_1} \cdots \s^{\mu_n}\;,\label{eq:general_int}
\end{equation}
where the measure, as introduced in the main text, is given by
\begin{equation}
\d S(k)\coloneqq S_0 \frac{m}{\varsigma \pi} \d^4 \s \delta(\s^2+\varsigma^2) \delta(k\cdot \s)\;.
\end{equation}
Given that the integral \eqref{eq:general_int} only depends on the momentum $k$ and has to be orthogonal to it in all indices, it can only depend on combinations of the projector $K^{\mu\nu}$, i.e.,
\begin{equation}
I^{\mu_1\cdots \mu_n} = 
\begin{cases}  I^{(n)} K^{(\mu_1\mu_2}\cdots K^{\mu_{n-1}\mu_{n})} &\;,\quad \text{if $n$ is even}\;,\\
0&\;,\quad \text{if $n$ is odd}\;.
\end{cases}
\end{equation}
Here, the round brackets denote the symmetrization in all indices,
\begin{equation}
\label{eq:symmetrization}
K^{(\mu_1\mu_2}\cdots K^{\mu_{n-1}\mu_{n})}:=\frac{1}{(n-1)!!}\sum_{\mathcal{P}} K^{\mu_1\mu_2}\cdots K^{\mu_{n-1}\mu_{n}}\;,
\end{equation}
where the factor $(n-1)!!$ counts the terms in the sum over permutations, which is denoted by $\mathcal{P}$.
At this point, one only needs to compute $I^{(n)}$. Since it holds that
\begin{equation}
K_{\mu_1\mu_2}\cdots K_{\mu_{n-1}\mu_n}K^{(\mu_1\mu_2}\cdots K^{\mu_{n-1}\mu_{n})}=n+1\;,
\end{equation}
the integral $I^{(n)}$ can be computed as
\begin{equation}
I^{(n)}=\frac{1}{n+1}K_{\mu_1\mu_2}\cdots K_{\mu_{n-1}\mu_n} I^{\mu_1\cdots\mu_n}=\frac{(-\varsigma^2)^{n/2} }{n+1} \int \d S(k)\;.
\end{equation}
The last integral is a Lorentz scalar and easily calculated in the particle-rest frame, where $\delta(k\cdot \s) = \delta(\s^0)/m$,
\begin{equation}
\int \d S(k) = \frac{S_0}{\varsigma \pi} \int \d^3 \s \delta(\s^2 +\varsigma^2) = \frac{S_0}{2\varsigma^2 \pi} \int \d^3 \s \delta(\s +\varsigma)  = 2S_0\;.
\end{equation}
In total one thus finds
\begin{equation}
\int \d S(k) \s^{\mu_1} \cdots \s^{\mu_{n}}= S_0 \frac{2(-\varsigma^2)^{n/2} }{n+1} K^{(\mu_1\mu_2}\cdots K^{\mu_{n-1}\mu_{n})}\;.
\label{eq:sol_simple}
\end{equation}
Comparing this result for $n=0$ and $n=2$ to Eq. \eqref{eq:spin_ints}, it becomes apparent that the constants $S_0$ and $\varsigma$ have to be chosen as $\varsigma^2=3$ and $S_0=1$ for spin-$\sfrac{1}{2}$ particles, whereas $\varsigma^2=2$ and $S_0=3/2$ in the spin-$1$ case.

\subsection{Thermodynamic integrals}
\label{app:therm_int}
The basic thermodynamic integral one has to evaluate is given by [cf. Eq. \eqref{eq:def_Inq}]
\begin{equation}
    I_{nq}=\frac{1}{(2q+1)!!} \int \d \Gamma E_\k^{n-2q} \left(E_\k^2-m^2\right)^{q}f_{0\k}\;.
\end{equation}
Since this integral is a scalar, it can be evaluated in any frame, which most conveniently is chosen to be the fluid-rest frame, where $u^\mu\equiv(1,\mathbf{0})$ and $E_\k\equiv k^0=\sqrt{k^2+m^2}$. Switching to spherical coordinates and performing the angular integrations, one finds
\begin{align}
I_{nq}&=\frac{g}{(2q+1)!!}\frac{1}{2\pi^2} \int_0^\infty \d k\, k^{2q+2} (k^2+m^2)^{(n-1)/2-q} \left(e^{\beta_0\sqrt{k^2+m^2}-\alpha_0}+a\right)^{-1} \nonumber\\
&=\frac{g}{(2q+1)!!}\frac{\beta_0^{-n-2}}{2\pi^2} \int_z^\infty \d y\, \left(y^2-z^2\right)^{q+1/2} y^{n-2q} \left(e^{y-\alpha_0}+a\right)^{-1}\;,
\end{align}
where $g=2\sigma+1$ is the degeneracy factor for spin-$\sigma$ particles and the substitution $y\coloneqq \beta_0\sqrt{k^2+m^2}$ was performed in the second equality. Furthermore, $z\coloneqq m\beta_0$ is the ratio of mass over temperature. In the following I consider classical statistics, i.e., $a=0$. The remaining task then consists of evaluating an integral of the form
\begin{equation}
\mathcal{I}_{ab}(z) \coloneqq \int_z^\infty \d y\, \left(y^2-z^2\right)^{b-1/2} y^{a} e^{-y}\;, \label{eq:def_cal_I}
\end{equation}
with $a= n-2q$ and $b= q+1$.
In the ultrarelativistic limit $z\to 0$, the integral is evaluated straightforwardly as
\begin{equation}
\mathcal{I}_{ab}(0)=\int_0^\infty \d y\, y^{2b+a-1} e^{-y}=\Gamma(a+2b)\;.
\end{equation} 
In the case where $z>0$, it is advantageous to first define $x\coloneqq y/z$, such that
\begin{equation}
\mathcal{I}_{ab}(z)=z^{a+2b}\int_1^\infty \d x\, \left(x^2-1\right)^{b-1/2} x^{a} e^{-zx}\;.
\end{equation} 
Note that the following recursion holds,
\begin{equation}
\mathcal{I}_{ab}(z)=\mathcal{I}_{a+2,b-1}(z)-z^2\mathcal{I}_{a,b-1}(z)\;,
\end{equation}
which can be applied iteratively to obtain
\begin{equation}
\mathcal{I}_{ab}(z)=\sum_{j=0}^{b} (-1)^j  \binom{b}{j} z^{2j} \mathcal{I}_{a+2b-2j,0}(z)\;.\label{eq:sum_cal_I}
\end{equation}
The integrals $\mathcal{I}_{a+2b-2j,0}(z)$ can be evaluated in terms of the Bickley function \cite{NIST:DLMF}
\begin{equation}
\mathrm{Ki}_r(z)\coloneqq \int_0^\infty \d \tau \cosh^{-r}\tau e^{-z\cosh\tau} =\int_1^\infty \d x\, \frac{x^{-r}}{\sqrt{x^2-1}} e^{-zx} \equiv z^{r}\mathcal{I}_{-r,0}(z)\;,
\end{equation}
which then yields
\begin{equation}
\mathcal{I}_{ab}(z)=z^{a+2b}\sum_{j=0}^{b} (-1)^j \binom{b}{j}  \mathrm{Ki}_{2j-2b-a}(z)\;.\label{eq:sum_cal_I_Ki}
\end{equation}
Summarizing, one has
\begin{equation}
I_{nq}=\frac{g e^{\alpha_0}}{(2q+1)!!} \frac{\beta_0^{-n-2}}{2\pi^2}\times 
\begin{cases}
\displaystyle{z^{n+2}\sum_{j=0}^{q+1} (-1)^j \binom{q+1}{j} \mathrm{Ki}_{2j-2-n}(z)} \;,\quad &z>0\\
\Gamma(n+2)\;,\quad &z=0
\end{cases}\;. \label{eq:Inq_general_solution}
\end{equation}
In order to implement Eq. \eqref{eq:Inq_general_solution} efficiently in the case $z>0$, one may use the following recursion relation for $r\geq 2$,
\begin{equation}
r\mathrm{Ki}_{r+1}(z)=(r-1)\mathrm{Ki}_{r-1}(z)-z\mathrm{Ki}_r(z)+z\mathrm{Ki}_{r-2}(z)\;,
\end{equation}
with the starting values given by \cite{Florkowski:2014sfa}
\begin{subequations}
\begin{align}
\mathrm{Ki}_0(z)&=K_0(z)\;,\\
\mathrm{Ki}_1(z)&=\frac{\pi}{2}\left\{1-z\left[  \mathbf{L}_{-1}(z)K_0(z)+\mathbf{L}_0(z)K_1(z) \right]  \right\}\;,
\end{align}
\end{subequations}
where $K_r(z)$ is the modified Bessel function of the second kind and $\mathbf{L}_r(z)$ denotes the modified Struve function. In the cases where the index is negative, the Bickley function (with $r\geq 0$) can be expressed as
\begin{equation}
\mathrm{Ki}_{-r}(z)=(-1)^r \frac{\d^{r}}{\d z^{r}} K_0(z)\;.
\end{equation}
The result \eqref{eq:Inq_general_solution} is the general solution that will hold for any value of $n$ and $q$, as long as the integral converges, which is ensured by demanding that $q>-3/2$ for $z>0$ and $n>-2$ for $z=0$.
For completeness, note that the thermodynamic integrals fulfill the following identities \cite{Ambrus:2022vif},
\begin{subequations}
\begin{align}
\beta_0 J_{nq}&= I_{n-1,q-1}+(n-2q)I_{n-1,q}\;,\\
\beta_0 K_{nq}&= J_{n-1,q-1}+(n-2q)J_{n-1,q}\;.
\end{align}
\end{subequations}

\section{Calculations}
\label{app:comp}
In this appendix, I show some explicit computations and definitions that are not necessary for the understanding of the main text and thus omitted therein for brevity.
In particular, in Subsec. \ref{app:defs} the definitions of the transition rates and spacetime shifts are restated for particles of different spin. Subsec. \ref{app:Tmunu} is concerned with deriving the form of the dissipative part of $T^{[\mu\nu]}$, Eq. \eqref{eq:delta_TA}, whereas Subsec. \ref{app:mom} provides additional remarks on the moment equations \eqref{eq:taudot_0}--\eqref{eq:eom_t_tensor}. Finally, Subsec. \ref{app:coll} deals with the derivation of the collision terms \eqref{eqs:C_1_full}.

\subsection{Transition rates and spacetime shifts}
\label{app:defs}
For spin-$\sfrac{1}{2}$ and spin-1 particles, the transition rates are defined as \cite{Wagner:2022amr} \footnote{Note that in comparison to the reference there is a difference of a factor of two due to a different convention for the prefactor of the collision integral.}
\begin{equation}
\W^{(\sfrac{1}{2})}=\frac{m^4}{2} M^{\alpha_1\alpha_2 \beta_1\beta_2}M^{\gamma_1\gamma_2 \delta_1 \delta_2} h_{1,\beta_1\gamma_1}h_{2,\beta_2\gamma_2}h'_{\delta_2\alpha_2}\left(\bar{h}_{\delta_1\beta}h_{\beta\alpha_1}+h_{\delta_1\beta}\bar{h}_{\beta\alpha_1}\right)\label{eq:W_12}
\end{equation}
and \cite{Wagner:2023cct} \footnote{In comparison to the reference, there is a change of prefactor of $(2\pi\hbar)^3$. This is because, in the current work, this phase-space factor has been absorbed into the distribution function, like it has been done in Ref. \cite{Wagner:2022amr}.}
\begin{equation}
\W^{(1)}= \frac{1}{32} M^{\mu_1\mu_2\nu_1\nu_2} M^{\zeta_1\zeta_2 \eta_1\eta_2} h_{1,\zeta_1 \nu_1} h_{2,\zeta_2\nu_2} h'_{\mu_2\eta_2} \left(\bar{h}_{\mu_1}{}^\alpha H_{\alpha\eta_1}+H_{\mu_1}{}^\alpha \bar{h}_{\alpha\eta_1}\right)
\label{eq:W_1}
\end{equation}
respectively, with the tensors $M$ being the (tree-level) vertices of the theory. Note that in Eq. \eqref{eq:W_12} the indices are in Dirac space, while they are spacetime indices in Eq. \eqref{eq:W_1}. Furthermore, the quantities
\begin{subequations}
\begin{align}
h_{\alpha\beta}(k,\s)&\coloneqq \frac12 \left(\mathds{1}+\gamma_5 \slashed{\s}\right)\frac{\slashed{k}+m}{2m}\;,\\
h^{\mu\nu}(k,\s) &\coloneqq \frac13 K^{\mu\nu}+\frac{i}{2}\epsilon^{\mu\nu\alpha\beta}\frac{k_\alpha}{m}\s_\beta+ K^{\mu\nu}_{\alpha\beta} \s^\alpha \s^\beta \;,\\
H^{\mu\nu}(k,\s) &\coloneqq \frac13 K^{\mu\nu}+\frac{i}{2}\epsilon^{\mu\nu\alpha\beta}\frac{k_\alpha}{m}\s_\beta+ \frac58 K^{\mu\nu}_{\alpha\beta} \s^\alpha \s^\beta \;,
\end{align}
\end{subequations}
which constitute matrices in Dirac space and spacetime, have been introduced, and abbreviated as, e.g., $h_{1,\alpha\beta}\coloneqq h_{\alpha\beta}(k_1,\s_1)$.
In a similar manner, the spacetime shifts pertaining to spin-$\sfrac{1}{2}$ systems are given by
\begin{subequations}\label{eq:def_Deltas_Dirac}
\begin{align}
\Delta_{1}^\mu &\coloneqq-\frac{i\hbar}{8m} \frac{m^4}{\mathcal{W}^{(\sfrac{1}{2})}}  M^{\alpha_1 \alpha_2 \beta_1 \beta_2}M^{\gamma_1\gamma_2 \delta_1 \delta_2} h_{2,\beta_2 \gamma_2}  h'_{\delta_2 \alpha_2} h_{\delta_1 \alpha_1} \left[h_1,\gamma^\mu\right]_{\beta_1 \gamma_1}\;,\label{eq:def_Delta_1_Dirac}\\
\Delta_{2}^\mu &\coloneqq-\frac{i\hbar}{8m} \frac{m^4}{\mathcal{W}^{(\sfrac{1}{2})}}  M^{\alpha_1 \alpha_2 \beta_1 \beta_2}M^{\gamma_1\gamma_2 \delta_1 \delta_2} h_{1,\beta_1\gamma_1}   h'_{\delta_2 \alpha_2} h_{\delta_1 \alpha_1} \left[h_2,\gamma^\mu\right]_{\beta_2 \gamma_2}\;,\label{eq:def_Delta_2_Dirac}\\
\Delta'^\mu &\coloneqq-\frac{i\hbar}{8m} \frac{m^4}{\mathcal{W}^{(\sfrac{1}{2})}}  M^{\alpha_1 \alpha_2 \beta_1 \beta_2}M^{\gamma_1\gamma_2 \delta_1 \delta_2} h_{1,\beta_1\gamma_1} h_{2,\beta_2 \gamma_2} h_{\delta_1 \alpha_1} \left[h',\gamma^\mu\right]_{\delta_2 \alpha_2}\;,\label{eq:def_Delta_prime_Dirac}\\
\Delta^\mu &\coloneqq-\frac{i\hbar}{8m} \frac{m^4}{\mathcal{W}^{(\sfrac{1}{2})}}  M^{\alpha_1 \alpha_2 \beta_1 \beta_2}M^{\gamma_1\gamma_2 \delta_1 \delta_2}h_{1,\beta_1\gamma_1}h_{2,\beta_2 \gamma_2}   h'_{\delta_2 \alpha_2}  \left[h,\gamma^\mu\right]_{\delta_1 \alpha_1}\;,\label{eq:def_Delta_Dirac}
\end{align}
\end{subequations}
while the respective spin-1 quantities read
\begin{subequations}\label{eq:def_Deltas_Proca}
\begin{align}
\Delta_1^\mu &\coloneqq \frac13 \frac{i\hbar}{32m^2} \frac{1}{\mathcal{W}^{(1)}} M^{\mu_1\mu_2\nu_1\nu_2}  M^{\zeta_1\zeta_2\eta_1\eta_2} h_{2,\zeta_2\nu_2}h'_{\mu_2\eta_2}H_{\mu_1\eta_1}\left(h_1^\mu{}_{\nu_1}k_{1,\zeta_1}-h_{1,\zeta_1}{}^\mu k_{1,\nu_1}  \right)\label{eq:def_Delta_1_Proca}\;,\\
\Delta_2^\mu &\coloneqq \frac13 \frac{i\hbar}{32m^2} \frac{1}{\mathcal{W}^{(1)}} M^{\mu_1\mu_2\nu_1\nu_2}  M^{\zeta_1\zeta_2\eta_1\eta_2} h_{1,\zeta_1\nu_1}h'_{\mu_2\eta_2}H_{\mu_1\eta_1}\left(h_2^\mu{}_{\nu_2}k_{2,\zeta_2}-h_{2,\zeta_2}{}^\mu k_{2,\nu_2}  \right)\label{eq:def_Delta_2_Proca}\;,\\
\Delta'^\mu &\coloneqq \frac13 \frac{i\hbar}{32m^2} \frac{1}{\mathcal{W}^{(1)}} M^{\mu_1\mu_2\nu_1\nu_2}  M^{\zeta_1\zeta_2\eta_1\eta_2} h_{1,\zeta_1\nu_1} h_{2,\zeta_2\nu_2}H_{\mu_1\eta_1} \left(h'^\mu{}_{\eta_2}k'_{\mu_2}-h'_{\mu_2}{}^\mu k'_{\eta_2}  \right)\label{eq:def_Delta_prime_Proca}\;,\\
\Delta^\mu &\coloneqq \frac13 \frac{i\hbar}{32m^2} \frac{1}{\mathcal{W}^{(1)}} M^{\mu_1\mu_2\nu_1\nu_2}  M^{\zeta_1\zeta_2\eta_1\eta_2} h_{1,\zeta_1\nu_1} h_{2,\zeta_2\nu_2}h'_{\mu_2\eta_2} \left(H^\mu{}_{\eta_1}k_{\mu_1}-H_{\mu_1}{}^\mu k_{\eta_1}  \right)\label{eq:def_Delta_Proca}\;.
\end{align}
\end{subequations}
In the case of a simple four-point interaction in the spin-$\sfrac{1}{2}$ case, $\mathcal{L}_\mathrm{int}=G(\overline{\psi}\psi)^2$, the vertex reads 
\begin{subequations}
\begin{equation}
M^{\alpha_1\alpha_2\beta_1\beta_2}=\frac{2 G}{\hbar} \left(\delta^{\alpha_1 \beta_1} \delta^{\alpha_2\beta_2}-\delta^{\alpha_1 \beta_2} \delta^{\alpha_2\beta_1}\right)
\end{equation}
while in the spin-$1$ case, when $\mathcal{L}_\mathrm{int}=\hbar G(V^\dagger_\mu V^\mu)^2$, it is given by
\begin{equation}
M^{\mu_1\mu_2\nu_1\nu_2}=2\hbar^2 G \left(g^{\mu_1\nu_1}g^{\mu_2\nu_2}+g^{\mu_1\nu_2}g^{\mu_2\nu_1}\right)\;.
\end{equation}
\end{subequations}

\subsection{The antisymmetric part of the energy-momentum tensor}
\label{app:Tmunu}
Since the antisymmetric part of the energy-momentum tensor is determined by the collision term, it can be split into a local part $T_\ell^{[\mu\nu]}$, and a nonlocal part $T_{n\ell}^{[\mu\nu]}$. In Ref. \cite{Wagner:2024fhf}, it has been shown that these contributions read\footnote{In the mentioned reference, the identity has been proven for spin-$\sfrac{1}{2}$ particles, but can be shown to hold in the case of spin 1 as well, using the same techniques.}
\begin{equation}
T^{[\mu\nu]}_\ell= \frac12 \int \d\Gamma_1 \, \d \Gamma_2 \, \d\Gamma' \, \d \Gamma \, \d \bar{S}(k) (2\pi\hbar)^4 \delta^{(4)}(k+k'-k_1-k_2) \mathcal{W} \Delta^{[\mu} k^{\nu]}\left(f_1 f_2 \ft' \ft - \ft_1 \ft_2 f' f\right)\label{eq:T_loc_1}
\end{equation}
and 
\begin{subequations}
\begin{align}
T_{n\ell,1}^{[\mu\nu]}&\coloneqq \frac{\hbar}{4} \int \d\Gamma_1 \, \d \Gamma_2 \, \d\Gamma' \, \d \Gamma \, \d \bar{S}(k) (2\pi\hbar)^4 \delta^{(4)}(k+k'-k_1-k_2) \mathcal{W} \Sigma_\s^{\mu\nu} \Delta \cdot \partial \left(f_1 f_2 \ft' \ft - \ft_1 \ft_2 f' f\right)\;,\\
T_{n\ell,2}^{[\mu\nu]}&\coloneqq -\frac{\hbar}{4}\int \d\Gamma_1 \, \d \Gamma_2 \, \d\Gamma' \, \d \Gamma \, \d \bar{S}(k) (2\pi\hbar)^4 \delta^{(4)}(k+k'-k_1-k_2) \mathcal{W} \Sigma_\s^{\mu\nu} \bigg\{ \Delta_1^\lambda \left[ (\partial_\lambda f_1) f_2 \ft' \ft - (\partial_\lambda \ft_1) \ft_2 f' f\right]\nonumber\\
&\quad+\Delta_2^\lambda \left[ f_1 (\partial_\lambda f_2)  \ft' \ft - \ft_1 (\partial_\lambda \ft_2) f' f\right]
+\Delta'^\lambda \left[ f_1 f_2 (\partial_\lambda \ft') \ft - \ft_1 \ft_2 (\partial_\lambda f') f\right]
+\Delta^\lambda \left[ f_1 f_2 \ft' (\partial_\lambda \ft) - \ft_1 \ft_2 f' (\partial_\lambda f)\right] \bigg\}\,,
\end{align}
\end{subequations}
respectively, with $T_{n\ell}^{[\mu\nu]}=T_{n\ell,1}^{[\mu\nu]}+T_{n\ell,2}^{[\mu\nu]}$.

\subsubsection{Local-equilibrium contribution}
Upon inserting the local-equilibrium distribution function, the local contribution \eqref{eq:T_loc_1} becomes
\begin{equation}
T_{\ell,\mathrm{eq}}^{[\mu\nu]}=\frac{\hbar}{4} \int_f \Delta^{[\mu}k^{\nu]}\Omega_{0,\alpha\beta}\left(\Sigma_{\s_1}^{\alpha\beta}+\Sigma_{\s_2}^{\alpha\beta}-\Sigma_{\s'}^{\alpha\beta}-\Sigma_{\s}^{\alpha\beta}\right)=\mathcal{U}^{\mu\nu\alpha\beta}\Omega_{0,\alpha\beta}\;,
\end{equation}
with
\begin{equation}
\mathcal{U}^{\mu\nu\alpha\beta}\coloneqq \frac{\hbar}{4} \int_f \Delta^{[\mu}k^{\nu]}\left(\Sigma_{\s_1}^{\alpha\beta}+\Sigma_{\s_2}^{\alpha\beta}-\Sigma_{\s'}^{\alpha\beta}-\Sigma_{\s}^{\alpha\beta}\right)\;.
\end{equation}
On the other hand, one finds that the first nonlocal contribution $T_{n\ell,1,\mathrm{eq}}^{[\mu\nu]}$ vanishes. Using the relation
\begin{equation}
\partial_\lambda f_0 = f_0 \ft_0 \left[\partial_\lambda \alpha_0 + k^\rho \left(\varpi_{\lambda\rho}-\xi_{\lambda\rho}\right)\right]\;,
\end{equation}
with the thermal shear $\xi^{\mu\nu}\coloneqq \frac12 \partial^{(\mu} \beta_0 u^{\nu)}$, the second nonlocal contribution becomes
\begin{equation}
T_{n\ell,2,\mathrm{eq}}^{[\mu\nu]}=-\frac{\hbar}{2} \int_f \Delta^\alpha \left(\Sigma_{\s_1}^{\mu\nu}+\Sigma_{\s_2}^{\mu\nu}-\Sigma_{\s'}^{\mu\nu}-\Sigma_\s^{\mu\nu}\right)\left[I_\alpha+k^\beta\left(\varpi_{\alpha\beta}-\xi_{\alpha\beta}\right)\right]=\mathcal{X}^{\mu\nu\alpha\beta}\varpi_{\alpha\beta}-\mathcal{Y}^{\mu\nu\alpha\beta}\xi_{\alpha\beta}+\mathcal{Z}^{\mu\nu\alpha}I_\alpha\;,
\end{equation}
with 
\begin{subequations}
\begin{align}
\mathcal{X}^{\mu\nu\alpha\beta}&\coloneqq  -\frac{\hbar}{4} \int_f \Delta^{[\alpha} k^{\beta]} \left(\Sigma_{\s_1}^{\mu\nu}+\Sigma_{\s_2}^{\mu\nu}-\Sigma_{\s'}^{\mu\nu}-\Sigma_\s^{\mu\nu}\right) \;,\\
\mathcal{Y}^{\mu\nu\alpha\beta}&\coloneqq  -\frac{\hbar}{4} \int_f \Delta^{(\alpha} k^{\beta)} \left(\Sigma_{\s_1}^{\mu\nu}+\Sigma_{\s_2}^{\mu\nu}-\Sigma_{\s'}^{\mu\nu}-\Sigma_\s^{\mu\nu}\right) \;,\\
\mathcal{Z}^{\mu\nu\alpha}&\coloneqq  -\frac{\hbar}{2} \int_f \Delta^{\alpha}  \left(\Sigma_{\s_1}^{\mu\nu}+\Sigma_{\s_2}^{\mu\nu}-\Sigma_{\s'}^{\mu\nu}-\Sigma_\s^{\mu\nu}\right) \;.
\end{align}
\end{subequations}
Note that $\mathcal{X}^{\mu\nu\alpha\beta}=-\mathcal{U}^{\alpha\beta\mu\nu}$. From considering the possible tensor structures \cite{Wagner:2024fhf},
\begin{equation}
\mathcal{X}^{\mu\nu\alpha\beta}=\mathcal{X}_1 u^{[\alpha}\Delta^{\beta][\mu}u^{\nu]}+ \mathcal{X}_2 g^{[\beta}_\rho \Delta^{\alpha][\mu}\Delta^{\nu]\rho}\;,\qquad 
\mathcal{Y}^{\mu\nu\alpha\beta}=\mathcal{Y}_1 u^{(\alpha}\Delta^{\beta)[\mu}u^{\nu]}\;,\qquad 
\mathcal{Z}^{\mu\nu\alpha}= \mathcal{Z}_1 u^{[\mu}\Delta^{\nu]\alpha}\;,
\end{equation}
it becomes clear that $\mathcal{X}^{\mu\nu\alpha\beta}=\mathcal{X}^{\alpha\beta\mu\nu}$ and thus $\mathcal{X}^{\mu\nu\alpha\beta}=-\mathcal{U}^{\mu\nu\alpha\beta}$. The coefficients \eqref{eqs:Gammas} are obtained from projecting the respective tensors,
\begin{alignat}{3}
\Gamma^{(\omega)}&= -\frac{4\mathcal{X}_2}{\hbar^2}=-\frac{1}{3\hbar^2} \Delta_{\mu\alpha}\Delta_{\nu\beta} \mathcal{X}^{\mu\nu\alpha\beta}\;,\qquad 
&&\Gamma^{(\kappa)}= -\frac{2\mathcal{X}_1}{\hbar^2}=-\frac{2}{3\hbar^2} u_\mu u_\beta \Delta_{\nu\alpha} \mathcal{X}^{\mu\nu\alpha\beta}\;,\nonumber\\
\Gamma^{(a)}&= \frac{2\mathcal{Y}_1}{\hbar^2}=-\frac{2}{3\hbar^2} u_\mu u_\beta \Delta_{\nu\alpha} \mathcal{Y}^{\mu\nu\alpha\beta}\;,\qquad
&&\Gamma^{(I)}= \frac{\mathcal{Z}_1}{\hbar^2}=\frac{1}{3\hbar^2} u_\mu  \Delta_{\nu\alpha} \mathcal{Z}^{\mu\nu\alpha}\;,
\end{alignat}
and one arrives at Eq. \eqref{eq:TA0}.

\subsubsection{Dissipative contribution}
Given that dissipative contributions from the nonlocal part of the collision term are neglected, only those dissipative parts that originate from the local part $\delta T_\ell^{[\mu\nu]}$ have to be considered. 
Linearizing in the deviation from equilibrium and using the symmetries of the transition rate (meaning that only the parts of the distribution functions linear in a spin vector contribute), one obtains
\begin{equation}
\delta T^{[\mu\nu]}_\ell = \sum_{\ell=0}^\infty \sum_{n=0}^{N_\ell^{(1)}} \left(\mathcal{T}_n^{(\ell)}\right)^{\mu\nu}_{\alpha,\nu_1\cdots\nu_\ell} \tau_n^{\langle\alpha\rangle,\nu_1\cdots\nu_\ell}\;,
\label{eq:delta_Tmunu1}
\end{equation}
where the moment expansion \eqref{eq:delta_f_expl_2} has been used, and the following tensors have been introduced,
\begin{align}
\left(\mathcal{T}^{(\ell)}_n\right)^{\mu\nu}_{\alpha,\nu_1\cdots\nu_\ell}&\coloneqq -\int_f \Delta^{[\mu} k^{\nu]}  \left(
\mathcal{H}^{(1,\ell)}_{\k_1 n} \Xi_{1,\alpha\beta}\s_1^\beta k_{1,\langle\nu_1} \cdots k_{1,\nu_\ell\rangle} 
+\mathcal{H}^{(1,\ell)}_{\k_2 n} \Xi_{2,\alpha\beta}\s_2^\beta k_{2,\langle\nu_1} \cdots k_{2,\nu_\ell\rangle}\right. \nonumber\\
&\hspace{2.2cm} \left.-\mathcal{H}^{(1,\ell)}_{\k' n} \Xi'_{\alpha\beta}\s'^\beta k'_{\langle\nu_1} \cdots k'_{\nu_\ell\rangle} 
-\mathcal{H}^{(1,\ell)}_{\k n} \Xi_{\alpha\beta}\s^\beta k_{\langle\nu_1} \cdots k_{\nu_\ell\rangle} 
 \right)\;.
\end{align}
While this expression is very general, not all terms can survive. Since the final expression $\delta T^{[\mu\nu]}_\ell$ needs to be an antisymmetric rank-two tensor, it can only take the forms $u^{[\mu}X^{\nu]}$ or $\epsilon^{\mu\nu\alpha\beta}u_\alpha Y_\beta$, with $X^\mu$ and $Y^\mu$ being a vector and an axial vector, respectively. This immediately excludes all irreducible moments of tensor-rank higher than two in momentum.
On the other hand, from the decompositions \eqref{eq:decomp_tau2} and \eqref{eq:decomp_tau3} it is clear that the only axial vectors at hand are $\tau_r^{\langle\mu\rangle}$ and $t_r^\mu$, while $w_r^\mu$ constitutes the only vector-valued quantity. Then, omitting the subscript $\ell$, Eq. \eqref{eq:delta_Tmunu1} can be rewritten as
\begin{equation}
\delta T^{[\mu\nu]} = \sum_{n=0}^{N_0^{(1)}} \left(\mathcal{T}^{(0)}_n\right)^{\mu\nu}_{\alpha} \tau_n^{\langle\alpha\rangle}
+\frac12 \sum_{n=0}^{N_1^{(1)}} \left(\mathcal{T}^{(1)}_n\right)^{\mu\nu}_{\alpha,\beta} \epsilon^{\alpha\beta\gamma\delta}u_\gamma w_{n,\delta}
+\frac35\sum_{n=0}^{N_2^{(1)}} \left(\mathcal{T}^{(2)}_n\right)^{\mu\nu}_{\alpha,\beta\gamma}\Delta^{\alpha\beta} t_n^\gamma\;.
\label{eq:delta_Tmunu2}
\end{equation}
Again, from the possible structures of $\delta T^{[\mu\nu]}_\ell$ and the symmetries of the tensors $\mathcal{T}^{(\ell)}$ one can deduce that they must take the form 
\begin{equation}
\left(\mathcal{T}^{(0)}_n\right)^{\mu\nu}_{\alpha} = \mathcal{T}^{(0)}_n \epsilon^{\mu\nu}{}_{\lambda\alpha}u^\lambda\;,\qquad \left(\mathcal{T}^{(1)}_n\right)^{\mu\nu}_{\alpha,\beta} = \mathcal{T}^{(1)}_n \epsilon^{\mu\nu}{}_{\alpha\beta}\;,\qquad \frac35\left(\mathcal{T}^{(2)}_n\right)^{\mu\nu}_{\alpha,\beta \gamma} \Delta^{\alpha\beta} = \mathcal{T}^{(2)}_n \epsilon^{\mu\nu}{}_{\lambda\gamma}u^\lambda\;.
\end{equation}
From this relation and factoring out a factor of $\hbar$ then follows Eq. \eqref{eq:delta_TA},
\begin{equation}
\delta T^{[\mu\nu]}=\hbar \left[\epsilon^{\mu\nu\alpha\beta}u_\alpha\left(\sum_{n=0}^{N_0^{(1)}}\gamma_n^{(0)}\tau_{n,\beta}+\sum_{n=0}^{N_2^{(1)}}\gamma_n^{(2)}t_{n,\beta}\right)-u^{[\mu}\sum_{n=0}^{N_1^{(1)}}\gamma_n^{(1)}w^{\nu]}_n\right]\;,
\end{equation}
with the coefficients
\begin{subequations}
\begin{align}
\gamma_n^{(0)} &\coloneqq \frac{1}{\hbar}\mathcal{T}^{(0)}_n = -\frac{1}{6\hbar} \epsilon_{\mu\nu}{}^{\rho\alpha}u_\rho \left(\mathcal{T}^{(0)}_n\right)^{\mu\nu}_{\alpha}\;,\\
\gamma_n^{(1)} &\coloneqq \frac{1}{\hbar}\mathcal{T}^{(1)}_n = -\frac{1}{24\hbar} \epsilon_{\mu\nu}{}^{\alpha\beta} \left(\mathcal{T}^{(1)}_n\right)^{\mu\nu}_{\alpha,\beta}\;,\\
\gamma_n^{(2)} &\coloneqq \frac{1}{\hbar}\mathcal{T}^{(2)}_n = -\frac{1}{10\hbar} \epsilon_{\mu\nu}{}^{\rho\gamma}u_\rho\left(\mathcal{T}^{(2)}_n\right)^{\mu\nu}_{\alpha,\beta\gamma} \Delta^{\alpha\beta}\;.
\end{align}
\end{subequations}

\subsection{Moment equations}
\label{app:mom}
The equations of motion for the irreducible moments of $\delta f_{\k \s}$, i.e., both for the moments of spin-ranks zero and one, $\rho_r^{\mu_1\cdots\mu_\ell}$ and $\tau_r^{\mu,\mu_1\cdots\mu_\ell}$, follow from the Boltzmann equation \eqref{eq:Boltzmann_decomp}. 
Since the moments $\rho_r^{\mu_1\cdots\mu_\ell}$ and $\tau_r^{\mu,\mu_1\cdots\mu_\ell}$ differ only by an additional index tied to a spin vector (which itself is independent of spacetime), their equations of motion will be very similar. Indeed, it will be sufficient to re-evaluate the terms pertaining to local equilibrium, cf. Eq. \eqref{eq:Boltzmann_decomp}, while the other contributions are identical to their spinless counterparts given e.g. in Ref. \cite{Denicol:2012cn}, save for an additional spin index.
Recalling the definition of $\widetilde{\Omega}_0^{\mu\nu}$ \eqref{eq:def_Omegatilde}, the local-equilibrium contribution to the equations of motion for the moments of tensor-rank zero in momentum is
\begin{align}
&\quad-\Delta^\mu_\lambda\int \d \Gamma \s^\lambda  E_\k^{r-1}\left(E_\k\dot{f}_{\text{eq}}+k\cdot \nabla f_{\text{eq}}\right)\nonumber\\
&=\frac{\sigma \hbar}{m}\Delta^\mu_\lambda \int \d K E_\k^{r-1}\left(E_\k u\cdot \partial +k\cdot \nabla \right)f_{0\k}\widetilde{f}_{0\k} \widetilde{\Omega}_0^{\lambda\nu}k_\nu\nonumber\\
&=\frac{\sigma \hbar}{gm} \left(J_{r+1,0}\dot{\widetilde{\Omega}}{}^{\langle\mu\rangle\nu}_{0}u_\nu-J_{r+1,1}\Delta^\mu_\lambda\nabla_\nu \widetilde{\Omega}_0^{\lambda\nu}\right)+\frac{\sigma \hbar}{m}\widetilde{\Omega}_0^{\langle\mu\rangle\nu} \int \d K E_\k^{r-1} k_\nu\left(E_\k u\cdot \partial +k\cdot \nabla \right)f_{0\k}\widetilde{f}_{0\k} \nonumber\\
&=\frac{\sigma \hbar}{gm}  \bigg\{J_{r+1,0}\dot{\widetilde{\Omega}}{}^{\langle\mu\rangle\nu}_{0}u_\nu-J_{r+1,1}\Delta^\mu_\lambda\nabla_\nu \widetilde{\Omega}_0^{\lambda\nu} 
+2\omega_0^\mu \left[K_{r+1,0}\dot{\alpha}_0-K_{r+2,0}\dot{\beta}_0+\theta (J_{r+1,0}+rJ_{r+1,1})\right] \nonumber\\
&\quad + \beta_0 K_{r+2,1} \widetilde{\Omega}_0^{\langle\mu\rangle\nu}\dot{u}_\nu -\widetilde{\Omega}_0^{\langle\mu\rangle\nu}\left(K_{r+1,1}I_\nu -K_{r+2,1}\nabla_\nu \beta_0\right) \bigg\}\nonumber\\
&=\frac{2\sigma \hbar}{gm}  \bigg(J_{r+1,0}\dot{\omega}_0^{\langle\mu\rangle}+J_{r+1,1}\left(\sigma^{\mu\nu}-\omega^{\mu\nu}\right)\omega_{0,\nu}
+\omega_0^\mu \left\{K_{r+1,0}\dot{\alpha}_0-K_{r+2,0}\dot{\beta}_0+\theta \left[J_{r+1,0}+\left(r-\frac23\right)J_{r+1,1}\right]\right\}
 \nonumber\\
&\quad
+J_{r+1,1}\epsilon^{\mu\nu\alpha\beta}u_\nu \nabla_\alpha \kappa_{0,\beta}
 + \epsilon^{\mu\nu\alpha\beta} u_\nu \left[ J_{r+1,0}\dot{u}_\alpha   +K_{r+1,1}I_\alpha -K_{r+2,1}\left(\nabla_\alpha \beta_0+\beta_0 \dot{u}_\alpha\right)\right]\kappa_{0,\beta} \bigg)\;.
\label{eq:eq_01}
\end{align}
Here, the definition of the local-equilibrium distribution function \eqref{eq:f0} and the spin-space integrals \eqref{eq:spin_ints} have been used. Note that, in comparison to Ref. \cite{Weickgenannt:2022zxs}, the terms in the fourth line proportional to $\omega_0^\mu$ and $\dot{\widetilde{\Omega}}{}_0^{\mu\nu}$ have opposite sign.
Then, the equation of motion for these moments is
\begin{align}
\dot{\tau}^{\langle\mu\rangle}_r-C^{\langle\mu\rangle}_{r-1}&= 
\frac{2\sigma\hbar}{gm}\bigg(\omega_0^\mu  \left\{K_{r+1,0}\dot{\alpha}_0-K_{r+2,0}\dot{\beta}_0+\left[J_{r+1,0}+\left(r-\frac23\right)J_{r+1,1}\right]\theta\right\}-J_{r+1,1}\left(\omega^{\mu\nu}-\sigma^{\mu\nu}\right)\omega_{0,\nu}\nonumber\\
&\quad+J_{r+1,0}\dot{\omega}_0^{\langle\mu\rangle}+J_{r+1,1}\epsilon^{\mu\nu\alpha\beta}u_\nu \nabla_\alpha \kappa_{0,\beta}+\epsilon^{\mu\nu\alpha\beta}u_\nu \kappa_{0,\beta}\left[J_{r+1,0}\dot{u}_\alpha+K_{r+1,1}I_\alpha-K_{r+2,1}(\nabla_\alpha \beta_0+\beta_0 \dot{u}_\alpha)\right]\bigg)\nonumber\\
&\quad+ r \dot{u}_\nu \tau_{r-1}^{{\langle\mu\rangle},\nu}+(r-1)\sigma_{\alpha\beta} \tau_{r-2}^{{\langle\mu\rangle},\alpha\beta}
-\Delta^\mu_\lambda\nabla_\nu \tau^{\lambda,\nu}_{r-1}-\frac13\left[(r+2)\tau_r^{\langle\mu\rangle}
-(r-1)m^2\tau_{r-2}^{\langle\mu\rangle} \right]\theta \;.
\label{eq:taudot_0_app}
\end{align}
In the case of momentum-rank one, the local-equilibrium term is
\begin{align}
&\quad-\Delta^\mu_\lambda\int \d \Gamma  \s^\lambda E_\k^{r-1} k^{\avg{\nu}} \left(E_\k\dot{f}_{\text{eq}}+  k\cdot \nabla f_{\text{eq}}\right)\nonumber\\
&= \frac{\sigma\hbar}{m} \Delta^\mu_\lambda\int \d K E_\k^{r-1} k^{\avg{\nu}} \left(E_\k u\cdot \partial +k\cdot \nabla\right)f_{0\k} \widetilde{f}_{0\k} \widetilde{\Omega}_0^{\lambda\rho} k_\rho \nonumber\\
&= -\frac{\sigma \hbar}{gm} \left( J_{r+2,1} \dot{\widetilde{\Omega}}{}_0^{\langle\mu\rangle\langle\nu\rangle} 
+J_{r+2,1}u_\rho\Delta^\mu_\lambda \nabla^\nu \widetilde{\Omega}_0^{\lambda\rho} \right)+\widetilde{\Omega}_0^{\langle\mu\rangle\rho} \frac{\sigma\hbar}{m} \int \d K E_\k^{r-1} k_\rho k^{\avg{\nu}} \left(E_\k u\cdot \partial +k\cdot \nabla\right)f_{0\k} \widetilde{f}_{0\k} \nonumber\\
&=\frac{\sigma \hbar}{gm} \bigg\{ -J_{r+2,1} \dot{\widetilde{\Omega}}{}_0^{\langle\mu\rangle\langle\nu\rangle} 
-J_{r+2,1}u_\rho \Delta^\mu_\lambda \nabla^\nu \widetilde{\Omega}_0^{\lambda\rho} +\widetilde{\Omega}_0^{\langle\mu\rangle\langle\nu\rangle}\left(-K_{r+2,1}\dot{\alpha}_0+K_{r+3,1} \dot{\beta}_0-\frac53 \beta_0 K_{r+3,2}\theta\right)\nonumber\\
&\quad-2\beta_0 K_{r+3,2}\widetilde{\Omega}_0^{\langle\mu\rangle\rho}\sigma^\nu{}_\rho +2\omega_0^\mu \left[-K_{r+2,1}I^\nu  +K_{r+3,1}\left(\nabla^\nu \beta_0+\beta_0 \dot{u}^\nu\right)\right]\bigg\}\nonumber\\
&=\frac{2\sigma \hbar}{gm} \bigg\{ -J_{r+2,1} \left(\epsilon^{\mu\nu\alpha\beta}u_\alpha \dot{\kappa}_{0,\beta}- \dot{u}^{[\mu}\omega_0^{\nu]}\right) 
-\epsilon^{\mu\lambda\alpha\beta}u_\alpha \kappa_{0,\beta}\left[\left(2\beta_0 K_{r+3,2}-J_{r+2,1}\right)\sigma^\nu{}_\lambda - J_{r+2,1}\omega^\nu{}_\lambda\right]-J_{r+2,1}\Delta^\mu_\lambda \nabla^\nu \omega_{0}^\lambda \nonumber\\
&\quad -\epsilon^{\mu\nu\alpha\beta}u_\alpha \kappa_{0,\beta}\left[K_{r+2,1}\dot{\alpha}_0-K_{r+3,1} \dot{\beta}_0+\frac13\left(5 \beta_0 K_{r+3,2}-J_{r+2,1}\right)\theta\right] -\omega_0^\mu \left[K_{r+2,1}I^\nu  -K_{r+3,1}\left(\nabla^\nu \beta_0+\beta_0 \dot{u}^\nu\right)\right]\bigg\}\;.
\end{align}
Comparing the fourth and fifth lines to Ref. \cite{Weickgenannt:2022zxs}, the term proportional to $\dot{\Omega}_0^{\langle\mu\rangle\langle\nu\rangle}$ has a different sign.
This then leads to the following equation of motion,
\begin{align}
\dot{\tau}^{\langle\mu\rangle,\langle\nu\rangle}_r-C^{\langle\mu\rangle,\langle\nu\rangle}_{r-1}
&=-\frac{2\sigma \hbar}{gm} \bigg\{ J_{r+2,1} \left(\epsilon^{\mu\nu\alpha\beta}u_\alpha \dot{\kappa}_{0,\beta}-\dot{u}^{[\mu} \omega_0^{\nu]}\right)+\omega_0^\mu \left[K_{r+2,1}I^\nu -K_{r+3,1}\left(\nabla^\nu \beta_0+\beta_0 \dot{u}^\nu\right)\right] \nonumber\\
&\quad +J_{r+2,1}\Delta^\mu_\lambda \nabla^\nu \omega_0^\lambda
+\left[\left(2\beta_0 K_{r+3,2}-J_{r+2,1}\right)\sigma^\nu{}_\lambda-J_{r+2,1}\omega^\nu{}_\lambda\right]\epsilon^{\mu\lambda\alpha\beta}u_\alpha \kappa_{0,\beta}\nonumber\\
&\quad  +\epsilon^{\mu\nu\alpha\beta}u_\alpha \kappa_{0,\beta}\left[K_{r+2,1}\dot{\alpha}_0-K_{r+3,1} \dot{\beta}_0+\frac13\left( 5\beta_0 K_{r+3,2}-J_{r+2,1}\right)\theta\right]\bigg\}\nonumber\\
&\quad+\omega^\nu{}_{\rho}\tau^{\langle\mu\rangle,\rho}_r+\frac13\left[ (r-1)m^2\tau_{r-2}^{{\langle\mu\rangle},\nu}
-(r+3)\tau_r^{{\langle\mu\rangle},\nu}\right]\theta\nonumber\\
&\quad+\frac15\left[(2r-2)m^2\tau^{{\langle\mu\rangle},\lambda}_{r-2}
-(2r+3)\tau^{{\langle\mu\rangle},\lambda}_r \right] \sigma^\nu{}_{\lambda}
+\frac13 \dot{u}^\nu \left[ m^2 r \tau^{\langle\mu\rangle}_{r-1}-(r+3)\tau^{\langle\mu\rangle}_{r+1} \right]\nonumber\\
&\quad-\frac13\Delta^\mu_\lambda \nabla^\nu \left( m^2 \tau_{r-1}^\lambda-\tau_{r+1}^\lambda\right)+r \dot{u}_\rho \tau_{r-1}^{{\langle\mu\rangle},\nu\rho}-\Delta^\nu_\lambda  \Delta^\mu_\alpha \nabla_\rho  
\tau^{\alpha,\lambda\rho}_{r-1}
+(r-1)\sigma_{\lambda\rho}\tau_{r-2}^{{\langle\mu\rangle},\nu\lambda\rho}\;.
\label{eq:taudot_1_app}
\end{align}
Lastly, in the case of momentum-rank two, the term pertaining to local equilibrium reads
\begin{align}
&\quad -\Delta^\mu_\rho\int \d \Gamma \s^\rho E_\k^{r-1} k^{\langle\nu}k^{\lambda\rangle} \left(E_\k\dot{f}_{\text{eq}}+  k\cdot \nabla f_{\text{eq}}\right)\nonumber\\
&=\frac{\sigma \hbar}{m}  \Delta^\mu_\rho \int \d K  E_\k^{r-1} k^{\langle\nu}k^{\lambda\rangle} k_\alpha \left(E_\k u\cdot \partial + k\cdot \nabla\right) f_{0\k}\widetilde{f}_{0\k}\widetilde{\Omega}_0^{\rho\alpha}\nonumber\\
&= \frac{2\sigma \hbar}{gm} \Big\{ K_{r+3,2}\Delta^\mu_\rho\Delta_{\alpha\beta}^{\nu\lambda} \nabla^\alpha \widetilde{\Omega}^{\rho\beta} +\widetilde{\Omega}^{\langle\mu\rangle\langle\nu} \left[K_{r+3,2}I^{\lambda\rangle}-K_{r+4,2}\left(\nabla^{\lambda\rangle}\beta_0+\beta_0 \dot{u}^{\lambda\rangle}\right)\right]-2\beta_0 K_{r+4,2} \omega_0^\mu \sigma^{\nu\lambda} \Big\}\nonumber\\
&= \frac{4\sigma \hbar}{gm} \bigg\{ -K_{r+3,2}\left(\sigma^{\mu\langle\nu}-\omega^{\mu\langle\nu}+\frac13 \theta \Delta^{\mu\langle\nu}\right)\omega_0^{\lambda\rangle} -K_{r+3,2}\left(\nabla^{\langle\nu}\kappa_{0,\beta}\right)\epsilon^{\lambda\rangle\mu\alpha\beta}u_\alpha\nonumber\\
&\quad- \left[K_{r+3,2}I^{\langle\nu}-K_{r+4,2}\left(\nabla^{\langle\nu}\beta_0+\beta_0 \dot{u}^{\langle\nu}\right)\right]\epsilon^{\lambda\rangle\mu\alpha\beta}u_\alpha \kappa_{0,\beta}+\left(K_{r+3,2}-\beta_0 K_{r+4,2}\right) \omega_0^\mu \sigma^{\nu\lambda} \bigg\}\;,
\end{align}
which gives rise to the following evolution equation,
\begin{align}
\dot{\tau}_r^{{\langle\mu\rangle},\langle\nu\lambda\rangle}-C_{r-1}^{{\langle\mu\rangle},\langle\nu\lambda\rangle}
&= \frac{4\sigma \hbar}{gm} \bigg\{ -\left[K_{r+3,2}I^{\langle\lambda}-K_{r+4,2}\left(\nabla^{\langle\lambda}\beta_0+\beta_0 \dot{u}^{\langle\lambda}\right)\right]\epsilon^{\nu\rangle\mu\alpha\beta}u_\alpha \kappa_{0,\beta}-K_{r+3,2}\left(\nabla^{\langle\nu}\kappa_{0,\beta}\right)\epsilon^{\lambda\rangle\mu\alpha\beta}u_\alpha\nonumber\\
&\quad 
-K_{r+3,2}\left[\left(\sigma^{\mu\langle\nu}-\omega^{\mu\langle\nu}\right)\omega_0^{\lambda\rangle}+\frac{\theta}{3}\Delta^{\mu\langle\nu}\omega_0^{\lambda\rangle}\right]
+\left(K_{r+3,2}-\beta_0 K_{r+4,2}\right) \omega_0^\mu \sigma^{\nu\lambda} \bigg\}+r\dot{u}_\rho \tau^{{\langle\mu\rangle},\nu\lambda\rho}_{r-1}\nonumber\\
&\quad +\frac25\left[r m^2\tau_{r-1}^{{\langle\mu\rangle},\langle\nu}
-(r+5) \tau_{r+1}^{{\langle\mu\rangle},\langle\nu}\right]\dot{u}^{\lambda\rangle} -\Delta^\mu_\gamma\Delta^{\nu\lambda}_{\alpha\beta} \nabla_\rho 
\tau_{r-1}^{\gamma,\alpha\beta\rho}+\Delta^\mu_\rho\frac25 \Delta^{\nu\lambda}_{\alpha\beta}
\nabla^\beta\left(\tau_{r+1}^{\rho,\alpha}-m^2 \tau_{r-1}^{\rho,\alpha} \right)\nonumber\\
&\quad+ \frac13\left[(r-1)m^2\tau_{r-2}^{{\langle\mu\rangle},\nu\lambda}-(r+4)\tau_r^{{\langle\mu\rangle},\nu\lambda} \right]\theta
+(r-1)\sigma_{\rho\tau}\tau_{r-2}^{{\langle\mu\rangle},\nu\lambda\rho\tau}\nonumber\\
&\quad +\frac27 \left[2(r-1)m^2\tau_{r-2}^{{\langle\mu\rangle},\rho\langle\nu}-(2r+5)\tau_r^{{\langle\mu\rangle},\rho\langle \nu} \right] 
\sigma^{\lambda\rangle}_{\ \rho}+2\tau_r^{{\langle\mu\rangle},\rho\langle\nu}\omega^{\lambda\rangle}_{\ \rho} \nonumber\\
&\quad+\frac{2}{15} \left[ (r-1)m^4 \tau^{\langle\mu\rangle}_{r-2}-(2r+3)m^2\tau_r^{\langle\mu\rangle}
+(r+4)\tau^{\langle\mu\rangle}_{r+2}\right]\sigma^{\nu\lambda}\;.
\label{eq:taudot_2_app}
\end{align}
Contracting the equations of motion \eqref{eq:taudot_1_app} and \eqref{eq:taudot_2_app} appropriately as described in Eqs. \eqref{eqs:contractions_twtdot} and applying the approximate relations \eqref{eq:asymp_rest} then leads to Eqs. \eqref{eq:eom_t_axial}, \eqref{eq:eom_w}, and \eqref{eq:eom_t_tensor}.

\subsection{Collision terms}
\label{app:coll}
The collision terms treated in Sec. \ref{sec:coll} consist of two contributions: the local part, which in its linearized form is given by Eq. \eqref{eq:L_moms}, and the nonlocal part, which, neglecting dissipative contributions, is defined in Eq. \eqref{eq:C_0_def}. 
First, I treat the linearized local collision terms.
Orienting on symmetries, it makes sense to start with the collision terms $L_{\ell,r-1}^{\langle\mu\rangle}$ and $L_{\ell,t,r-1}^\mu\coloneqq \Delta_{\nu\lambda}L_{\ell,r-1}^{\langle\nu\rangle,\langle\mu\lambda\rangle}$, which are axial-vector valued, and thus can be proportional to either $\tau_r^{\langle\mu\rangle}$ or $t_r^\mu$. This means that one must have
\begin{subequations}
\begin{align}
L_{\ell,r-1}^{\langle\mu\rangle}&= \sum_{n=0}^{N_0^{(1)}} \left(\mathcal{B}^{(00)}_{rn}\right)^{\mu}_{\nu} \tau_n^{\langle\nu\rangle} + \frac35 \sum_{n=0}^{N_2^{(1)}} \left(\mathcal{B}^{(02)}_{rn}\right)^{\mu}_{\nu,\nu_1\nu_2}  \Delta^{\nu_1\nu}t_{n}^{\nu_2}\;,\\
L_{\ell,t,r-1}^{\mu}&= \sum_{n=0}^{N_0^{(1)}} \left(\mathcal{B}^{(20)}_{rn}\right)^{\mu_1,\mu\mu_2}_{\nu} \Delta_{\mu_1\mu_2}\tau_n^{\langle\nu\rangle} + \frac35 \sum_{n=0}^{N_2^{(1)}} \left(\mathcal{B}^{(22)}_{rn}\right)^{\mu_1,\mu\mu_2}_{\nu,\nu_1\nu_2} \Delta_{\mu_1\mu_2} \Delta^{\nu_1\nu}t_{n}^{\nu_2}\;,
\end{align}
where the tracelessness of the tensors $(\mathcal{B}^{(\ell\ell')}_{rn})^{\mu,\mu_1\cdots\mu_\ell}_{\nu,\nu_1\cdots\nu_{\ell'}}$ in the second pair of indices was used.
Similarly, the vector-valued collision term $L_{\ell,w,r-1}^\mu\coloneqq \epsilon^\mu{}_{\nu\alpha\beta}u^\nu L_{\ell,r-1}^{\langle\alpha\rangle,\langle\beta\rangle}$ can only be proportional to the vector-valued moment $w_r^\mu$, and the tensor-valued collision term $L_{\ell,t,r-1}^{\mu\nu}\coloneqq \epsilon^{\delta}{}_{\alpha\beta\rho}u^\rho \Delta^{\mu\nu}_{\gamma\delta}L_{\ell,r-1}^{\langle\alpha\rangle,\langle\beta\gamma\rangle}$ has to be proportional to $t_r^{\mu\nu}$. Then, one finds
\begin{align}
L_{\ell,w,r-1}^\mu &=  \frac12\sum_{n=0}^{N_1^{(1)}} \epsilon^\mu{}_{\nu\alpha\beta}u^\nu\frac14 \left(\mathcal{B}^{(11)}_{rn}\right)^{[\alpha,\beta]}_{[\gamma,\delta]}\epsilon^{\gamma\delta\rho\sigma}u_\rho w_{r,\sigma}  \;,\\
L_{\ell,t,r-1}^{\mu\nu} &= -\frac23\sum_{n=0}^{\overline{N}_2^{(1)}} \epsilon^{\delta}{}_{\alpha\beta\rho}u^\rho \Delta^{\mu\nu}_{\gamma\delta}\frac14\left(\mathcal{B}^{(22)}_{rn}\right)^{[\alpha,\beta]\gamma}_{[\sigma,\eta]\zeta} t_{r,\lambda}{}^{\zeta}\epsilon^{\eta\sigma\kappa\lambda}u_\kappa  \;,
\end{align}
\end{subequations}
where the appropriate antisymmetries due to contraction with Levi-Cività symbols were made explicit.
The appropriate tensor structures of the quantities $(\mathcal{B}^{(\ell\ell')}_{rn})^{\mu,\mu_1\cdots\mu_\ell}_{\nu,\nu_1\cdots\nu_{\ell'}}$ are
\begin{subequations}
\begin{alignat}{3}
\left(\mathcal{B}^{(00)}_{rn}\right)^{\mu}_{\nu} &= \mathcal{B}^{(0)}_{rn} \Delta^\mu_\nu \;,& \qquad 
\frac35\left(\mathcal{B}^{(02)}_{rn}\right)^{\mu}_{\nu,\nu_1\nu_2}\Delta^{\nu_1\nu}&= \mathcal{B}^{(02)}_{rn} \Delta^{\mu}_{\nu_2}\;,\\
\left(\mathcal{B}^{(20)}_{rn}\right)^{\mu_1,\mu\mu_2}_{\nu}\Delta_{\mu_1\mu_2}&= \mathcal{B}^{(20)}_{rn} \Delta^{\mu}_\nu\;,&\qquad 
\frac35\left(\mathcal{B}^{(22)}_{rn}\right)^{\mu_1,\mu\mu_2}_{\nu,\nu_1\nu_2}\Delta_{\mu_1\mu_2}\Delta^{\nu_1\nu}&= \mathcal{B}^{(2)}_{rn} \Delta^{\mu}_{\nu_2}\;,\\
\frac14\left(\mathcal{B}^{(11)}_{rn}\right)^{[\alpha,\beta]}_{[\gamma,\delta]} &= \frac12 \mathcal{B}^{(1)}_{rn} \Delta^{[\alpha}_\gamma \Delta^{\beta]}_\delta\;,\\
\frac14\left(\mathcal{B}^{(22)}_{rn}\right)^{[\alpha,\beta]\gamma}_{[\sigma,\eta]\zeta} &=\frac14\overline{\mathcal{B}}^{(2)}_{rn}\Delta^{[\alpha}_{[\sigma}\Delta^{\beta]\gamma}_{\eta]\zeta}\;,
\end{alignat}
\end{subequations}
which then gives rise to the respective terms in Eqs. \eqref{eqs:C_1_full}.
The respective coefficients are subsequently obtained as
\begin{alignat}{2}
\B^{(0)}_{rn}&=\frac13 \Delta^\nu_\mu (\B^{(00)}_{rn})^\mu_\nu \;,\qquad 
&&\B^{(02)}_{rn}= \frac15 \Delta^{\nu_2}_\mu \Delta^{\nu_1\nu} (B^{(02)}_{rn})^\mu_{\nu,\nu_1\nu_2}\;,\nonumber\\
\B^{(20)}_{rn}&=\frac13 \Delta^\nu_\mu \Delta_{\mu_1\mu_2} (\B^{(20)}_{rn})^{\mu_1,\mu\mu_2}_\nu \;,\qquad 
&&\B^{(2)}_{rn}= \frac15 \Delta^{\nu_2}_\mu \Delta^{\nu_1\nu} \Delta_{\mu_1\mu_2}(\B^{(22)}_{rn})^{\mu_1,\mu\mu_2}_{\nu,\nu_1\nu_2}\;,\nonumber\\
\B^{(1)}_{rn}&= \frac16 \Delta^{[\nu}_\mu \Delta^{\nu_1]}_{\mu_1} (\B^{(11)}_{rn})^{\mu,\mu_1}_{\nu,\nu_1} \;,\qquad 
&&\overline{\B}^{(2)}_{rn}= \frac{3}{40} \Delta^{[\nu}_{[\mu} \Delta^{\nu_1]\nu_2}_{\mu_1]\mu_2}(\B^{(22)}_{rn})^{\mu,\mu_1\mu_2}_{\nu,\nu_1\nu_2}\;.
\end{alignat}

In the following, I concentrate on the nonlocal contributions  \eqref{eq:C_0_def}, which can be treated in a similar way.
In particular, it can be seen that the only axial vector which may enter the expression is given by $\omega_0^\mu+\beta_0 \omega^\mu$. On the other hand, under the approximation of omitting dissipative contributions to the nonlocal part of the collision term, the only possible vectors are $I^\mu$ and $\kappa_0^\mu+\beta_0 \dot{u}^\mu$. Note that here the pressure gradient $F^\mu$ does not appear because it can be expressed through the acceleration $\dot{u}^\mu$, using the equation of motion \eqref{eq:eom_u}, with the dissipative terms being neglected. For the same reason, the vector-valued quantity arising from the symmetrized derivative of the inverse four-temperature, $u_\mu \Delta^\nu_\lambda \partial^{(\mu}\beta_0 u^{\lambda)}=\beta_0\dot{u}^\nu + \nabla^\nu \beta_0 \simeq I^\mu/h_0$, can be expressed through the gradient of chemical potential over temperature.
Lastly, the only traceless symmetric tensor that may appear is given by $\beta_0\sigma^{\mu\nu}$.
Defining the relevant projections $C_{n\ell,t,r-1}^\mu$, $C_{n\ell,w,r-1}^\mu$, and $C_{n\ell,t,r-1}^{\mu\nu}$ as before, they can be written as
\begin{subequations}
\begin{align}
C_{n\ell,r-1}^{\langle\mu\rangle} &=  X^{\mu\nu}_{r}\left(\omega_{0,\nu}+\beta_0 \omega_\nu\right) \;,\\
C_{n\ell,t,r-1}^{\mu} &=X_{t,r}^{\mu\nu}\left(\omega_{0,\nu}+\beta_0 \omega_\nu\right)		\;,\\
C_{n\ell,w,r-1}^{\mu} &=	X_{w,r}^{\mu\nu}\left(\kappa_{0,\nu}+\beta_0 \dot{u}_\nu\right)+ Y^{\mu\nu}_{w,r} I_\nu	\;,\\
C_{n\ell,t,r-1}^{\mu\nu} &=	X_{t,r}^{\mu\nu,\alpha\beta}\beta_0 \sigma_{\alpha\beta}	\;,
\end{align}
\end{subequations}
with the definitions
\begin{subequations}
\begin{align}
X^{\mu\nu}_r &\coloneqq \frac{\sigma\hbar}{m}\int_f \W E_\k^{r-1} \s^{\langle\mu\rangle}u_\alpha \left(k_1^{[\alpha}\s_1^{\nu]}+k_2^{[\alpha}\s_2^{\nu]}-k^{[\alpha}\s^{\nu]}-k'^{[\alpha}\s'^{\nu]}\right)  \;,\\
X_{t,r}^{\mu\nu} & \coloneqq \frac{\sigma\hbar}{m}\int_f \W E_\k^{r-1} \s_\beta k^{\langle\beta}k^{\mu\rangle}   u_\alpha \left(k_1^{[\alpha}\s_1^{\nu]}+k_2^{[\alpha}\s_2^{\nu]}-k^{[\alpha}\s^{\nu]}-k'^{[\alpha}\s'^{\nu]}\right)\;,\\
X_{w,r}^{\mu\nu} & \coloneqq -\frac{\sigma\hbar}{m}\epsilon^{\mu\lambda\alpha\beta}u_\lambda \int_f \W E_\k^{r-1} \s_\alpha k_\beta \epsilon^{\gamma\delta\rho\nu}u_\rho \left(k_{1,\gamma}\s_{1,\delta}+k_{2,\gamma}\s_{2,\delta}-k_\gamma \s_\delta- k'_\gamma \s'_\delta\right) \;,\\
Y_{w,r}^{\mu\nu} & \coloneqq \epsilon^{\mu\lambda\alpha\beta}u_\lambda \int_f \W E_\k^{r-1} \s_\alpha k_\beta \left[\Delta_1^\nu+\Delta_2^\nu-\Delta^\nu-\Delta'^\nu-\frac{1}{h_0} \left(\Delta_1^{\nu}E_{\k_1}+\Delta_2^{\nu}E_{\k_2}-\Delta^{\nu}E_{\k}-\Delta'^{\nu}E_{\k'}\right)\right]\;,\label{eq:def_Ymunu}\\
X_{t,r}^{\mu\nu,\alpha\beta} & \coloneqq -\Delta^{\mu\nu,\rho\sigma} \epsilon_\sigma{}^{\gamma\delta\lambda}u_\lambda \int_f \W E_\k^{r-1}\s_\gamma k_{\langle\delta}k_{\rho\rangle} \left(\Delta_1^\alpha k_1^\beta+\Delta_2^\alpha k_2^\beta-\Delta^\alpha k^\beta- \Delta'^\alpha k'^\beta\right)\;.
\end{align}
\end{subequations}
Note that in order to arrive at Eq. \eqref{eq:def_Ymunu} the conservation of the total angular momentum $J^{\mu\nu}=\sigma \hbar \Sigma_\s^{\mu\nu} + \Delta^{[\mu} k^{\nu]}$ has been used.
Then, the only possible structures of these tensors are
\begin{equation}
X^{\mu\nu}_r = g_r^{(0)}\Delta^{\mu\nu}\;,\quad 
X^{\mu\nu}_{t,r} = g_r^{(2)}\Delta^{\mu\nu}\;,\quad
X^{\mu\nu}_{w,r} = g_r^{(\kappa)}\Delta^{\mu\nu}\;,\quad
Y^{\mu\nu}_{w,r} = g_r^{(I)}\Delta^{\mu\nu}\;,\quad
X^{\mu\nu,\alpha\beta}_{t,r} = h_r^{(2)}\Delta^{\mu\nu,\alpha\beta}\;,
\end{equation}
leading to the first terms in Eqs. \eqref{eqs:C_1_full}. The coefficients are subsequently extracted as
\begin{align}
g_r^{(0)}=\frac13 \Delta_{\mu\nu} X_r^{\mu\nu}\;,\quad
g_r^{(2)}=\frac13 \Delta_{\mu\nu} X_{t,r}^{\mu\nu}\;,\quad
g_r^{(\kappa)}=\frac13 \Delta_{\mu\nu} X_{w,r}^{\mu\nu}\;,\quad
g_r^{(I)}=\frac13 \Delta_{\mu\nu} Y_{w,r}^{\mu\nu}\;,\quad
h_r^{(2)}=\frac15 \Delta_{\mu\nu,\alpha\beta} X_{t,r}^{\mu\nu,\alpha\beta}\;.
\end{align}

\section{Transport coefficients}
\label{app:coeff}
\begingroup
\allowdisplaybreaks
The transport coefficients appearing in the term $\mathcal{J}_\omega^\mu$ in Eq. \eqref{eq:def_Jomega} and thus in the equation of motion for $\omega_0^\mu$ \eqref{eq:eom_omega_hydro} read
\begin{subequations}
\begin{align}
\tau_\omega &\coloneqq \T_{A}^{(\omega)} \frac{2\sigma^2 }{gm^2}(J_{30}-J_{31})-\sum_{n=0}^{N_0^{(1)}} \T_{A,n}^{(\omega\tau)} \frac{2\sigma\hbar}{gm} J_{n+1,0} \;,\\
\ell_{\omega \kappa} &\coloneqq -\T_{A}^{(\omega)} \frac{2\sigma^2}{gm^2}J_{31}+\sum_{n=0}^{N_0^{(1)}} \T_{A,n}^{(\omega\tau)}\frac{2\sigma \hbar}{gm}J_{n+1,1}  + \sum_{n=0}^{N_2^{(1)}}\T_{A,n}^{(\omega t)}\frac{10\sigma \hbar}{3gm}K_{n+3,2} \;,\\
\tau_{\omega \kappa} &\coloneqq \T_{A}^{(\omega)} \frac{2\sigma^2}{gm^2}\left(3J_{31}-\beta_0 K_{41}\right)+\sum_{n=0}^{N_0^{(1)}} \T_{A,n}^{(\omega\tau)}\frac{2\sigma \hbar}{gm}J_{n+1,0}=-\tau_\omega \;,\\
\lambda_{\omega \kappa} &\coloneqq -\T_{A}^{(\omega)} \frac{2\sigma^2}{gm^2}\left(K_{31}-\frac{K_{41}}{h_0}\right)+\sum_{n=0}^{N_0^{(1)}} \T_{A,n}^{(\omega\tau)}\frac{2\sigma \hbar}{gm}\left(K_{n+1,1}-\frac{K_{n+2,1}}{h_0}\right) + \sum_{n=0}^{N_2^{(1)}}\T_{A,n}^{(\omega t)}\frac{10\sigma \hbar}{3gm}\left(K_{n+3,2}-\frac{K_{n+4,2}}{h_0}\right)\;,\\
\ell_{\omega n} &\coloneqq  \sum_{n=0}^{N_0^{(1)}} \T_{A,n}^{(\omega\tau)}\frac{B_{n-1}}{2} + \sum_{n=0}^{N_2^{(1)}}\T_{A,n}^{(\omega t)}\frac{B_{n+1}-m^2 B_{n-1}}{6}\;,\\
\tau_{\omega n} &\coloneqq -\sum_{n=0}^{N_0^{(1)}} \T_{A,n}^{(\omega\tau)}\left(\frac12 \frac{\partial B_{n-1}}{\partial \ln \beta_0}+n\frac{B_{n-1}}{2}\right)  + \sum_{n=0}^{N_2^{(1)}}\T_{A,n}^{(\omega t)}\frac16 \left[nm^2B_{n-1}-(n+5)B_{n+1}-\frac{\partial B_{n+1}}{\partial \ln \beta_0}+m^2 \frac{\partial B_{n-1}}{\partial \ln \beta_0}\right]\;,\\
\lambda_{\omega n} &\coloneqq \sum_{n=0}^{N_0^{(1)}} \T_{A,n}^{(\omega\tau)}\frac12\left(\frac{\partial }{\partial \alpha_0}+\frac{1}{h_0}\frac{\partial }{\partial \beta_0}\right)B_{n-1}  + \sum_{n=0}^{N_2^{(1)}}\T_{A,n}^{(\omega t)}\frac16\left(\frac{\partial}{\partial \alpha_0}+\frac{1}{h_0}\frac{\partial}{\partial \beta_0}\right)\left(B_{n+1}-m^2 B_{n-1}\right) \;,\\
\delta_{\omega \omega} &\coloneqq -\T_{A}^{(\omega)} \frac{2\sigma^2}{gm^2}\left[\left(K_{30}-K_{31}\right)\mathcal{H}-\left(K_{40}-K_{41}\right)\overline{\mathcal{H}}+J_{30}-\frac{J_{31}}{3}\right]\nonumber\\
&\quad +\sum_{n=0}^{N_0^{(1)}} \T_{A,n}^{(\omega\tau)}\frac{2\sigma \hbar}{gm}\left[K_{n+1,0}\mathcal{H}-K_{n+2,0}\overline{\mathcal{H}}+J_{n+1,0}+\left(n-\frac{2}{3}\right)J_{n+1,1}\right] - \sum_{n=0}^{N_2^{(1)}}\T_{A,n}^{(\omega t)}\frac{20\sigma \hbar}{9gm}K_{n+3,2} \;,\\
\lambda_{\omega \omega} &\coloneqq \T_{A}^{(\omega)} \frac{2\sigma^2}{gm^2}J_{31} +\sum_{n=0}^{N_0^{(1)}} \T_{A,n}^{(\omega\tau)}\frac{2\sigma \hbar}{gm}J_{n+1,1}+ \sum_{n=0}^{N_2^{(1)}}\T_{A,n}^{(\omega t)}\frac{4\sigma \hbar}{gm}\left(\frac56 K_{n+3,2}-\beta_0 K_{n+4,2}\right)\;,\\
\lambda_{\omega \t} &\coloneqq \T_{A}^{(\omega)} \frac{\sigma}{m\hbar}  - \sum_{n=0}^{N_2^{(1)}}\T_{A,n}^{(\omega t)}D_n\;.
\end{align}
\end{subequations}
The coefficients that are contained in the term $\mathcal{J}_\kappa^\mu$ in Eq. \eqref{eq:def_Jkappa} and thus occur in the equation of motion for $\kappa_0^\mu$, Eq. \eqref{eq:eom_kappa_hydro}, are defined as
\begin{subequations}
\begin{align}
\mathfrak{b} &\coloneqq \T_{V}^{(\kappa)}\widetilde{\Gamma}^{(I)}+\sum_{n=0}^{N_1^{(1)}}  \T_{V,n}^{(\kappa w)} g_n^{(I)} \;,\\
\tau_\kappa &\coloneqq \T_{V}^{(\kappa)} \frac{4\sigma^2}{gm^2}J_{31}+\sum_{n=0}^{N_1^{(1)}}  \T_{V,n}^{(\kappa w)}\frac{4\sigma \hbar}{gm}J_{n+2,1} \;,\\
\ell_{\kappa\omega} &\coloneqq \T_{V}^{(\kappa)} \frac{2\sigma^2}{gm^2}J_{31}+\sum_{n=0}^{N_1^{(1)}}  \T_{V,n}^{(\kappa w)}\frac{2\sigma \hbar}{gm}J_{n+2,1}=\frac{\tau_\kappa}{2}\;,\\
\tau_{\kappa\omega} &\coloneqq -\T_{V}^{(\kappa)} \frac{2\sigma^2}{gm^2}\left(J_{30}-\beta_0 K_{41}\right)+\sum_{n=0}^{N_1^{(1)}}  \T_{V,n}^{(\kappa w)}\frac{4\sigma \hbar}{gm}J_{n+2,1}=\tau_\kappa\;,\\
\lambda_{\kappa\omega} &\coloneqq \T_{V}^{(\kappa)} \frac{2\sigma^2}{gm^2}\left(K_{31}-\frac{K_{41}}{h_0}\right)+\sum_{n=0}^{N_1^{(1)}}  \T_{V,n}^{(\kappa w)}\frac{2\sigma \hbar}{gm}\left(K_{n+2,1}-\frac{K_{n+3,1}}{h_0}\right)\;,\\
\delta_{\kappa\kappa} &\coloneqq -\T_{V}^{(\kappa)} \frac{4\sigma^2}{gm^2}\left(K_{31}\mathcal{H}-K_{41}\overline{\mathcal{H}}+\frac43 J_{31}\right)-\sum_{n=0}^{N_1^{(1)}}  \T_{V,n}^{(\kappa w)}\frac{4\sigma \hbar}{gm}\left[K_{n+2,1}\mathcal{H}-K_{n+3,1}\overline{\mathcal{H}}+\frac13 \left(5\beta_0 K_{n+3,2}-J_{n+2,1}\right)\right]\;,\\
\lambda_{\kappa\kappa} &\coloneqq \T_{V}^{(\kappa)} \frac{2\sigma^2}{gm^2}J_{31}+\sum_{n=0}^{N_1^{(1)}}  \T_{V,n}^{(\kappa w)}\frac{2\sigma \hbar}{gm}\left(2\beta_0 K_{n+3,2}-J_{n+2,1}\right)\;,\\
\tau_{\kappa\kappa} &\coloneqq \T_{V}^{(\kappa)} \frac{2\sigma^2}{gm^2}J_{31}+\sum_{n=0}^{N_1^{(1)}}  \T_{V,n}^{(\kappa w)}\frac{2\sigma \hbar}{gm}J_{n+2,1}=\frac{\tau_\kappa}{2} \;,\\
\delta_{\kappa n} &\coloneqq \sum_{n=0}^{N_1^{(1)}}  \T_{V,n}^{(\kappa w)}\bigg\{\frac13\left[(n-1)m^2 B_{n-2}-(n+3)B_n\right]-\left(\mathcal{H}\frac{\partial}{\partial \alpha_0}+\overline{\mathcal{H}}\frac{\partial}{\partial \beta_0}\right)B_n\bigg\}\;,\\
\lambda_{\kappa n} &\coloneqq \sum_{n=0}^{N_1^{(1)}}  \T_{V,n}^{(\kappa w)}\frac{1}{10}\left[(2n-2)m^2 B_{n-2}-(2n+3)B_n\right]\;,\\
\tau_{\kappa n} &\coloneqq \sum_{n=0}^{N_1^{(1)}}  \T_{V,n}^{(\kappa w)}\frac12B_n\;,\\
\ell_{\kappa n} &\coloneqq -\sum_{n=0}^{N_1^{(1)}}  \T_{V,n}^{(\kappa w)}B_n\;,\\
\tau_{\kappa \t} &\coloneqq \T_{V}^{(\kappa)} \frac{\sigma}{\hbar m}+\sum_{n=0}^{N_1^{(1)}}  \T_{V,n}^{(\kappa w)} \left(n D_{n-1}+\frac{\partial D_{n-1}}{\partial \ln \beta_0}\right)\;,\\
\lambda_{\kappa \t} &\coloneqq -\sum_{n=0}^{N_1^{(1)}}  \T_{V,n}^{(\kappa w)} \left(\frac{\partial}{\partial \alpha_0}+\frac{1}{h_0}\frac{\partial }{\partial \beta_0}\right)D_{n-1}\;,\\
\ell_{\kappa \t} &\coloneqq -\T_{V}^{(\kappa)} \frac{\sigma}{\hbar m}-\sum_{n=0}^{N_1^{(1)}}  \T_{V,n}^{(\kappa w)} D_{n-1}\;.
\end{align}
\end{subequations}
Lastly, the coefficients appearing in the term $\mathcal{J}_\t^{\mu\nu}$ in Eq. \eqref{eq:def_Jt} and the equation of motion for $\t^{\mu\nu}$ \eqref{eq:eom_t_hydro} read
\begin{subequations}
\begin{align}
\mathfrak{d} &\coloneqq \sum_{n=0}^{\overline{N}_2^{(1)}} \T^{(t)}_{T,0n} h_n^{(2)}\;,\\
\tau_\t &\coloneqq \sum_{n=0}^{\overline{N}_2^{(1)}} \T^{(t)}_{T,0n} D_n\;,\\
\delta_{\t\t} &\coloneqq \sum_{n=0}^{\overline{N}_2^{(1)}} \T^{(t)}_{T,0n}\left\{\frac13\left[(n-1)m^2 D_{n-2}-(n+4)D_n\right]-\left(\mathcal{H}\frac{\partial}{\partial \alpha_0}+\overline{\mathcal{H}}\frac{\partial}{\partial \beta_0}\right)D_n\right\}\;,\\
\lambda_{\t\t} &\coloneqq \sum_{n=0}^{\overline{N}_2^{(1)}} \T^{(t)}_{T,0n}\frac17\left[2(n-1)m^2 D_{n-2}-(2n+5)D_n\right]\;,\\
\tau_{\t\t} &\coloneqq \sum_{n=0}^{\overline{N}_2^{(1)}} \T^{(t)}_{T,0n}\frac53 D_n=\frac{5}{3}\tau_\t\;,\\
\ell_{\t\kappa} &\coloneqq \sum_{n=0}^{\overline{N}_2^{(1)}} \T^{(t)}_{T,0n}\frac{6\sigma \hbar}{gm}K_{n+3,2}\;,\\
\lambda_{\t\kappa} &\coloneqq \sum_{n=0}^{\overline{N}_2^{(1)}} \T^{(t)}_{T,0n}\frac{6\sigma \hbar}{gm}\left(K_{n+3,2}-\frac{K_{n+4,2}}{h_0}\right)\;,\\
\ell_{\t n} &\coloneqq \sum_{n=0}^{\overline{N}_2^{(1)}} \T^{(t)}_{T,0n}\frac{3}{10}\left(B_{n+1}-m^2 B_{n-1}\right)\;,\\
\tau_{\t n} &\coloneqq \sum_{n=0}^{\overline{N}_2^{(1)}} \T^{(t)}_{T,0n}\frac{3}{10}\left[nm^2 B_{n-1}-(n+5)B_{n+1}-\frac{\partial}{\partial \ln \beta_0}\left(B_{n+1}-m^2 B_{n-1}\right)\right]\;,\\
\lambda_{\t n} &\coloneqq \sum_{n=0}^{\overline{N}_2^{(1)}} \T^{(t)}_{T,0n}\left(\frac{\partial}{\partial \alpha_0}+\frac{1}{h_0}\frac{\partial}{\partial \beta_0}\right)\left(B_{n+1}-m^2 B_{n-1}\right)\;,\\
\tau_{\t \omega} &\coloneqq \sum_{n=0}^{\overline{N}_2^{(1)}} \T^{(t)}_{T,0n}\frac{6\sigma \hbar}{gm}K_{n+3,2}\;,\\
\lambda_{\t \omega} &\coloneqq \sum_{n=0}^{\overline{N}_2^{(1)}} \T^{(t)}_{T,0n}\frac{4\sigma \hbar}{gm}\left(\frac32 K_{n+3,2}-\beta_0 K_{n+4,2}\right)
\;.
\end{align}
\end{subequations}
\endgroup

\bibliographystyle{apsrev}
\bibliography{bib_full,bib_spinwaves}

\end{document}